\documentclass[%
aps,
twocolumn, prd,
showpacs,preprintnumbers,nofootinbib,
amsmath,amssymb,
aps,
10pt,
superscriptaddress]{revtex4-1}

\RequirePackage{lineno}
\usepackage{float}
\usepackage{graphicx}
\usepackage{dcolumn}
\usepackage{bm}
\usepackage{color}
\usepackage{array}
\usepackage{multirow}
\usepackage[normalem]{ulem}
\usepackage{epstopdf}
\usepackage{hyperref}
\hypersetup{
    colorlinks=true,
    linkcolor=blue,
    filecolor=magenta,      
    urlcolor=cyan,
}

\renewcommand{\Re}{\operatorname{\mathbb{R}e}}
\renewcommand{\Im}{\operatorname{\mathbb{I}m}}

\begin{document}

\title{Searching for the full symphony of black hole binary mergers}

\author{Ian~Harry} 
\email{ian.harry@aei.mpg.de}
\affiliation{Max Planck Institute for Gravitational Physics (Albert Einstein Institute),
  Am M\"uhlenberg 1, D-14476 Potsdam-Golm, Germany}
  
\author{Juan Calder\'on~Bustillo} 
\email{juan.bustillo@physics.gatech.edu}
\affiliation{Center for Relativistic Astrophysics and School of Physics,
             Georgia Institute of Technology, Atlanta, GA 30332}

\author{Alexander~Nitz} 
\email{alex.nitz@aei.mpg.de}
\affiliation{Max-Planck-Institut f¨ur Gravitationsphysik, Albert-Einstein-Institut, D-30167
             Hannover, Germany}


\begin{abstract}
Current searches for the gravitational-wave signature of compact binary mergers rely on matched-filtering
data from interferometric observatories with sets of modelled gravitational waveforms. These searches
currently use model waveforms that do not include the higher-order mode content of the gravitational-wave signal.
Higher-order modes are important for many compact binary mergers and their omission reduces the sensitivity
to such sources. In this work we explore the sensitivity loss incurred from omitting higher-order modes.
We present a new method for searching for compact binary mergers using waveforms that include higher-order mode
effects, and evaluate the sensitivity increase that using our new method would allow. We find that, when 
evaluating sensitivity at a constant rate-of-false alarm, and when including the fact that signal-consistency
tests can reject some signals that include higher-order mode content, we observe a sensitivity increase of up
to a factor of 2 in volume for high mass ratio, high total-mass systems. For systems with equal mass, or with
total mass $\sim 50 M_{\odot}$, we see more modest sensitivity increases, $< 10\%$, which indicates that the
existing search is already performing well. Our new search method is also directly
applicable in searches for generic compact binaries.
\end{abstract}

\maketitle

\section{Introduction}

The Advanced LIGO gravitational-wave observatories~\cite{TheLIGOScientific:2014jea} have observed
multiple black hole binary mergers in Advanced LIGO's first two observing
runs~\cite{Abbott:2016blz,Abbott:2016nmj,Abbott:2017vtc}. Many
additional black hole binary mergers are expected to be observed in the coming years~\cite{Abbott:2017vtc},
with additional detectors in Italy~\cite{TheVirgo:2014hva}, Japan~\cite{Aso:2013eba} and India~\cite{LigoIndia}
helping to improve coverage of the gravitational-wave sky~\cite{Aasi:2013wya}.
The continued observation of black hole binary mergers will allow for a better understanding
of the rate of such mergers~\cite{Abbott:2016nhf}, and give a sense of the mass and component spin distribution of
black hole binary systems~\cite{TheLIGOScientific:2016wfe,TheLIGOScientific:2016pea}.
This will in turn allow a better understanding of how such systems
form~\cite{AmaroSeoane:2009ui,AmaroSeoane:2012tx,Gair:2010dx,Sesana:2009wg,Volonteri:2012ig,Volonteri:2012tp}.

Searches for compact binary mergers rely on matched filtering the data taken from
gravitational-wave observatories with theoretical filter
waveforms~\cite{Nitz:2017svb,Usman:2015kfa,Canton:2014ena,Cannon:2011vi,Cannon:2012zt}. The set of filter
waveforms is chosen such that a signal occurring anywhere within the parameter space of interest
can be recovered well by at least one of the waveforms in the set of
filters~\cite{Sathyaprakash:1991mt,Babak:2006ty,Ajith:2012mn,Brown:2012qf,Capano:2016dsf}.
It is critical that the waveform models being used as filters are accurate representations
of the signals that will be produced from compact binary mergers in the Universe. Much work
in the years leading up to Advanced LIGO's first discovery focused on modelling waveforms using numerical \cite{Pretorius:2005gq,Szilagyi:2009qz,Scheel:2008rj, Frauendiener:2011zz, Hannam:2010ec,ChuThesis,Vaishnav:2007nm, Herrmann:2006ks,Healy:2009zm}
and analytical~\cite{Blanchet:1995ez,Buonanno:1998gg} techniques, and on combining these methods together to create waveform models
accurate at all stages of the merger \cite{Taracchini:2012ig,Bohe:2016gbl,Ajith:2007kx,Khan:2015jqa} .

However, a number of assumptions are made about the emitted gravitational-wave signal to simplify
the search and reduce the search parameter space. Specifically, current
searches for compact binary mergers neglect any affect due to precession of the orbital plane~\cite{Apostolatos:1994mx},
orbital eccentricity~\cite{Damour:2004bz} or neutron-star equation-of-state~\cite{Flanagan:2007ix,Read:2013zra}.
Current searches also neglect the affect
of the so-called higher-order modes of gravitational-wave emission~\cite{Blanchet2008}, and it is on
this topic that we will focus in this manuscript. Making these simplifying assumptions does not affect the ability
to observe the majority of compact binary mergers, as evident from the current observations, but can
mean that the detection efficiency is not optimal. It would also induce an observational
bias against compact binary mergers that are not well described by these assumptions,
and these kind of systems can be the ones of most value, astrophysically, as the additional information
from features such as higher-order modes allows more precise measurement of the various source
parameters~\cite{OShaughnessy:2014shr}.

Several studies have shown that the omission of higher-order modes in searches can lead to a reduction
in detection rate~\cite{McKechan:2011ps, Pekowsky:2012sr, Capano:2013raa,Varma:2014jxa,Bustillo:2015qty,Bustillo:2016gid}.
Specifically, the omission of higher modes is likely to reduce the detection of systems 
with large mass ratio $q=m_1/m_2 \geq 4$ and large total mass
$M=m_1+m_2 > 100M_{\odot}$ \cite{Pekowsky:2012sr, Capano:2013raa,Varma:2014jxa,Bustillo:2015qty,Bustillo:2016gid}. To date Advanced 
LIGO has only detected systems with mass ratios and total masses lower than these values. However, we still
know very little about the mass distribution of compact binary mergers and searches should
be capable of observing any possible system. For such high-mass systems, more loosely modelled search
techniques~\cite{Klimenko:2008fu,Sutton:2009gi,Cornish:2014kda,Lynch:2015yin,Klimenko:2015ypf}
also offer the ability to observe short duration compact binary mergers~\cite{Mazzolo:2014kta}.
For very high-mass systems the sensitivity
of such searches can become comparable to that of modelled search methods, and could potentially exceed the
sensitivity of modelled searches in the case that the system is not well described by the model.
Nevertheless, in cases where the waveform model is well understood, searches that use that knowledge
\emph{should} be more sensitive than those that don't.

In this study, we present the first end-to-end modelled search method for black hole binary mergers using filter waveforms that
include the higher-order modes of the gravitational-wave signals and demonstrate the improvement in sensitivity
that can be obtained when using this method to search for compact binary mergers in Advanced LIGO data.
Our method involves including the source
orientation angles in the list of parameters that are sampled over when creating a set of filter waveforms
to use in the search. At the time of writing, the only waveform model that is available including both
the full inspiral, merger and ringdown components of the waveform \emph{and} including higher-order modes,
is the non-spinning effective-one-body model presented in \cite{Pan:2011gk}. We must therefore restrict
ourselves in this work to only considering waveforms that do not include the effects of the components' spins.
However, the methods we describe here are directly applicable to the case of aligned-spin waveform models,
which are now in an advanced stage of development~\cite{London:2017bcn,COTESTA}.
Indeed the search method presented here is fully generic, and could be run on eccentric,
precessing waveforms including higher-order modes.

The layout of this paper is as follows.
In section \ref{sec:paramranges} we motivate the parameter
range that we choose to consider in this work.
In section \ref{sec:waveforms} we give a brief reminder of how
the presence of higher-order modes will affect the emitted gravitational-wave signal, and discuss the
waveform models used in this work. In section \ref{sec:homsearch} we introduce the fully
generic search method that we will use. In section \ref{sec:sensitivity} we assess the benefit of
deploying a search including the higher-order mode components of the gravitational-wave signal.
In section \ref{sec:realdataproblems} we also explore how signal-based consistency tests, necessary
in searches of real data to distinguish real signals from instrumental transients, can sometimes
falsely reject real gravitational-wave signals containing higher-order modes, and how this problem
is alleviated when using higher-order mode waveforms as filters. Finally we conclude in section
\ref{sec:conclusions}.

\section{Parameter space considerations}
\label{sec:paramranges}

In this section we motivate and describe the black hole binary parameter space that we will use in the rest
of this work. While we will use a specific parameter space here, we stress again that the search methods we
will describe can be applied for any parameter space of interest.

It was recently reported that no
gravitational-wave signals were observed in a search of Advanced LIGO data targeting
``intermediate mass black hole binaries'', which are defined to be black hole binaries with total mass
$M \geq 100 M_{\odot}$~\cite{TheLIGOScientific:2016wfe}.

These sources are important from an astrophysical point of view. They are proposed to
be precursors of supermassive black holes in some hierarchical formation
scenarios~\cite{AmaroSeoane:2009ui,AmaroSeoane:2012tx,Gair:2010dx,Sesana:2009wg,Volonteri:2012ig,Volonteri:2012tp}.
However, there is not yet evidence for their existence. Detection of higher-order modes would allow 
for more detailed tests of General Relativity in the strong field regime. Examples of this include studies
of the quasi-normal ringdown modes \cite{Gossan:2011ha,Yang:2017zxs} and studies evaluating the mass
of the graviton via searching for a dispersion relation in the speed of propagation of gravitational-waves~\cite{Berti:2015itd}.
For the kind of sources considered in that work, higher-modes are believed to have a significant impact on
gravitational-wave signals~\cite{Pekowsky:2012sr,Varma:2014jxa,Bustillo:2015qty,Bustillo:2016gid}.

The effect of higher-order modes was not studied in the search for
``intermediate mass black hole binaries'' reported in \cite{TheLIGOScientific:2016wfe}; both the waveforms
used in the search, and the simulations to assess its sensitivity, did not include higher-order modes. It is
therefore interesting to explore the same parameter space here and assess whether neglecting higher-order
modes is a fair assumption in such studies, and to demonstrate the sensitivity increase that is possible if
higher-order modes are included. The matched-filter search used in~\cite{TheLIGOScientific:2016wfe} targeted
black hole binary mergers with total mass between $50 M_{\odot}$ and $600M_{\odot}$. At $600 M_{\odot}$ compact binary
mergers emit gravitational-wave signals that are for the most part too low frequency to be observed by Advanced
LIGO and the sensitive distance rapidly decreases.
It is also challenging to distinguish such signals from non-Gaussianities in the detector noise. Here we
choose to use a maximum total mass of $400 M_{\odot}$. Our constraints on the mass ratio, $q$, are limited by
constraints on the waveform model we use, which we will discuss in the next section.
We use a limit of $q \leq 10$, which also matches the limits chosen in~\cite{TheLIGOScientific:2016wfe}.
When we discuss masses in this work we will always refer to the masses of the signal observed by the observatory,
often referred to as ``detector frame masses''. Sources at cosmological distances will be redshifted with respect
to the observer, causing the signal to appear to have higher masses than the actual ones measured in the ``source frame''.

The sensitivity of Advanced LIGO has been improving since the beginning of Advanced LIGO's first observing run
and will continue to improve over the next years, before reaching its design sensitivity.
In order to obtain reasonable estimates
of the improvements derived from our search method, we will use two noise curves in this study. We will use a representative
measurement of the sensitivity curve from Advanced LIGO's first observing run (O1)~\cite{O1sens} and we will use Advanced LIGO's
``zero-detuned high-power'' design sensitivity curve~\cite{ZDHP}. For the former, we
set the lower frequency of our matched-filter to $f_{low}=20$Hz while for the latter we use $f_{low}=10$Hz.

\section{Gravitational-wave signals including higher-order modes}
\label{sec:waveforms}

The gravitational-wave emission from a non eccentric black hole binary merger depends on 15 parameters.
The individual masses, $m_i$, and dimensionless angular momenta (spins), $\vec\chi_i=\vec{s}_i/m_i$, of its two components
are parameters intrinsic to the source---collectively denoted $\Xi_i$. The time of the coalescence, measured in the
frame of the observer, is denoted $t_c$.
The remaining parameters describe the location and orientation of the observer, with respect to the source.
Consider a frame of reference in standard spherical coordinates $(D,\iota,\varphi)$ with origin in the
center of mass of the black hole binary. The polar angle $\iota$
is defined such that $\iota=0$ coincides with the total angular momentum of the binary\footnote{The origin of the $\varphi$
parameter is chosen to lie in the line connecting the two components at some fiducial time.}. The remaining 3
parameters are the sky-location of the source in the frame of the observer
$(\theta,\phi)$, and the polarisation $\psi$ of the signal.
A black hole binary is said to be ``face-on'' if $\iota=0$ or $\iota=\pi$ and ``edge-on'' if $\iota=\pi/2$.

For a generic gravitational-wave source observed by an interferometric detector
the observed strain $h(t)$ can be expressed as the sum of the two gravitational-wave polarizations weighted by the
sensitivity of the observer to each polarization
\begin{equation}
 h(t) = F_+(\theta,\phi,\psi) h_+(t) + F_{\times}(\theta,\phi,\psi) h_{\times}(t).
\end{equation}
Here $F_+$ and $F_{\times}$ denote the response function of the detector to each polarization~\cite{Finn:1992xs, Jaranowski:1998qm}.
The gravitational-wave polarizations $h_+(t)$ and $h_{\times}(t)$ can be expressed as
\begin{equation}
h_+(t) + i h_{\times}(t) = \sum_{\ell\geq 2}\sum_{m=-\ell}^{m=\ell}Y^{-2}_{\ell,m}(\iota,\varphi)h_{\ell,m}(t),
\label{gwmodes}
\end{equation}
where $Y^{-2}_{lm}$ denote the spherical harmonics of weight $-2$~\cite{sharm} and the $h_{\ell,m}(t)$ denote the various ``modes''
of gravitational-wave emission. For the case of compact binary mergers $h_{\ell,m}(t)$ will be a function of $\Xi_i$, $t_c$
and $D$ according to
\begin{equation}
 h_{\ell,m}(\Xi;t)=A_{\ell,m}(\Xi, D;t-t_c)e^{-i\upsilon_{\ell,m}(\Xi;t-t_c)}.
\end{equation}
Here $A_{\ell,m}$ is a real amplitude scaling for the various modes, and $\upsilon_{\ell,m}$ is a real
time-series giving the evolution of the phase of the various modes.

The black hole binaries detected by LIGO so far are characterized by a low mass ratio $q\leq 4$ and a total mass
$M < 100M_\odot$ \cite{TheLIGOScientific:2016wfe,Abbott:2016nmj,Abbott:2017vtc}. For such sources
the $(\ell,m)=(2,\pm 2)$ modes dominate the above sum for the vast majority
of the possible orientations of the source \cite{Pekowsky:2012sr}\footnote{We note that for systems with misaligned spins
the orbital plane precesses. Here the normal approach is to define $\iota$ and $\varphi$ in a stationary
source frame, and then if the orbital plane has precessed $\iota=0$ no longer corresponds to the direction of orbital angular
momentum and the \emph{other} $l=2$ modes can become dominant. One can alternatively consider
a source frame that tracks the precessing orbital angular momentum, and then $\iota$ and $\varphi$ will vary with time. In
this frame, the $(\ell,m)=(2,\pm 2)$ modes will again dominate the emitted gravitational-wave signal. We do not consider
precessing systems in this work.}. The rest of the modes, known as higher-order
modes, have only a small contribution during most of the inspiral and are only significant to the resulting
gravitational-wave signal
in the last few cycles and eventual merger of the black hole binary~\cite{Healy:2013jza,Bustillo:2015ova, Bustillo:2015qty}.
The amplitude of the higher-order modes grows as the mass ratio
$q$ of the system deviates from $1$, making their impact much stronger for large mass ratio black hole binaries. In addition,
the $Y^{-2}_{2,\pm 2}$ spherical
harmonics have maxima at $\iota=0$ and $\iota=\pi$ and a minimum at $\iota=\pi/2$. For many of the other harmonics
$\iota=\pi/2$ is a maximum. Therefore, the $(\ell,m)=
(2,\pm 2)$ modes will completely dominate the gravitational-wave signal
for face-on sources. For edge-on systems, especially ones with a high-mass ratio, the higher-order modes
are an important contribution to the full gravitational-wave
emission~\cite{McKechan:2011ps,Pekowsky:2012sr,Capano:2013raa,Varma:2014jxa,Bustillo:2015qty,Bustillo:2016gid}.
In addition, higher-order modes have a stronger affect in signals emitted by large total mass
sources. The phase of the $(\ell,m)$ mode scales, to good accuracy, as
$\upsilon_{\ell,m} \propto m \times \upsilon_{orb}/M$, where $\upsilon_{orb}$ denotes the orbital phase of the
binary. At high values of total mass, $M$, the dominant $(2, \pm2)$ modes can fall below the sensitive
band of the observatory, while the higher $m$ modes, at higher frequency, are still observable.

Waveform models that describe the full black-hole binary coalescence---through inspiral, merger and ringdown---can
be broadly divided into two approaches. The first is the ``effective-one-body'' approach, calibrated against numerical
relativity simulations~\cite{Buonanno:1998gg,Taracchini:2012ig,Damour:2012ky,Bohe:2016gbl}, the second is the various
phenomenological frameworks, also calibrated against numerical relativity
simulations~\cite{Santamaria:2010yb,Hannam:2013oca,Khan:2015jqa}. There are a number of different waveform models,
from both of these approaches, which have been used in the recent results papers from the LIGO and Virgo collaborations.
However, with the exception of numerically generated waveforms, which are currently impractical to use for searches with a wide
parameter space, these waveform models do not include the higher modes of the gravitational-wave emission, and consider solely
the dominant modes. The only waveform model available at the time of this study, which includes both higher-order modes
\emph{and} includes the merger and ringdown components of the gravitational-wave signal is an effective-one-body waveform
described in~\cite{Pan:2011gk}. This model includes the $(\ell,|m|)=(2,\pm 1),(2,\pm 2),(3,\pm 3),(4,\pm 4)$ and $(5,\pm 5)$
modes\footnote{This waveform model is known as EOBNRv2HM and is available in the LIGO Algorithm Library (LAL). We also use
a frequency-domain reduced-order model~\cite{Purrer:2014fza} of this waveform known as EOBNRv2HM\_ROM in LAL.}. The most
significant mode not included in this model is the $(3,\pm 2)$ mode, which can have comparable amplitude to the $(5,5)$
mode~\cite{Pan:2011gk,Healy:2013jza}. While the paper~\cite{Pan:2011gk} does demonstrate the accuracy of this model to generic
waveforms, it would of course be beneficial to have this, and other modes, included in future waveform models.
Unfortunately, this effective-one-body waveform model does not include the effect of the components' spins. Nevertheless this non-spinning
waveform model is sufficient to demonstrate the methodology described later in this work, and with it we can investigate
the sensitivity increase to non-spinning compact binary mergers if one searches with waveform filters that include higher-order modes.
We might expect that the relative sensitivity with non-spinning waveforms would be similar to that with aligned-spin
waveforms as the effects of spin inclusion and higher-order modes are largely orthogonal. However, it is true
that systems with anti-aligned spins have slightly stronger higher
modes than systems with aligned spins~\cite{Varma:2016dnf} and therefore it is possible that higher-modes would help more for 
negative spin sources than for positive ones. We will investigate this in the future when waveform models
are available to do so.

Finally, we note that during the writing of this manuscript a number of waveform models, in both the effective-one-body
and phenomenological frameworks are being developed that include higher-order modes, and allow for nonzero component
spins aligned with the orbital angular momentum~\cite{London:2017bcn,COTESTA}.
The methods described here can be applied directly to these waveform
models when they become available, and this would be a necessary step before utilising this methodology to search
for higher-order mode waveforms in real data. There is also work demonstrating that sets of numerical relativity
waveform might be used directly in a search~\cite{Kumar:2013gwa}, or that surrogate models could be created by interpolating between
a set of numerical relativity waveforms~\cite{Szilagyi:2015rwa,Blackman:2017dfb,Blackman:2017pcm}.
Such approaches might also present a way to use accurate aligned-spin, or even
precessing, higher-order mode waveforms in searches, but we do not explore that here.

\section{A modelled search for compact binary mergers with higher-order mode waveforms}
\label{sec:homsearch}

There are currently a number of different search methods being used to observe compact
binary coalescences using modelled waveforms in the data being collected by Advanced LIGO and Advanced
Virgo~\cite{Cannon:2011vi,Babak:2012zx,Cannon:2012zt,Canton:2014ena,Adams:2015ulm,Usman:2015kfa,Nitz:2017svb}.
The core of all of these different methods is the two-phase
matched-filter that was described in~\cite{Sathyaprakash:1991mt,Allen:2005fk}.
This two-phase matched-filter has proved to be very powerful in observing
compact binary mergers, but it does make a number of assumptions about the signal model, which
are not true generically, in particular when one is considering higher-order modes.
Specifically the method assumes that the normalized frequency domain representation
of the $+$ component of the gravitational wave signal $\tilde{h}_+$ is related to the frequency domain
representation of the $\times$ component of the gravitational wave signal $\tilde{h}_{\times}$ according
to $\tilde{h}_+ \propto i \tilde{h}_{\times}$. In addition, it is assumed that the ``extrinsic''
parameters of a gravitational-wave signal---the
sky-location, source orientation, polarization phase and distance---can all be absorbed by applying
a constant phase-shift, constant time-shift and a constant amplitude scaling to the observed waveform. With
these assumptions in place, one can analytically maximize over an overall amplitude and phase of the
signal, and use an inverse Fourier transform to quickly evaluate the statistic as a function of
time~\cite{Sathyaprakash:1991mt,Allen:2005fk}. Then
only the ``intrinsic'' parameters---the component masses and spins---are searched over by repeating
the search process with a well chosen discrete set of waveform models with varying values of the component
masses and spins, known as the ``template bank''. Physically, these assumptions hold if one assumes
that the sources being observed have no orbital eccentricity, no precession and no contribution from higher-order modes
to the gravitational-wave signal. However, these assumptions do not hold in the case here
where we wish to use waveforms including higher-order modes as filters in the search.

In \cite{Pan:2003qt,Harry:2016ijz} the authors explored relaxing the assumption that the system was not precessing
and developed search statistics that can be used in that case. In the method described in \cite{Pan:2003qt}
a complex maximization scheme was used to maximize over all non-intrinsic parameters, which was found to be computationally prohibitive if
forced to restrict to only physically possible values.
Whereas, in \cite{Harry:2016ijz} the authors included the inclination
of the source with respect to the observer as a parameter when constructing the template bank,
effectively considering this as an intrinsic parameter. However, this method cannot be applied to a generic
search because the assumption that the
$\varphi$ parameter (the azimuthal angle to the observer in the source frame) can
be modelled as an overall phase shift in the Fourier domain breaks down when considering
gravitational-wave signals with higher-order modes.

Nevertheless one can extend the method in \cite{Harry:2016ijz} in a reasonably trivial manner by relaxing
the assumption on $\varphi$ and also considering this as a parameter to
search over in the template bank. This is the approach we use in this work.
The resulting statistic is not new to this work, it also appears
in~\cite{Capano:2013raa, Schmidt:2014iyl}. This work, however, is the first case in which this has been applied in
an end-to-end search.

\subsection{A search statistic applicable for generic searches for compact binary mergers}

When searching for a signal $h$, with known form, but unknown amplitude, in Gaussian,
stationary noise $n$, with noise-power spectral
density $S_n(f)$ it can be demonstrated~\cite{Maggiore:1900zz,Creighton:2011zz} that
the optimal statistic for deciding whether a signal
$h$ is present, or not, in the data is given by
\begin{equation}
 \rho^2 \equiv \frac{\left( \Re \left[ \left\langle s | h \right\rangle \right] \right)^2}
    {\left\langle h | h \right\rangle} = \left(\Re [ \langle s | \hat{h} \rangle ]\right)^2,
\end{equation}
where $\rho$ defines a signal-to-noise ratio, $\hat{a}$ denotes a normalization of any filter
waveform $a$ such that
\begin{equation}
 \hat{a} = \frac{a}{\left\langle a | a \right\rangle^{\frac{1}{2}}},
\end{equation}
and we define the complex matched-filter
\begin{equation}
 \langle a|b\rangle = 4 \int^\infty_0 \frac{\tilde{a}(f)\tilde{b}^*(f)}{S_n(f)} df.
\end{equation}
For simplicity in what follows, we will distinguish between the complex matched-filter, and
the real component of the complex matched-filter by defining
\begin{equation}
\label{eq:inner_product}
 \left( a | b \right) = \Re \left[ \langle a|b\rangle \right],
\end{equation}
such that
\begin{equation}
 \label{eqn:matched_filter_snr_maxamp}
 \rho^2 = \left(\Re [ \langle s | \hat{h} \rangle ]\right)^2
 = ( s | \hat{h} )^2.
\end{equation}

As already mentioned in section \ref{sec:waveforms}, gravitational-wave signals observed
in an interferometric observatory such as LIGO or Virgo
can be expressed as a linear combination of the two gravitational-wave polarizations 
\begin{equation}
  h(t) = F_+(\theta, \phi, \psi)\, h_+(t) + F_{\times}(\theta, \phi, \psi)\,h_{\times}(t).
   \label{eqn:fplusfcross}
\end{equation}
As the amplitude of $h(t)$ is removed by normalization in equation \ref{eqn:matched_filter_snr_maxamp}
we can freely scale the amplitude when defining $h(t)$. It is convenient to combine $F_+$ and $F_{\times}$
into an overall amplitude rescaling and a single further parameter by defining
\begin{equation}
  \label{eq:udef}
  h(t) = A( u \hat{h}_+(t) + \hat{h}_{\times}(t)),
\end{equation}
where
\begin{equation}
  u = \frac{F_+}{F_{\times}}\sqrt{\frac{\left\langle \hat{h}_+ | \hat{h}_+ \right\rangle}
                                     {\left\langle \hat{h}_{\times} | \hat{h}_{\times} \right\rangle}}
\label{eq:u_def}                                     
\end{equation}
and
\begin{equation}
 A = F_{\times} \sqrt{\left\langle \hat{h}_{\times} | \hat{h}_{\times} \right\rangle}.
\end{equation}

One can then insert equation \ref{eq:udef} into equation \ref{eqn:matched_filter_snr_maxamp}, which removes
the amplitude term $A$, and from
there analytically maximize $\rho$ over $u$. This results in the following expression,
\begin{equation}
 \label{eqn:matched_filter_snr_hom}
 \max_{u} (\rho^2) = \frac{( s | \hat{h}_+ )^2 +
    ( s | \hat{h}_{\times} )^2 -
    2 ( s | \hat{h}_+ ) ( s | \hat{h}_{\times} ) 
      ( \hat{h}_+ | \hat{h}_{\times} )}
    {\left(1 - ( \hat{h}_+ | \hat{h}_{\times} )^2\right)}.
\end{equation}
Furthermore it is trivial to see that in the limit that $\tilde{h}_+ \propto i \tilde{h}_{\times}$ this will collapse to
the more familiar statistic used in current searches
\begin{equation}
\label{eqn:normal_matched_filter_snr}
\max_{u} (\rho^2) \simeq  \left\| \left\langle s | \hat{h}_+ \right\rangle \right\|^2.
\end{equation}

The statistic defined in equation \ref{eqn:matched_filter_snr_hom} is generic and can be applied
to any single detector search for compact binary coalescences. Physically, this statistic maximizes
over the $D$, $\theta$, $\phi$ and $\psi$ parameters---or the distance, sky location and polarization
phase---leaving all other parameters to be included in the template bank. For the case of eccentric,
precessing, higher-order mode waveforms, this will result in a very large dimension parameter space,
which may prove unfeasible in some situations. In such cases approaches such as the ones
explored in \cite{Pan:2003qt,Dhurandhar:2017rlr} might be useable to further shrink the dimensionality
of the parameter space by maximizing over the $Y^{-2}_{\ell,m}$ components, but this has yet to be successfully
applied to generic systems. However, as we explore below, our simple approach can successfully
be applied to the case of searching for higher-order mode signals in Advanced LIGO data.

\subsection{Exploring the necessity of the generic statistic for higher-order mode searches}

In equation \ref{eqn:matched_filter_snr_hom} we described a generic matched-filter statistic that
maximizes only over the amplitude, polarization phase and sky location of the signal. While this
statistic can be used generically, it is more computationally efficient to use the more commonly
used statistic in equation \ref{eqn:normal_matched_filter_snr} as it requires only one matched-filter
computation. Equation \ref{eqn:matched_filter_snr_hom} collapses to the form shown in
\ref{eqn:normal_matched_filter_snr} in the case when $\tilde{h}_+ \propto i \tilde{h}_{\times}$. It is therefore worth
investigating how well this relationship holds in the parameter space being considered to decide
whether it is possible to approximate equation \ref{eqn:matched_filter_snr_hom} with the more
efficient equation \ref{eqn:normal_matched_filter_snr}. It is also possible to use the more efficient
statistic in some part of the parameter space and swap over to the generic statistic only in the
regions of parameter space where it is needed.

\begin{figure}[t!bp]
\includegraphics[width=0.45\textwidth]{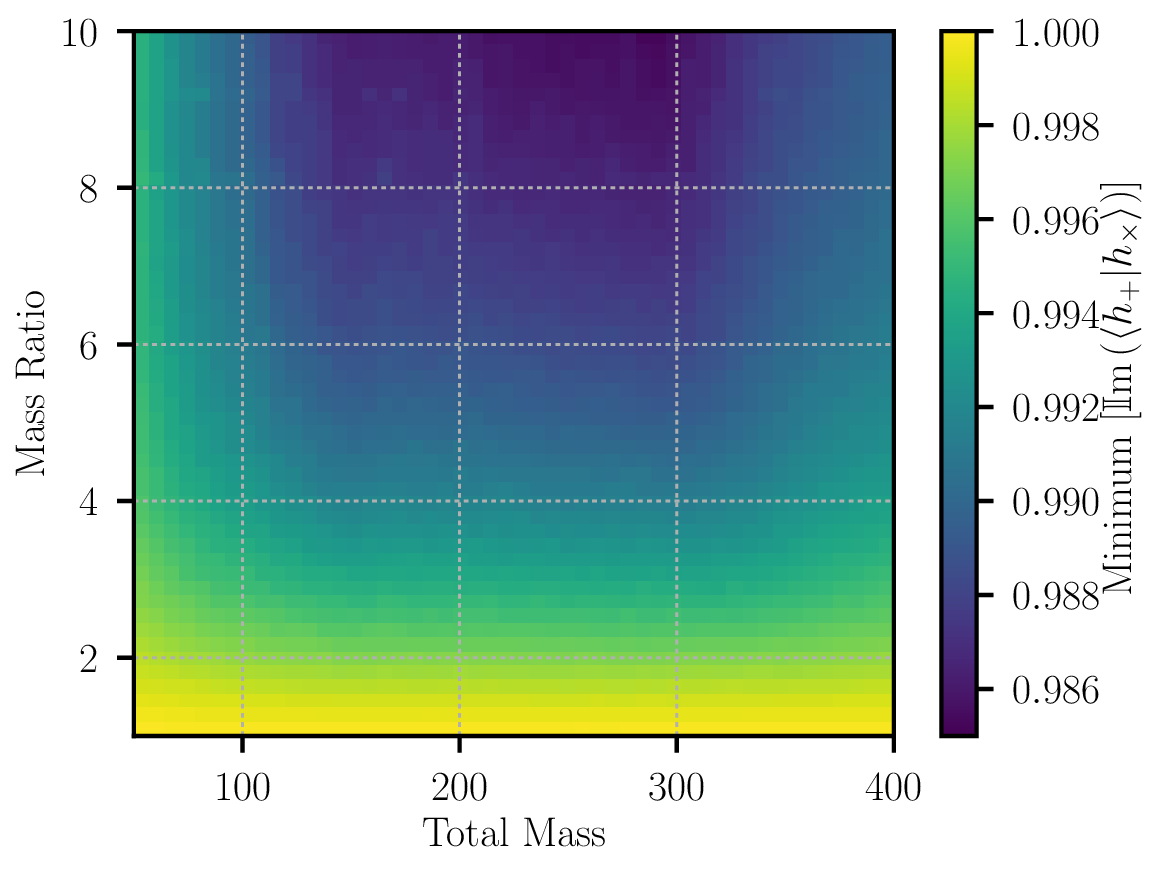}
\includegraphics[width=0.45\textwidth]{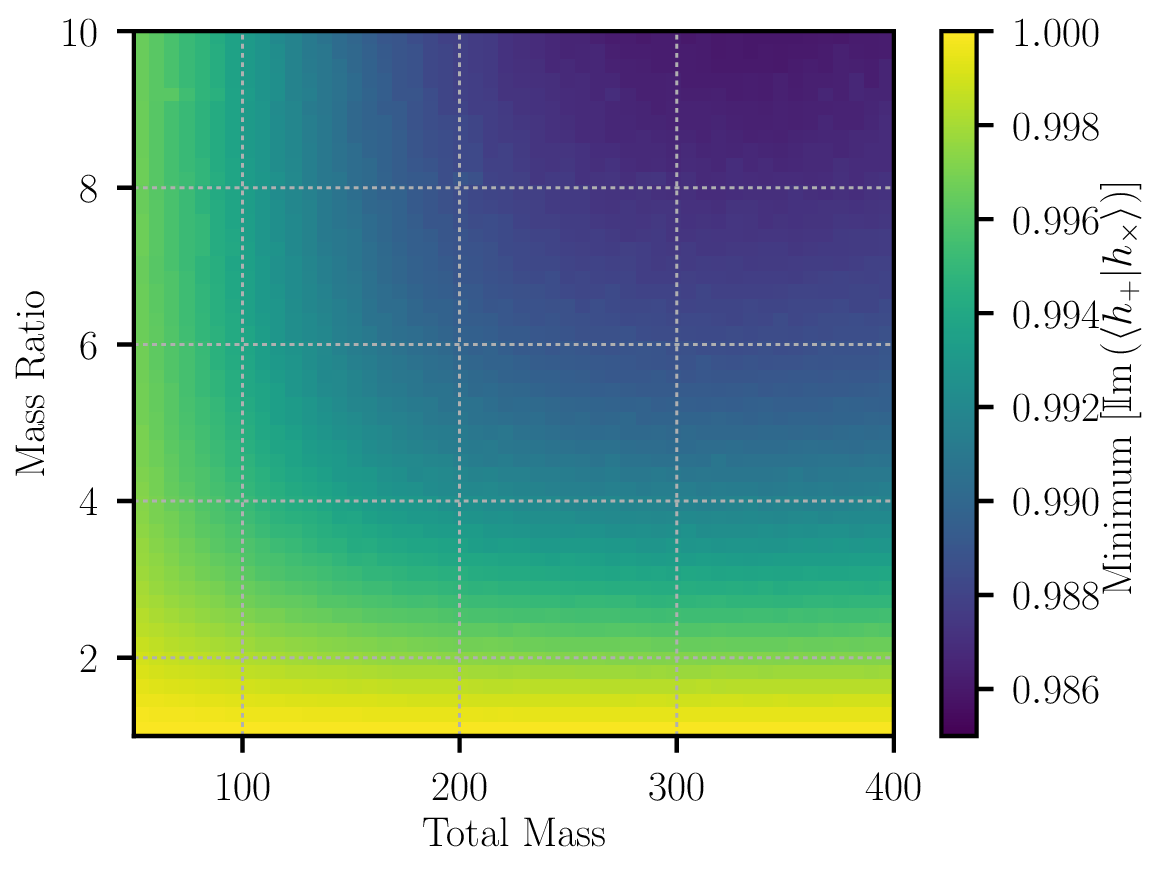}
\caption{\label{fig:pluscrosscorr}
The imaginary component of the complex overlap between $\hat{h}_+$ and $\hat{h}_{\times}$ for the representative
early Advanced LIGO noise curve (top), and for the predicted design sensitivity curve of Advanced LIGO (bottom).
The overlap is numerically minimized over inclination and reference orbital phase within each
of the pixels shown on these plots.}
\end{figure}

To investigate the possibility of using this approximation, we can simply calculate
the magnitude of the imaginary component of the overlap between $\hat{h}_+$ and
$\hat{h}_{\times}$ for systems within the parameter space defined in Section \ref{sec:paramranges}. To do this
we generate 5 million waveforms, with component masses uniformly chosen within our chosen
parameter space, and with isotropic distribution of inclination and reference orbital
phase\footnote{As we are only using $h_+$ and $h_{\times}$ we do not need to choose a sky location, polarization
phase or coalescence time for this set of signals.}. For each
of these waveforms we compute the imaginary component of the overlap between $\hat{h}_+$ and
$\hat{h}_{\times}$, $\Im \left( \langle \hat{h}_+ | \hat{h}_{\times} \rangle \right)$. The results are then binned
in terms of the total mass and mass ratio, and we show the \emph{minimum} value of this overlap, as a function
of the total mass and mass ratio in figure \ref{fig:pluscrosscorr}. We see that for both
the early and design Advanced LIGO sensitivity curves the minimum value of this overlap
is $\sim 0.985$. If we were instead to plot the \emph{average} value of this overlap the value would be larger
than 0.997 everywhere. These values indicate the largest loss of signal-to-noise ratio that is possible when
making the assumption that the polarization phase can be simplified.
For example in the case where the value of the imaginary component of the overlap
is 0.985, then $1.5\%$ of the maximum signal-to-noise ratio would be lost if the signal observed in the detector
is described by the $\times$ component of $h(t)$ and the $+$ component is used as a template. For other values of
polarization phase---such that the detector would observe a combination of the $+$ and $\times$ components---$> 98.5\%$
of the optimal signal-to-noise ratio would be recovered. Given that we will allow a $3\%$ loss of
signal-to-noise ratio due to the discreteness of the set of filter waveforms used,
this indicates that for the parameter spaces and noise curves that we consider in
this work it is sufficient to use the simple statistic in all regions of parameter space. We also verify this
claim later in the results sections. We emphasize
though that if this method is used in future searches using an extended region of parameter space,
or including effects of spin precession, this should be evaluated again.

\section{Assessing the sensitivity increase of a higher-order mode search}
\label{sec:sensitivity}

In the previous section we described a method that will allow the use of template waveforms that
include higher-order modes in searches for compact binary mergers. In this section we will assess
the increase in sensitivity that can be obtained by using this method to
search for compact binary mergers in Advanced LIGO data. We
begin by creating ``template banks'' of waveforms to cover the full parameter space described earlier
in section \ref{sec:paramranges}. From there we will explore the sensitivity increase that can
be obtained when using higher-order mode waveforms. We will first assess this by comparing sensitivities
above a constant signal-to-noise ratio threshold, for both the standard search, and our new method using
higher-order mode waveforms. We will also identify the points in parameter space for which the sensitivity
increases the most when including higher-order mode waveforms. Finally, we will use our new method in the \texttt{PyCBC}
analysis framework~\cite{Canton:2014ena,Usman:2015kfa,alex_nitz_2017_826910}
to analyse 5 days of Gaussian noise, colored to Advanced LIGO sensitivities.
This will allow us to assess the increase in the background rate when including the larger number of
templates that are needed to cover the higher-order mode signal parameter space. This will then enable us to 
compute the sensitivity increase at a constant false-alarm rate threshold between a search that includes the effects of higher-order modes,
and one that does not. 

\subsection{A template bank of filter waveforms including higher order modes}
\label{ssec:temp_banks}

The first step in assessing the sensitivity improvement that can be achieved by including the effects of
higher-order modes in the filter waveforms is to create the set of filter waveforms, or ``template
bank''. In this sub-section we describe the construction of the template banks that we will use, highlighting
any problems specific to construction of template banks of higher-order mode signals. 

\begin{figure*}[t!bp]
\includegraphics[width=\columnwidth]{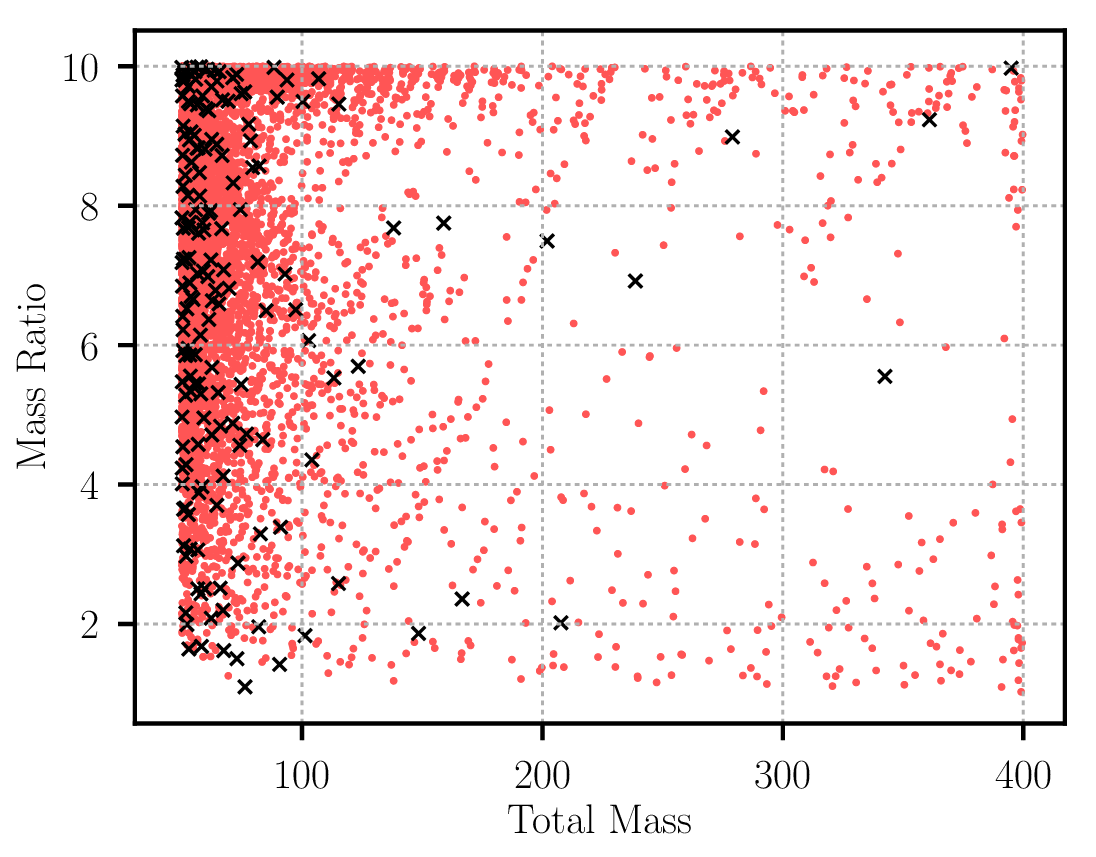}
\includegraphics[width=\columnwidth]{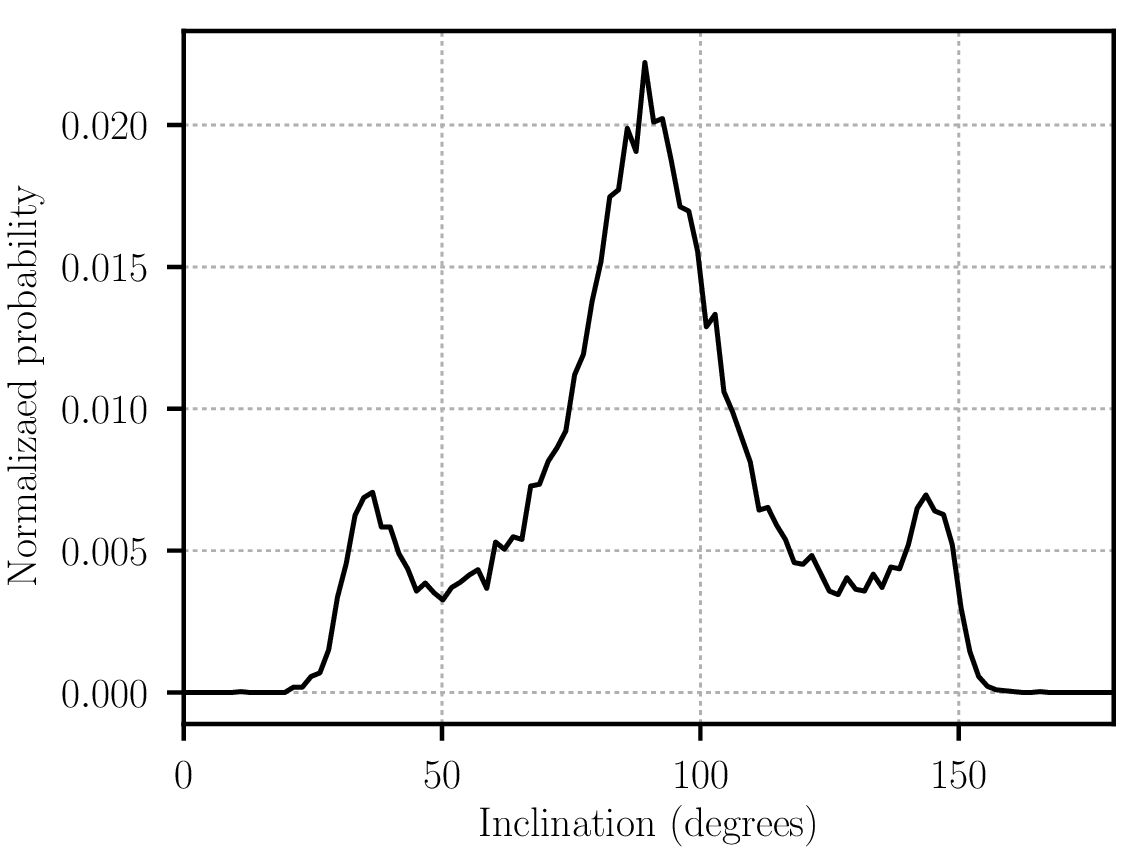}
\caption{\label{fig:template_bank_hom}
The left panel shows the distribution of templates, as a function of the total mass and mass ratio, for the template bank
created for the early Advanced LIGO noise curve. The black crosses indicate the templates that are needed
to cover the parameter space if higher-order modes are omitted, the red circles indicate the additional
templates that are required if higher-order modes are included in the search parameter space. The right
panel shows the distribution of the inclination angle of the higher-order mode waveforms in the template
bank created for the Advanced LIGO design noise curve. 
}
\end{figure*}

We begin by defining the overlap, $o(a,b)$, between a potential signal waveform $a$ and a filter waveform $b$
as the fraction of the optimal signal-to-noise-ratio $(a | a)$ of $a$ that is recovered when using $b$ as a filter
waveform,
\begin{equation}
 o(a,b) \equiv \max_{\Phi} \left( ( \hat{a} | \hat{b}(\Phi) ) \right),
\end{equation}
where $\Phi$ denotes the extrinsic parameters of $b$ that are not included as parameters in the template bank
and are maximized over. In this work we calculate overlaps by maximizing over the coalescence time and the parameter
$u$ defined in equation \ref{eq:u_def} using either equation \ref{eqn:matched_filter_snr_hom}
or equation \ref{eqn:normal_matched_filter_snr} as appropriate.

The ``fitting factor'' (often called ``effectualness'')~\cite{Apostolatos:1994mx}
is then defined as the maximum overlap between $a$ and all of the filter waveforms in the template bank $b_i$
\begin{equation}
 \mathrm{FF}(a, b_i) = \max_i o(a, b_i).
\end{equation}
When constructing template banks to use in analysis of gravitational-wave data the normal choice is to demand that
for any point in the parameter space that the template bank covers, the fitting factor with the template bank must
be greater than 0.97~\cite{Sathyaprakash:1991mt,Babak:2006ty,Babak:2012zx}.
That is to say that the maximum loss in signal-to-noise ratio due to discreteness
of the bank must not be greater than 3\% anywhere in the parameter space. However, as this parameter space explicitly
does not include the effect of higher-order modes---or precession, or eccentricity---fitting factors for real gravitational-wave
signals can be lower than this. For the case of template banks used in the most recent analyses of Advanced LIGO data,
the template bank is placed to cover a broad range of masses and spins~\cite{Capano:2016dsf,DalCanton:2017ala}.
When higher-order modes and precession
are not considered, the orientation of the source with respect to the observer is degenerate with $u$, which is analytically
maximized over, and so the template bank is placed in a 4-dimensional parameter space---the two masses, and two
component spins---using methods described in~\cite{Harry:2009ea,Ajith:2012mn,Brown:2012qf,Privitera:2013xza,Harry:2013tca,Capano:2016dsf}.

\renewcommand{\arraystretch}{1.2}
\begin{table}[tbp]
  \begin{tabular}{c|c|c}
     & O1 noise curve & Design noise curve \\ \hline 
    Waveforms without & \multirow{2}{*}{173} & \multirow{2}{*}{1745}\\
    higher-order modes & & \\ \hline
    Waveforms with & \multirow{2}{*}{6214} & \multirow{2}{*}{20500}\\
    higher-order modes & & \\
  \end{tabular}
    \caption{\label{tbl:tmpltbank_sizes}Sizes of the template banks considered in this work.
    The Advanced LIGO sensitivity curves and parameter spaces considered here are discussed in section~\ref{sec:paramranges}
    and the bank construction discussed in section~\ref{ssec:temp_banks}.}
\end{table}

In \cite{Harry:2016ijz} the authors discussed how to place a template bank for precessing waveforms,
and we follow a very similar approach here. Specifically, we use the ``stochastic''
placement algorithm described in~\cite{Harry:2009ea,Ajith:2012mn,Privitera:2013xza}, to create our template banks.
The basic idea of the stochastic placement algorithm
is that potential template
points are chosen randomly in the specified parameter space, and the fitting factor of these points computed with the
points currently accepted to the template bank. Points are added to the template bank if the fitting factor is smaller
than 0.97, and the process iterates until some pre-specified stopping condition is reached.
In the case of the higher-order mode template banks we use here, we place
waveforms in a 4-dimensional parameter space, the two component masses and the source orientation parameters ($\iota, \varphi$).
This method can be directly extended to cover the additional two-dimensions, describing the components' spins aligned
with the orbit, when waveforms including aligned-component spins \emph{and} higher-order modes become available.

For this work we compute template banks both for the representative early Advanced LIGO sensitivity curve
and the Advanced LIGO design noise curve, discussed earlier in section~\ref{sec:paramranges}.
The template banks are chosen to cover systems with total mass greater than 50
and less than 400 solar masses, with mass ratio limited to be less than 10, which we also discussed in section~\ref{sec:paramranges}. 
We compute template banks for waveforms that include higher-order mode effects,
and template banks including waveforms \emph{without} any higher-order modes. For each of the two sensitivity curves,
we begin by constructing a template bank of waveforms, covering our range of masses, that do not include higher-order modes.
We then take that template bank and add to it templates containing higher-order modes, using the stochastic process, until
we also have a template bank that is suitable for higher-order mode waveforms.
When placing the template bank of higher-order mode waveforms we use
equation \ref{eqn:matched_filter_snr_hom} to maximize over $u$. The sizes of these template banks are given
in Table~\ref{tbl:tmpltbank_sizes}. We note that the higher-order mode template banks are an order of magnitude
bigger than the standard template banks.

In Figure \ref{fig:template_bank_hom} we visualize the distribution of the waveforms in the template banks
that we have created. The left panel of Figure \ref{fig:template_bank_hom} shows the distribution of the
templates as a function of the total mass and mass ratio when including, and when not including, higher-order mode effects.
As well as being able to see that many more templates are needed when including higher-order modes, we observe
that we especially need many more templates at both high masses and high-mass ratios compared to the no
higher-order modes bank where these regions are sparsely populated. In the right panel of
Figure \ref{fig:template_bank_hom} we show the distribution of the inclination angle of higher-order mode
waveforms in the template bank. We can see that many more templates are needed for edge-on systems than
for face-on or face-away systems. We also observe two local maxima at $\sim 35$ and $\sim 135$ degrees.
These peaks are an artifact of the two-stage template bank creation process, and the fact that templates are added
to a set of non-higher-order mode waveforms, which will match well face-on and face-away systems.

\subsection{Sensitivity comparison at fixed signal-to-noise ratio}
\label{sec:banksimresults}

\begin{figure*}[t!bp]
\includegraphics[width=\columnwidth]{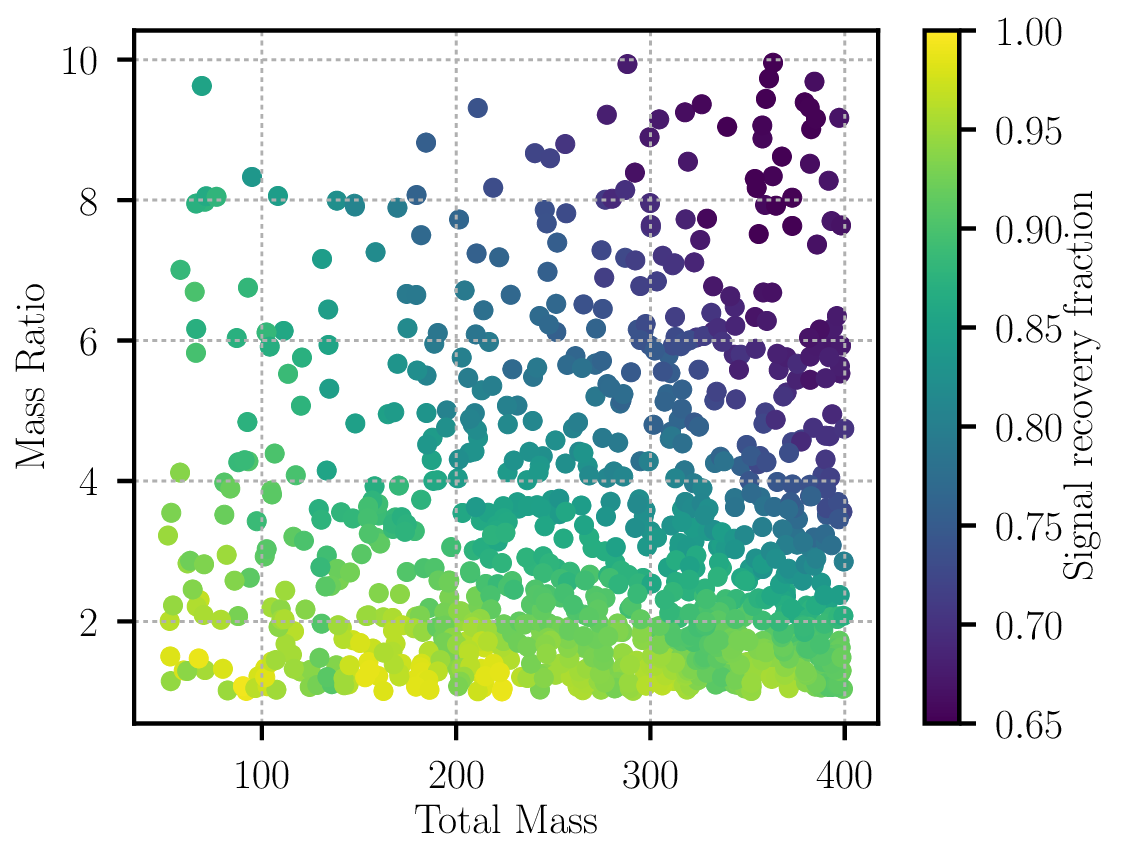}
\includegraphics[width=\columnwidth]{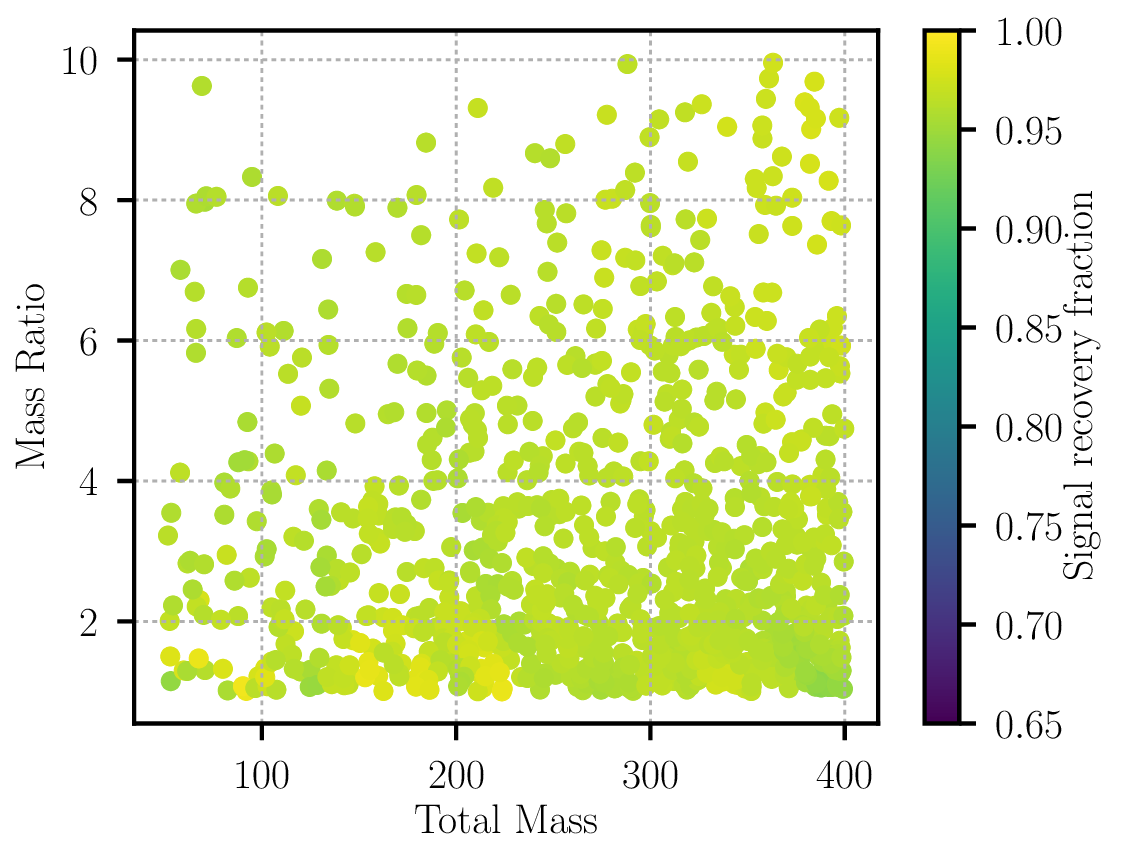}
\includegraphics[width=\columnwidth]{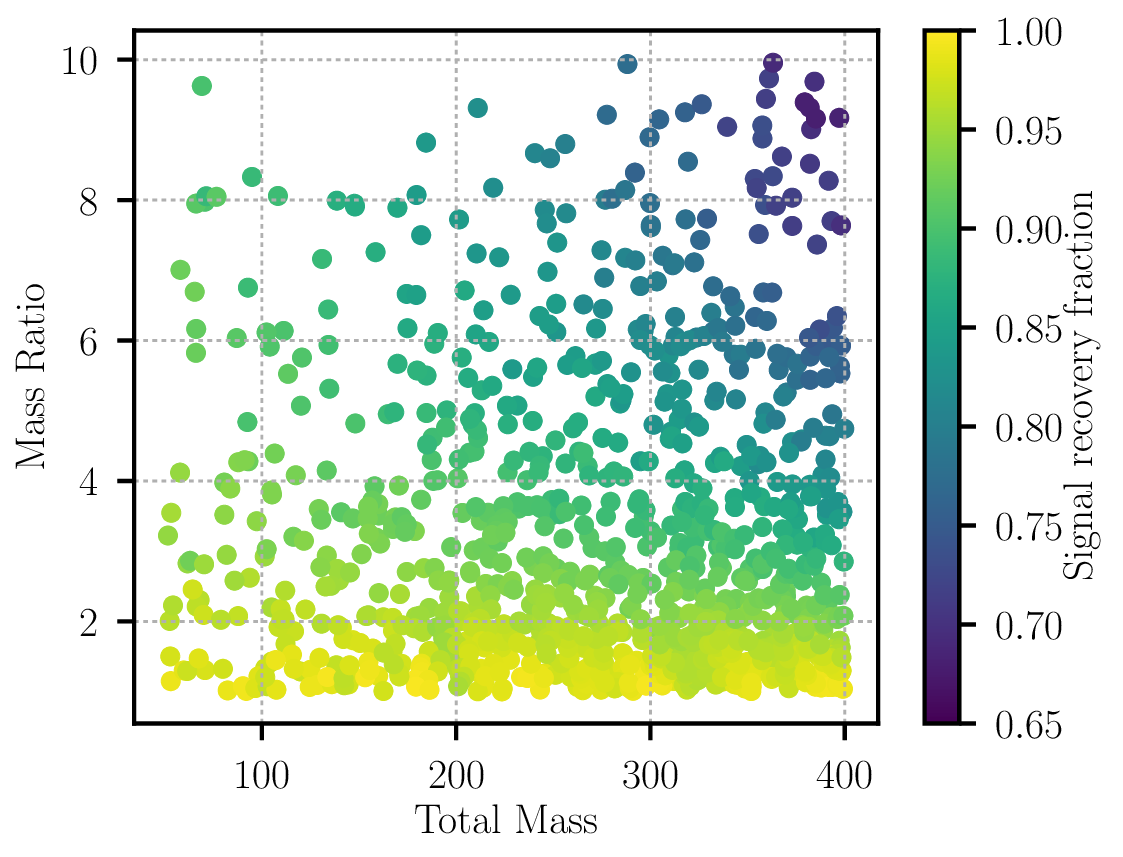}
\includegraphics[width=\columnwidth]{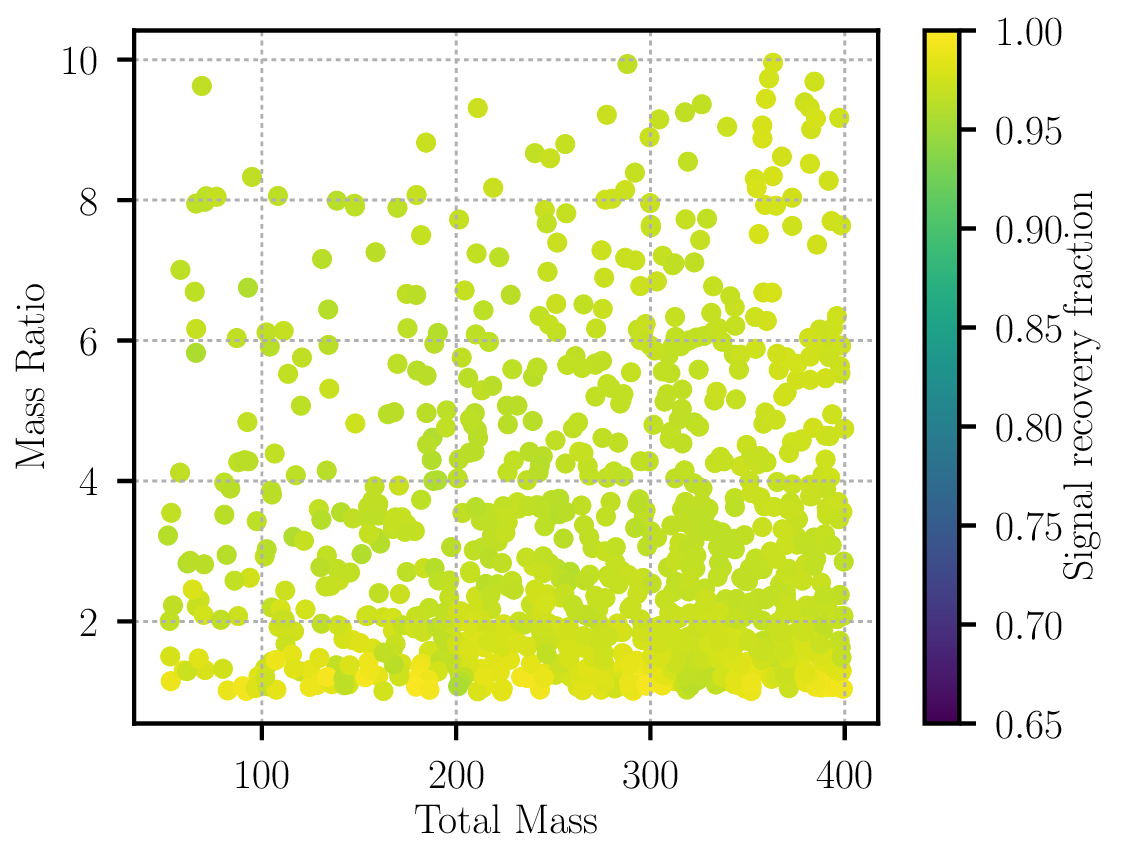}
\caption{\label{fig:signalrecfrac}
The signal recovery fraction plotted as a function of the total mass and mass ratio.
The top panels show results using a representative early Advanced LIGO sensitivity
curve, while the bottom panels use the design Advanced LIGO sensitivity. The left panels
are generated using appropriate template banks that do not include higher-order mode
waveforms, the right panels are generated with template banks that include higher-order
mode waveforms.}
\label{fig:signal_recovery_fraction}
\end{figure*}

\begin{figure*}[t!bp]
\includegraphics[width=\columnwidth]{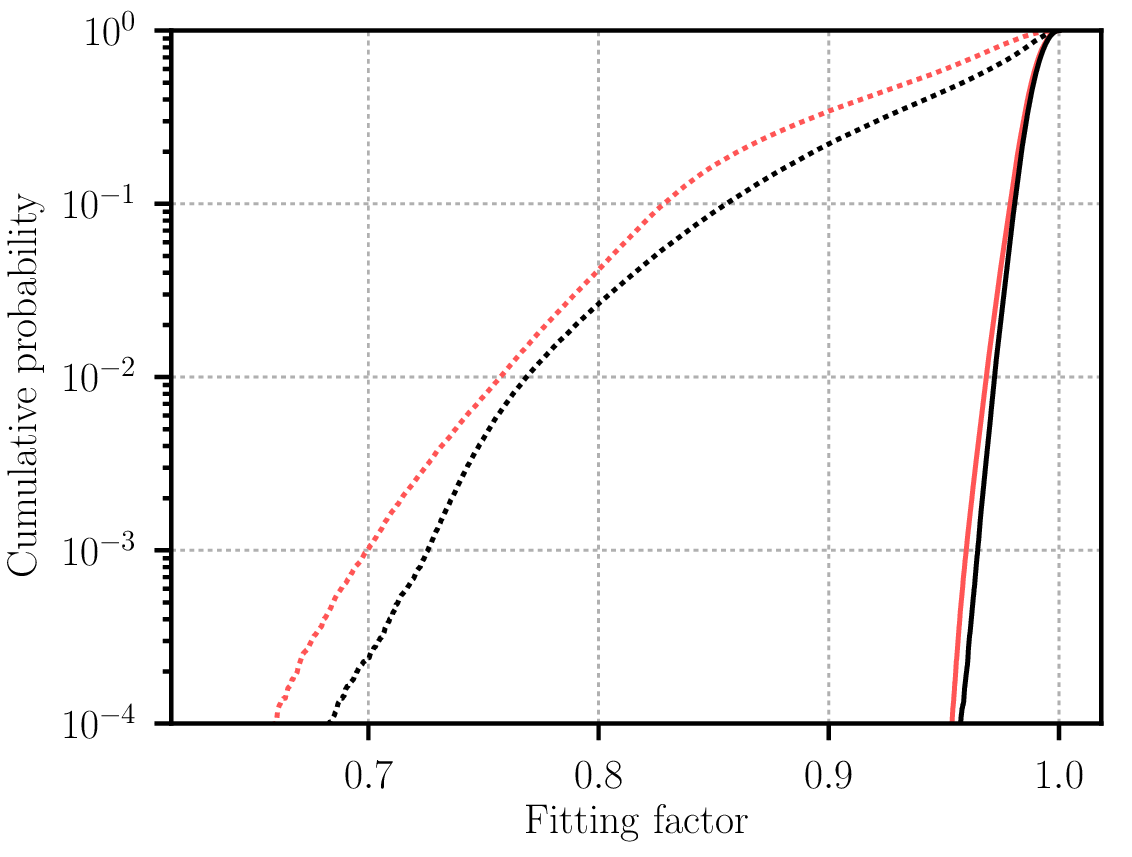}
\includegraphics[width=\columnwidth]{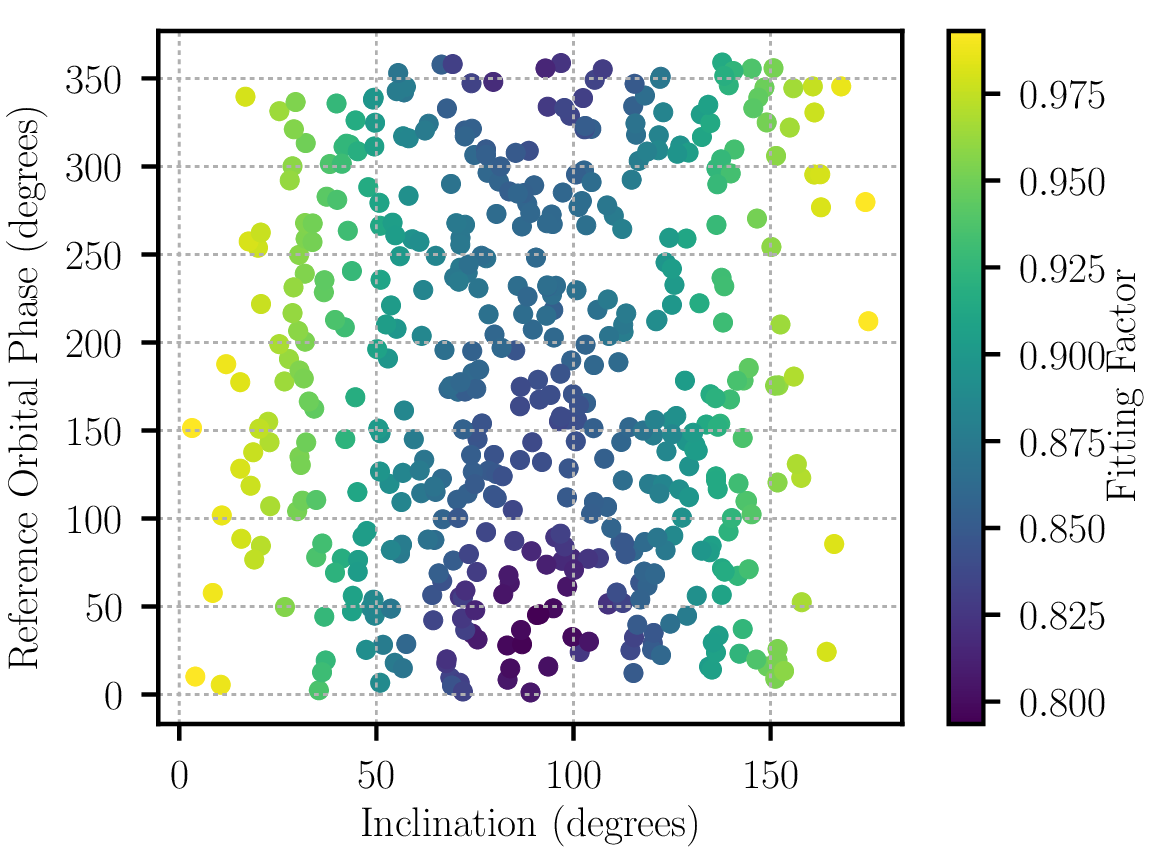}
\caption{\label{fig:fitting_factor_dist}
Left: Cumulative distribution of fitting factors for the 500,000 higher-order mode signal waveforms we describe in section
\ref{sec:banksimresults}. The black curves show results for the Advanced LIGO design sensitivity curve
and the red curves show results for the representative early Advanced LIGO sensitivity curve. Dotted
lines indicate results when using template banks that do not include higher-order modes, solid lines
show results using template banks including higher-order mode waveforms. Right: Distribution of fitting
factors as a function of the source orientation $(\iota,\varphi)$ for $500$ signal waveforms with total mass of $95 M_{\odot}$
and mass ratio of $8$.
}
\end{figure*}

We wish to evaluate and compare the sensitivity to a set of given signals using both our template banks containing
higher-order mode waveforms, and not containing them. This directly gives a measure of how much sensitivity
would be gained by using our new higher-order mode search method.
We first must define how this sensitivity will be computed.
To assess the sensitivity to a given set of waveforms, drawn from some stated distribution, we must, for each
waveform $g_i$ in the parameter space we consider, compute the fitting factor
that will be recovered using the given template bank, composed of waveforms $b_i$.
The distribution of fitting factors for the set of signals allows us to understand what fraction of signal-to-noise
ratio we will recover for each waveform, and identify regions of parameter space where sensitivity is poor.
However, it can often be misleading to only show the distribution of fitting factors, as often the systems for
which fitting factors are smallest are also those ones whose observable gravitational-wave signal is weaker.
To take into account the fact that different signals can be observed at different distances,
one can define the corresponding ``signal recovery fraction'' of a given template bank $b_i$ to a distribution
of signals $g_i$ as  
\begin{equation}
\mathrm{SRF} =\frac{\sum_i \left(FF(g_i, b_j)\right)^3 (g_i | g_i)^3}{ \sum_i (g_i | g_i)^3}.
\end{equation}
This was first introduced in terms of an ``effective fitting factor'' in~\cite{Buonanno:2002fy}. One can understand
the signal recovery fraction as the fraction of signals from a distribution $g_i$ that would be recovered above a
fiducial signal-to-noise ratio threshold with the template bank $b_i$ compared to a template bank with a
fitting factor of 1 for all $g_i$. For the plots shown in this section we compute the fitting factor, and then
signal recovery fractions, using 
equation~\ref{eqn:normal_matched_filter_snr} to maximize over $u$. We have also created the plots using
equation~\ref{eqn:matched_filter_snr_hom} and the numeric values agree to within $0.05\%$ with those in the plots shown. This
demonstrates again that it is sufficient to use the computationally simpler equation~\ref{eqn:normal_matched_filter_snr}
when performing a search using higher-order mode waveforms.

In Fig~\ref{fig:signal_recovery_fraction} we plot the signal recovery fraction as a function of the two component
masses for both the early and design Advanced LIGO sensitivity curves, and for template banks with and without higher-order
modes. For each point shown on this plot the signal recovery fraction is calculated by choosing a set of $g_i$
consisting of 500 waveforms. Each waveform has the same values of component masses, and the source orientations and
sky locations are chosen isotropically. We show 1000 unique points in these plots, so a total of $500000$ waveforms
are used in these simulations. We can clearly see in these plots that for equal mass systems the signal recovery
fractions are large for template banks with and without higher-order modes. However, as the mass ratios become
larger the signal recovery fraction can become as small as 0.65 when omitting higher-order modes,
implying that ignoring higher-order modes in
a search would result in a reduction in detection rate of up to 35\% for systems with those masses.
When we include higher-order mode waveforms, the signal recovery fractions are much more uniform, as
expected. Values of 0.95 are consistent with the loss expected due to discreteness of the template bank.
For the Advanced LIGO design sensitivity curve the effect of higher-order modes is smaller than that of the
representative early Advanced LIGO noise curve.
This is expected as the early Advanced LIGO noise curve is comparatively less sensitive at lower frequencies,
where higher-order modes
are less important. These results are consistent with earlier works exploring the effects of higher-order
modes~\cite{Pekowsky:2012sr,Bustillo:2015qty,Varma:2016dnf}, reinforcing that higher-order modes are important
for systems where the mass ratio and total mass is large.

In Fig~\ref{fig:fitting_factor_dist} we show the cumulative distribution of fitting factors for all of the $500000$
waveforms described above. This is shown for both the early and design Advanced LIGO sensitivity curves and for
template banks both including and not including higher-order mode waveforms. We can clearly see here that there
is a significant proportion of systems recovered with low fitting factors if higher-order modes are neglected.
Using our higher-order mode template banks completely removes the tail of low fitting factors. We also show fitting
factor as a function of the source orientation for all signals simulated at a total mass of $95 M_{\odot}$ and a mass
ratio of $8$. We can clearly see that the lowest fitting factors are obtained when the inclination angle is edge-on,
as expected.

\subsection{Sensitivity comparison at fixed false-alarm rate}
\label{sec:banksimfarweighted}

\begin{table}[tb]
  \begin{tabular}{c|c|c}
    \parbox[t]{2.5cm}{\hfill \break Bank \break } & \parbox[t]{2.5cm}{SNR threshold at false-alarm rate of $10^{-3} yr^{-1}$} &
           \parbox[t]{2.5cm}{SNR threshold at false-alarm rate of $0.5 \times 10^{-3} yr^{-1}$} \\ \hline
    Early, no HOM & 9.10 & 9.16 \\
    Early, with HOM & Not applicable & 9.68\\
    Design, no HOM & 9.31 & 9.37  \\
    Design, with HOM & Not applicable & 9.70 \\
  \end{tabular}
  \caption{\label{tbl:bkg_thresholds}Signal-to-noise ratio thresholds at the false-alarm rate values used in this study
  for the various template banks and sensitivity curves considered. ``Early'' refers to the representative sensitivity
  curve from Advanced LIGO's first observing run, ``Design'' refers to Advanced LIGO's design sensitivity. ``HOM'' stands
  for ``higher-order modes'', ``SNR'' stands for ``signal-to-noise ratio''.}
\end{table}

\begin{figure*}[t!bp]
\includegraphics[width=\columnwidth]{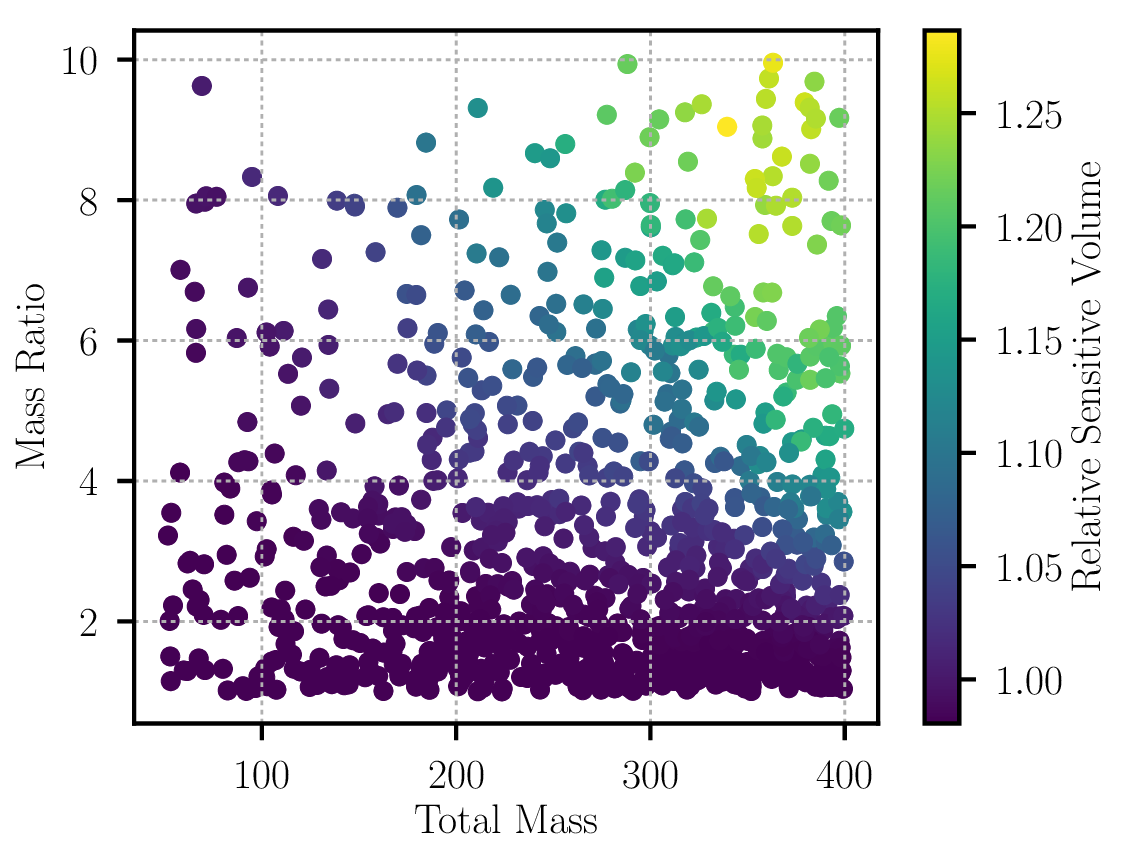}
\includegraphics[width=\columnwidth]{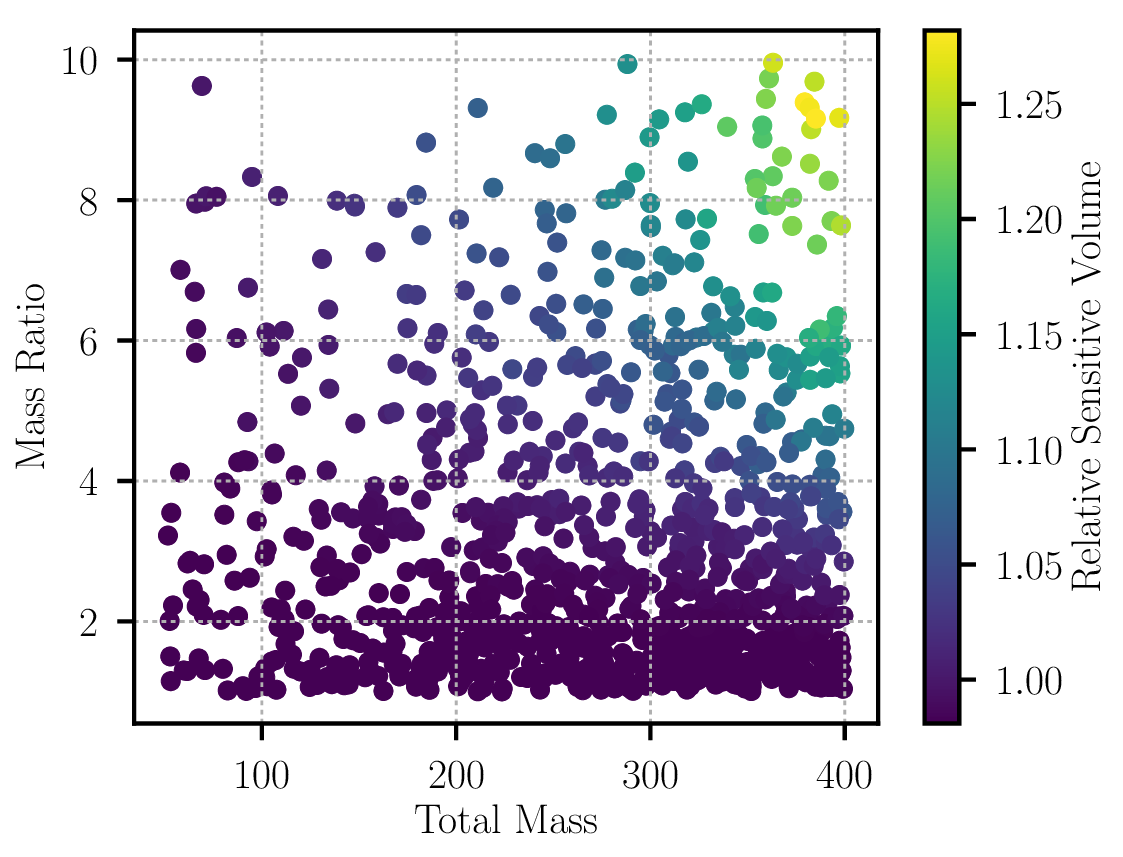}
\includegraphics[width=\columnwidth]{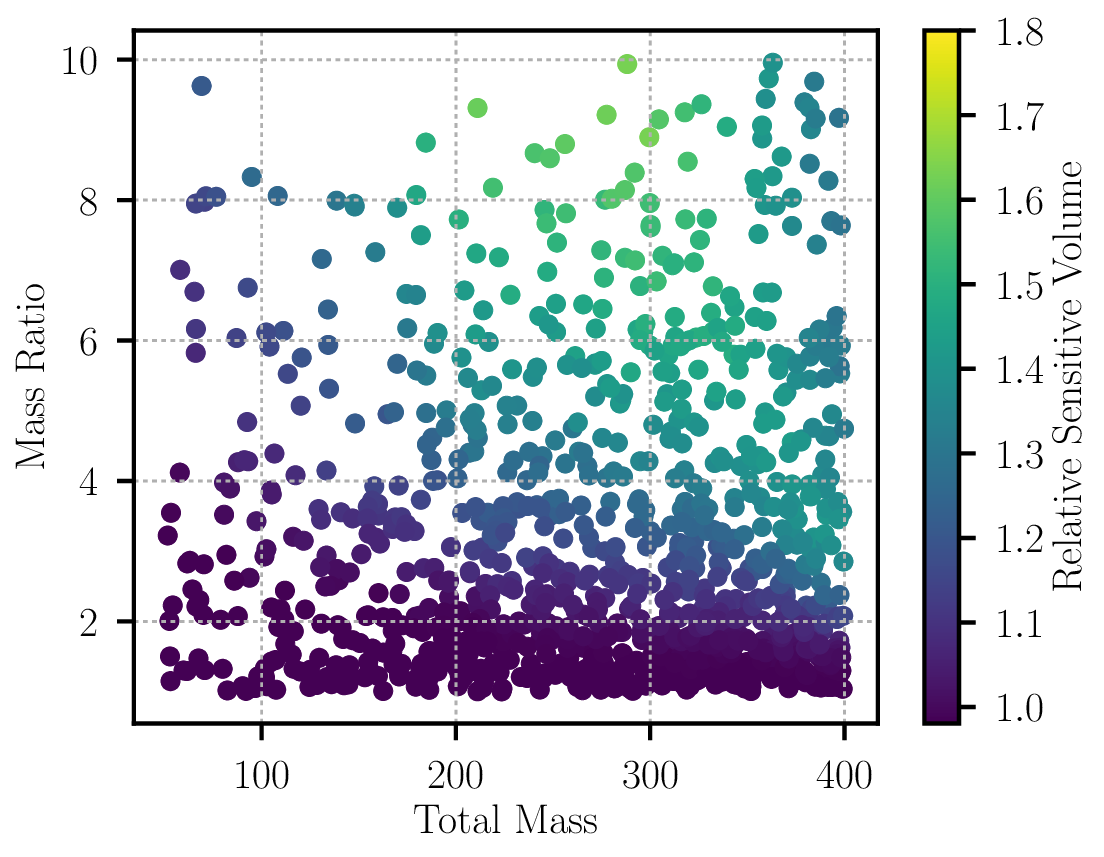}
\includegraphics[width=\columnwidth]{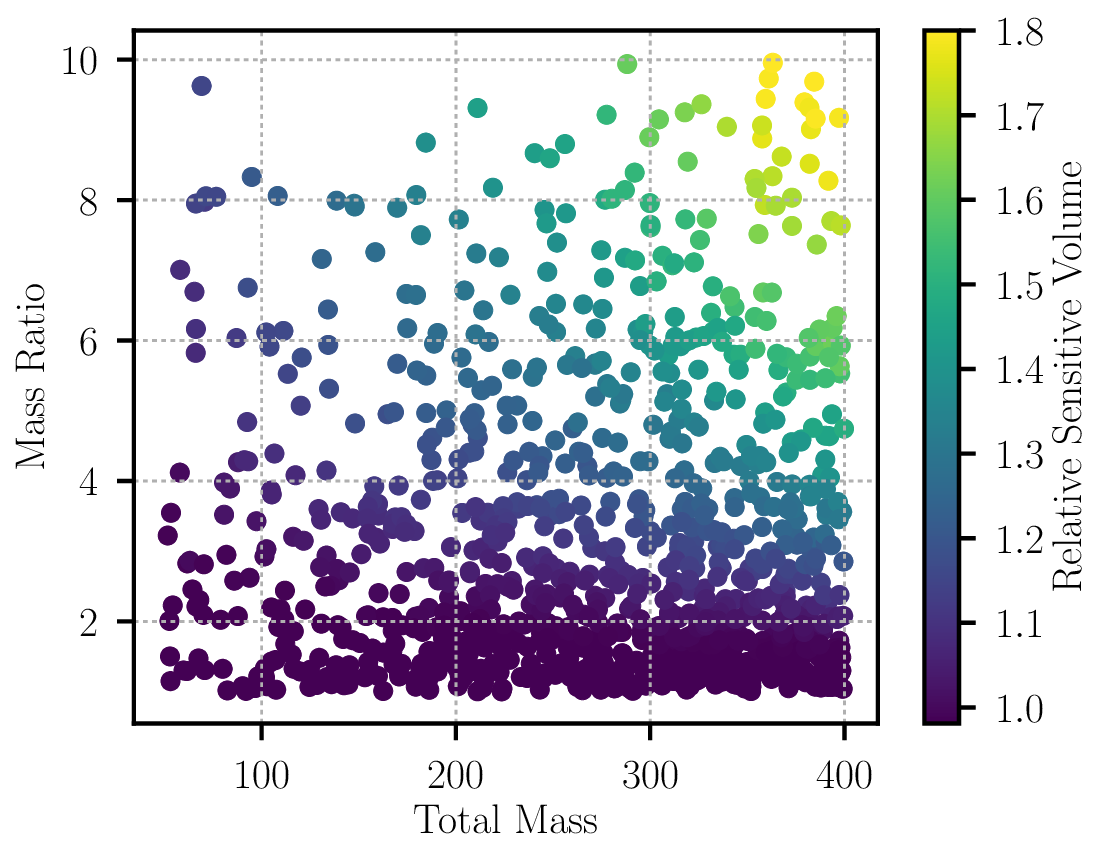}
\caption{\label{fig:sensratfar}
The sensitivity ratio between a search using higher-order mode waveforms as templates
and a search using non-higher-order-mode waveforms as templates, evaluating sensitivity using
only signal-to-noise ratio to rank potential events. Plotted for the early Advanced LIGO sensitivity
curve (left) and the Advanced LIGO design sensitivity curve (right). The plots on the top row consider
the full injection set, while plots on the bottom row consider only injections where the inclination ($\iota$)
is between 60 and 120 degrees. The sensitivity is evaluated at a false-alarm
rate of 1 per 1000 years, which includes the larger background that is present when using higher-order
mode waveforms as templates, as more template waveforms are needed. Values greater than one indicate
an increase in sensitivity, values less than one indicate a decrease in sensitivity.}
\end{figure*}

The results in section \ref{sec:banksimresults} demonstrate that when including higher-order mode effects
in the waveform filters used in a search the search efficiency will increase when evaluating efficiency
above a constant signal-to-noise ratio threshold. However, in a real search the signal-to-noise ratio threshold is a function
of the number of waveform templates, and the size of the parameter space covered. Our higher-order
mode template banks are roughly an order of magnitude larger than the corresponding non-higher-order mode
template banks. This increase in the number of templates will increase the rate of background events in
the search, and therefore a signal would require a larger signal-to-noise ratio to achieve the same significance,
when evaluated in terms of a false-alarm rate.
In this section we will assess the sensitivity increase that can be obtained when using our higher-order
mode template banks, at a constant false-alarm-rate threshold, which takes into account the increase in
background triggers from using a larger number of template waveforms.

The first step is to create a mapping between signal-to-noise ratio and false-alarm rate. We do this for
each of our template banks by simulating $\sim5$ days of Gaussian noise with either the representative early
Advanced LIGO sensitivity curve or the Advanced LIGO design sensitivity curve.
We then analyse this data with the various template banks using the \texttt{PyCBC}
analysis framework~\cite{Canton:2014ena,Usman:2015kfa,alex_nitz_2017_826910},
which allows us to directly map the signal-to-noise ratio to a false-alarm rate.
A false-alarm weighted relative signal-recovery fraction can then be computed according to
\begin{equation}
 \mathrm{SRF} =\frac{\sum_i \left(FF(g_i, b_j)\right)^3 (g_i | g_i)^3}{ \rho_{\mathrm{thresh}}^3 \sum_i (g_i | g_i)^3},
\end{equation}
where $\rho_{\mathrm{thresh}}$ is the signal-to-noise threshold corresponding to the desired false-alarm rate.
This doesn't have meaning as a statistic on it's own, but the ratio of this quantity computed for two different searches,
which will have different values of fitting factor and $\rho_{\mathrm{thresh}}$,
directly gives the relative sensitivity. One could compute this directly for our higher-order mode, and non-higher-order mode
template banks. However, this would result in a non-negligible decrease in sensitivity in the equal mass region of
parameter space. This is because this region of parameter space is already well recovered by non-higher-order mode waveforms
and the increased signal-to-noise ratio threshold, due to the higher-order mode waveforms, only causes a reduction
in sensitivity for equal-mass systems. Instead, we can choose to consider two separate searches, one with higher-order modes
and one without, and combine the results together, including the necessary trials factor of 2.
This would limit the decrease in sensitivity to $\sim1\%$ in regions where higher-order modes contribute
nothing while still allowing a sensitivity increase where higher-order modes are important.
Formally the false-alarm weighted relative signal-recovery fraction for this combined search would be computed according to
\begin{equation}
 \label{eq:combine_far}
 \mathrm{SRF}_{\mathrm{combined}} = \left( 
 \frac{ \sum_{i} (\mathrm{FF_{weighted}})_i  (g_i | g_i)^3 }{\sum_{i} (g_i | g_i)^3  } \right),
\end{equation}
where we define
\begin{equation}
 (\mathrm{FF_{weighted}})_i = \max_{j} \left\{(\rho_{\mathrm{thresh}})_j^{-3} \textrm{FF}_{i,j}^3\right\},
\end{equation}
where $j$ denotes the values for the two searches being performed.

In Table \ref{tbl:bkg_thresholds} we show the network signal-to-noise ratios corresponding to a false-alarm rate of $10^{-3} yr^{-1}$.
This is the threshold at which we choose to evaluate the relative sensitivity of our higher-order mode search. When computing
the combined search sensitivity we incorporate the trials factor by using a false-alarm rate of $0.5 \times 10^{-3} yr^{-1}$
in each search, and therefore $10^{-3} yr^{-1}$ in the combined search. To the accuracies quoted in this table, the threshold
obtained is the same if we use equation~\ref{eqn:normal_matched_filter_snr}
or equation~\ref{eqn:matched_filter_snr_hom} to maximize over $u$.
In Figure \ref{fig:sensratfar} we show the sensitivity increase between the higher-order mode and non-higher-order mode
searches as a function of the total mass and mass ratio. 
This is computed from the same set of waveforms as used in Figure~\ref{fig:signalrecfrac}.
We see that in both cases there is no increase in sensitivity for equal mass systems but an increase in sensitivity of up to
25\% for the systems with the highest total mass and mass ratio that we consider here. It is important to again emphasize
that while these averaged sensitivity increases are modest, even a single observation of a compact binary merger with measurable
higher-order mode emission would allow for much more precise measurement of source parameters than systems where only the
dominant gravitational-wave emission modes are observable~\cite{Varma:2014jxa,Graff:2015bba,Bustillo:2015qty,Varma:2016dnf}.
In this sense, we stress that we are presenting a gain in sensitivity
which is averaged over the possible orientations of the binary. These results are often dominated by face-on
binaries, which have a stronger emission, and for which the quadrupolar bank shows an excellent signal recovery. The
sensitivity gain is much larger for edge-on binaries, whose emission has a strong higher mode contribution, leading to a poor
signal recovery when a quadrupolar bank is used as demonstrated in Figure~\ref{fig:fitting_factor_dist}.
To emphasize this, in the lower panel of Figure~\ref{fig:signalrecfrac} we also show the sensitivity increase if considering
only those waveforms with $60^{\circ} < \iota < 120^{\circ}$, which are those waveforms oriented edge-on to the observer
and for which higher-modes are most important. Here we observe much higher sensitivity increases---up to 80\%---than with the
full set of simulated waveforms.

We note that equation~\ref{eq:combine_far} defines a simple measure for combining the higher-order mode and non-higher-order
mode searches, which does not reduce significantly the sensitivity to non-higher-order mode waveforms, while simultaneously
allowing a sensitivity increase for systems where higher-order modes are important. However, a more optimal method to
combine these two searches, would be to utilize a method similar to that defined in~\cite{Dent:2013cva}, which uses Bayesian
methodology to weight each template waveform according to its probability of observing a system. However, such a method
requires a good knowledge of the astrophysical distribution of systems, which is not known for intermediate-mass black
hole binary systems, and requires knowing relatively how often each template is to observe a signal, which is difficult to compute
with curved and degenerate parameter spaces where it can be difficult to determine what region of parameter space is best
covered by each template.

\section{Real data concerns for searches for high-mass waveforms}
\label{sec:realdataproblems}

In the previous sections we have evaluated the sensitivity increase when using filter waveforms containing higher-order
modes assuming that the detector noise is Gaussian and stationary and using only the signal-to-noise ratio to
evaluate the significance of events. In reality, data taken from gravitational-wave observatories is neither Gaussian
nor stationary and instrumental non-Gaussian noise transients will produce large values of signal-to-noise ratio in
a matched-filter search~\cite{Aasi:2012wd,Aasi:2014mqd,TheLIGOScientific:2016zmo}.
Therefore search strategies for compact binary coalescences must take into account such non-Gaussian
transients and be able to distinguish them from genuine astrophysical signals.
There are numerous works that have focused on this
problem~\cite{Allen:2005fk, Harry:2010fr, Babak:2012zx, Biswas:2013wfa, Messick:2016aqy,Nitz:2017svb}.
However, many of these tests were created considering
lower mass compact binary mergers than those considered here and these tests are known to be less efficient when searching
for intermediate-mass black hole binary mergers~\cite{Abbott:2017iws}. Some tests are beginning to focus on the
efficiency to higher mass black hole binary mergers, but are not yet able to separate all forms of transient noise, and are not
fully tuned for higher-order mode waveforms~\cite{Nitz:2017lco}.

Indeed for certain regions of the parameter space unmodelled search techniques have
been found to be more sensitive to compact binary mergers in data from LIGO's first observing run than modelled searches,
because they are better at removing instrumental artifacts~\cite{Mohapatra:2014rda,Bustillo:pycbccwb}.
Optimizing searches to better distinguish real astrophysical
signals from instrumental noise at high masses is an interesting topic that should be addressed, but this should be done
in a separate work, and we will not attempt to address this specific question here.

In this section we will explore how existing tests to separate real signals from noise artifacts can be applied when using
higher-order mode waveforms, we will demonstrate that these tests can misclassify genuine astrophysical signals with significant
higher-order mode contribution as instrumental artifacts, and that this problem is significantly mitigated when using higher-order mode
waveforms as filters in the search.

\subsection{Reweighted signal-to-noise ratio}
\label{ssec:newsnr}

\begin{figure*}[t!bp]
\includegraphics[width=\columnwidth]{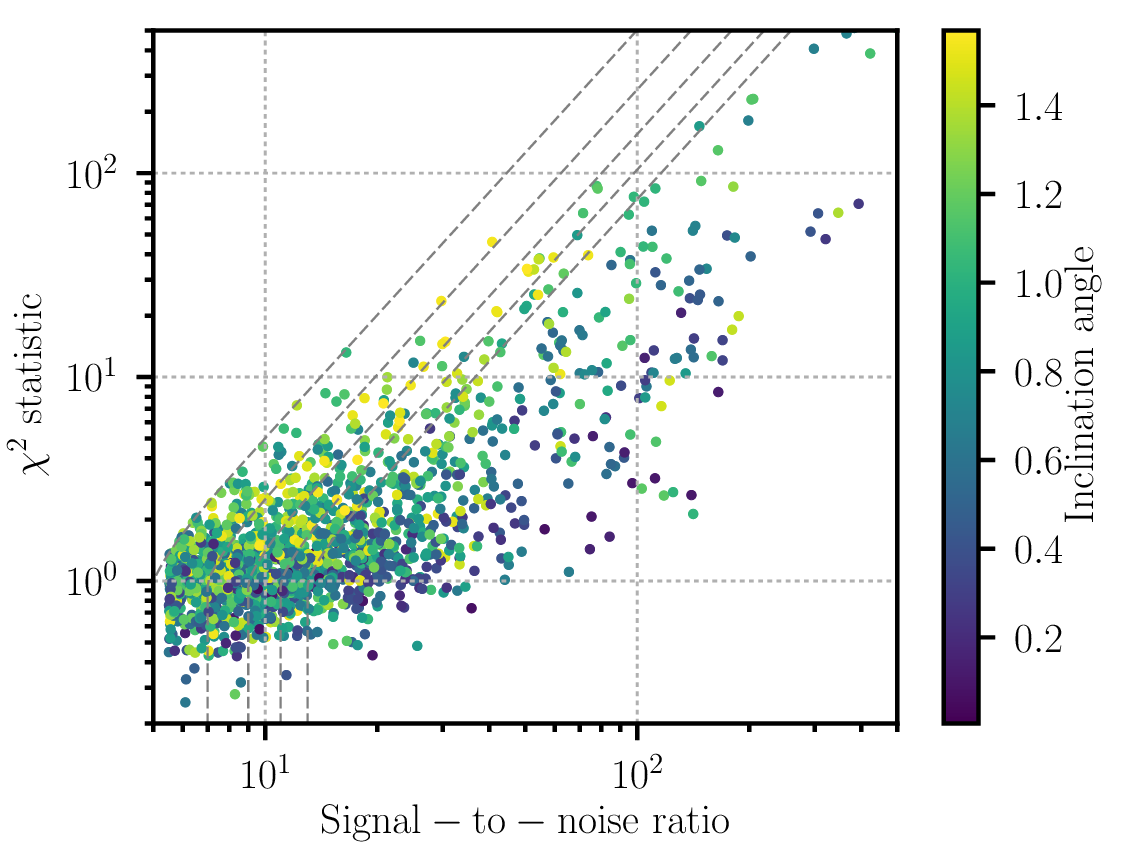}
\includegraphics[width=\columnwidth]{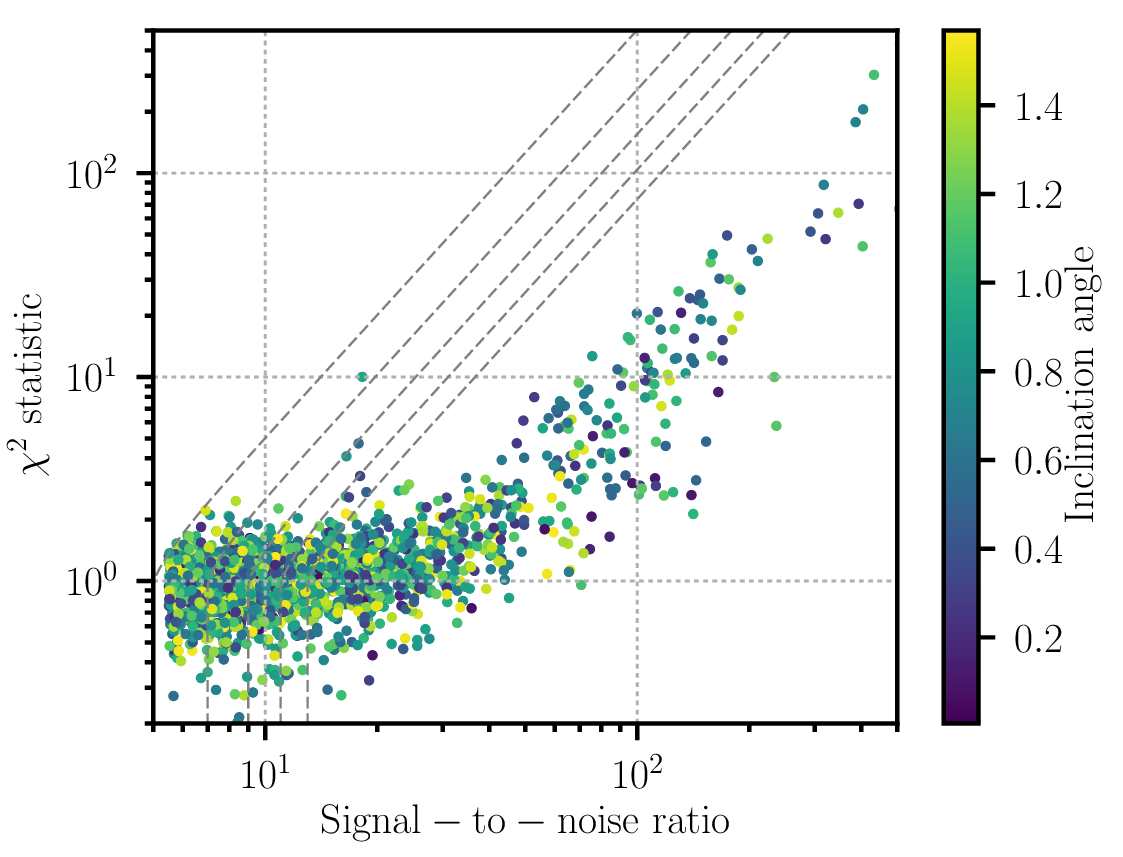}
\caption{\label{fig:chisq}
Impact of higher-modes on the $\chi^2$ signal consistency test. We plot $\chi^2$ against signal-to-noise ratio
for a randomly chosen subset of the simulated signals discussed in section~\ref{sec:realdataproblems}.
The inclination angle of the simulated signals are shown on the colorbar, here we restrict the inclination
angle to values between 0 and $\pi/2$ by taking $\pi/2 - |\pi/2 - \iota|$. The dashed lines denote contours
of equal $\rho_{\mathrm{reweighted}}$. The left panel shows the results
when searching with a template bank that does not include higher-order mode waveforms, the right panel shows
the results when searching with higher-order mode templates. Both panels are generated using the design Advanced
LIGO sensitivity curves.}
\end{figure*}

One of the most common methods for discriminating between gravitational wave triggers and noise artifacts is to check
whether the morphology of a potential signal in the data, $s$, is consistent with that of the filter waveform being
used, $h$. Several methods for doing this, testing different features of the potential signal's morphology,
have been proposed~\cite{Allen:2005fk, Harry:2010fr, Babak:2012zx, Messick:2016aqy}.
Of these, arguably the most
effective test is the one described in~\cite{Allen:2005fk}. In that test a number of filters are constructed from
the template waveform $h$ in the following way. A set of $N$ filters $h_i$ is chosen such that each $h_i$ is constructed,
in the frequency domain as
\begin{equation}
   \tilde{h}_i (f) = \begin{cases} 
        \tilde{h} (f) & \mathrm{for}\ f_L < f <= f_U \\
        0 & \mathrm{otherwise}
    \end{cases}.
\end{equation}
Each filter $h_i$ uses non-overlapping frequency windows, $f_L$ and $f_U$, such that $\sum_i h_i = h$. Also
$(h_i | h_i)= (h_j|h_j)$ for any value of $i$ and $j$ and $(h_i | h_j) = 0$ for any $i\neq j$.
By this definition if the data $s$ is a good match to the filter waveform
then each of the $h_i$ should recover the same signal-to-noise ratio, within deviations expected in Gaussian noise.
In contrast, noise artifacts are often well localized in time and would often produce a very large signal-to-noise
ratio in a small number of $h_i$ and a small signal-to-noise ratio in the rest. Therefore one can construct a
chi-squared test as
\begin{equation}
\chi^2 = \frac{N}{\langle h|h \rangle} \sum_{i=1}^{N} \left\|\langle s|h_i \rangle - \frac{\langle s|h \rangle }{N}\right\|^2.
\end{equation}
If $s$ is described by Gaussian noise with an added signal well modelled by $h$, this will follow
a $\chi^2$ distribution with $2N - 2$ degrees of freedom~\cite{Allen:2005fk}. For non-Gaussian artifacts
it has been empirically demonstrated that this will take larger values~\cite{Allen:2005fk, Babak:2012zx}, allowing
for separation between real signals and Gaussian artifacts. There are a number of different techniques
for combining the $\chi^2$ test with the signal-to-noise ratio to produce a ranking
statistic~\cite{Babak:2012zx,Adams:2015ulm,Messick:2016aqy,Nitz:2017svb},
we choose to use here the combination described in~\cite{Babak:2012zx}, which has been used to analyse Advanced
LIGO data with the \texttt{PyCBC} analysis method~\cite{Colaboration:2011np}. This ``reweighted signal-to-noise ratio''
is given by~\cite{Babak:2012zx}
\begin{equation}
 \rho_{\mathrm{reweighted}} = \begin{cases} 
        \rho & \mathrm{for}\ \chi^2 <= n_{\mathrm{d}} \\
        \rho\left[ \frac{1}{2} \left(1 + \left( \frac{\chi^2}{n_{\mathrm{d}}}\right)^3\right)\right]^{1/6} & 
        \mathrm{for}\ \chi^2 > n_{\mathrm{d}}
    \end{cases},
\end{equation}
where $n_{\mathrm{d}} = 2N - 2$.
We want to explore how well this reweighted signal-to-noise ratio performs when searching for higher-order
mode signals with and without higher-order mode filter waveforms.

\subsection{Sensitivity comparison at fixed false-alarm rate with reweighted signal-to-noise ratio}
\label{sec:senscompnewsnr}

\begin{figure*}[t!bp]
\includegraphics[width=\columnwidth]{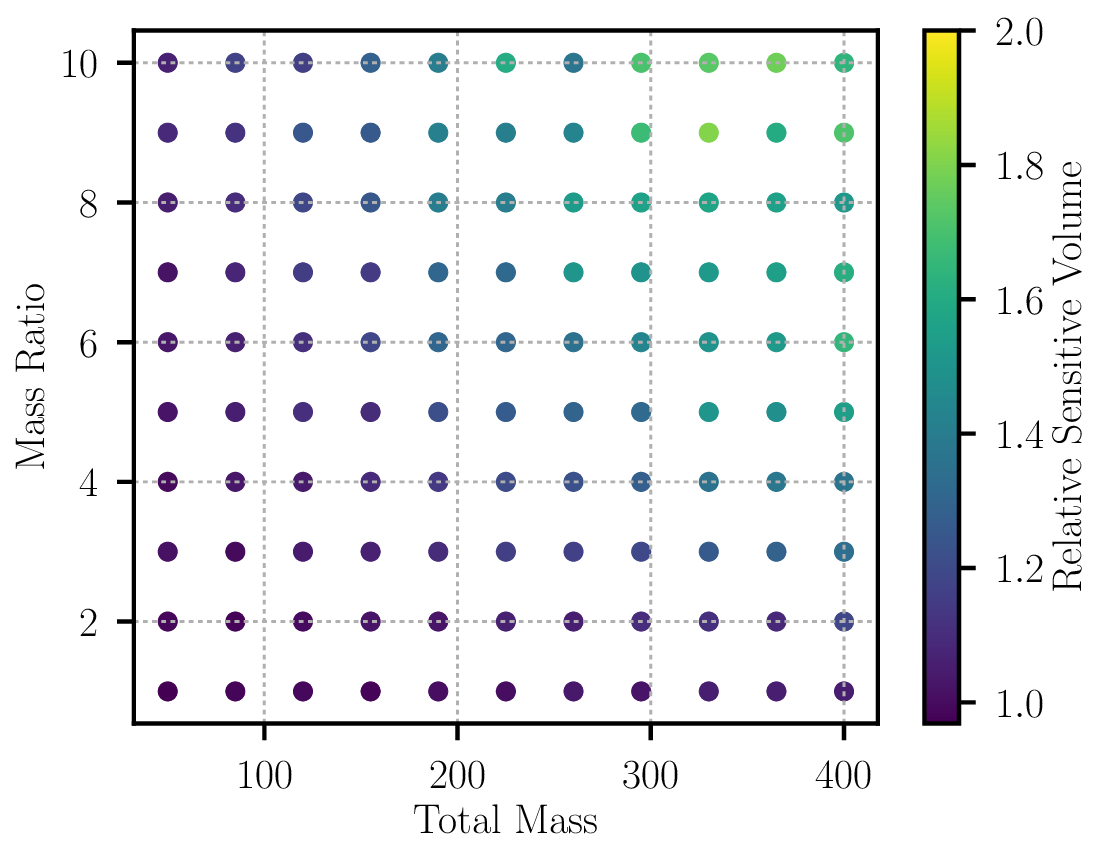}
\includegraphics[width=\columnwidth]{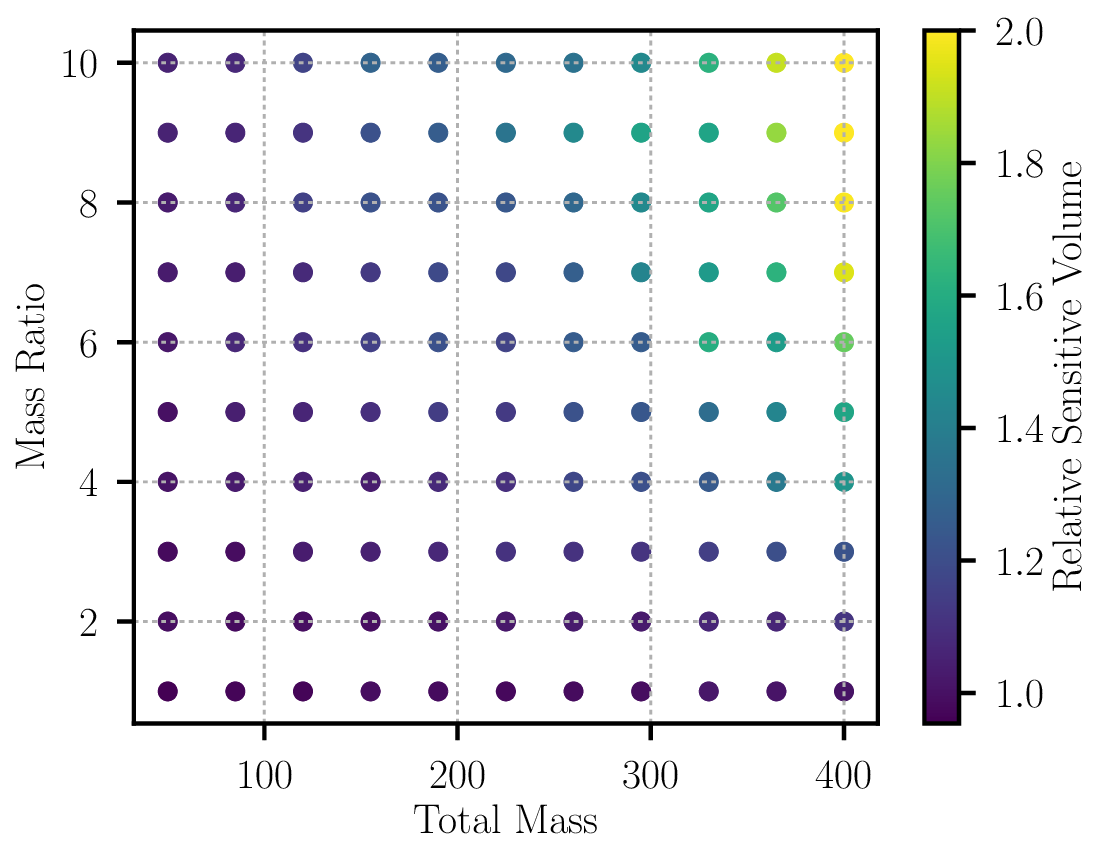}
\includegraphics[width=\columnwidth]{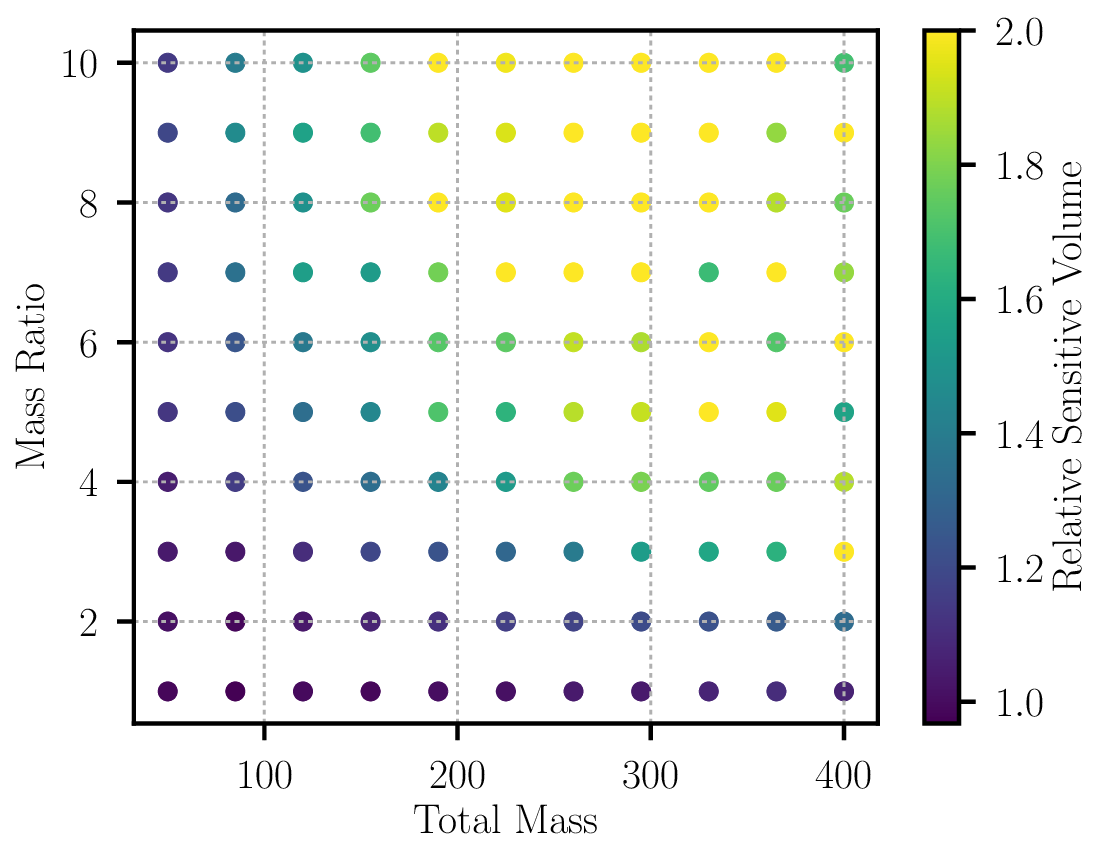}
\includegraphics[width=\columnwidth]{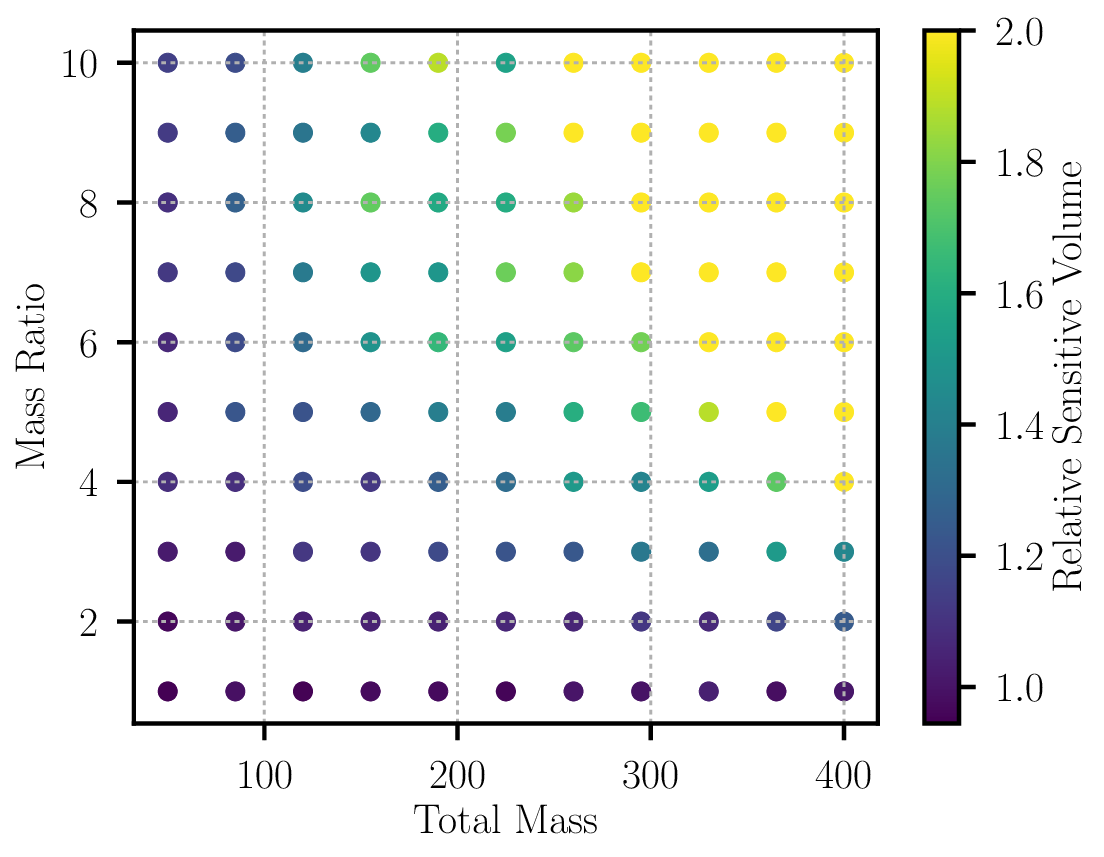}
\caption{\label{fig:sensratfarnewsnr}
The relative sensitivity between a search using higher-order mode waveforms as templates
and a search using non-higher-order-mode waveforms as templates when evaluating sensitivity using
$\rho_{\mathrm{reweighted}}$ at a constant rate of false alarm. Plotted for the representative early
Advanced LIGO sensitivity curve (left) and the design Advanced LIGO sensitivity curve (right).
The plots on the top row consider
the full injection set, while plots on the bottom row consider only injections where the inclination ($\iota$)
is between 60 and 120 degrees.
The sensitivity is evaluated at a false-alarm
rate of 1 per 1000 years, which includes the larger background that is present when using higher-order
mode waveforms as templates. Values greater than one indicate
an increase in sensitivity, values less than one indicate a decrease in sensitivity.
Here the color bars are capped at a value of 2 to aid distinguishability of values
between 1 and 2. However some values are larger than this, especially in the lower right panel, where the points with
highest mass and highest mass ratio have a relative sensitive volume of $\sim$ 4.}
\end{figure*}

We generate a large set of simulated intermediate-mass black hole binary waveforms
to assess the sensitivity of our higher-order mode search method when evaluating sensitivity using the reweighted
signal-to-noise ratio defined above.
We use the same distribution of parameters as described in the section~\ref{sec:sensitivity}, but as
the values of the $\chi^2$ test will depend on the amplitude of the signal, we include the distance and
sky location as
parameters when generating the simulation set. We also add the simulated signals to simulated Gaussian noise
when measuring the signal-to-noise ratio and $\chi^2$. The signals are added to noise simulating both
Advanced LIGO observatories and the quadrature sum of the recovered $\rho_{\mathrm{reweighted}}$ is used
to rank events. In total we choose to simulate signals with
$110$ unique masses and using $\sim 10000$ unique simulated signals for each mass.

In Figure~\ref{fig:chisq} we show the distribution of $\chi^2$ values as a function of
signal-to-noise ratio,
both with and without higher-order mode filter waveforms. The dashed lines show contours of constant
$\rho_{\mathrm{reweighted}}$; the $\rho_{\mathrm{reweighted}}$ increases as the signal-to-noise
ratio increases and as the value of the $\chi^2$ test decreases.
When searching for systems that are
oriented face-on to the observer using non-higher-order mode filter waveforms
the $\chi^2$ values tend to be low as the higher-order mode content is negligible. However, as
the inclination $\iota$ increases, higher-modes contribute more to the resulting signal, increasing the
mismatch between signal and template \emph{and} causing the $\chi^2$ to grow. When using higher-order mode
filter waveforms the $\chi^2$ values are lower, and lie away from the contour lines where non-Gaussian
artifacts would appear in real data.
The $\chi^2$ test would therefore cause an additional loss in sensitivity over that considered in section
\ref{sec:sensitivity} if searching for higher-order mode waveforms using filters that neglect higher-order modes.

We therefore reproduce the figures shown in Figure~\ref{fig:sensratfar}, but using $\rho_{\mathrm{reweighted}}$
to rank potential events instead of using signal-to-noise ratio.
This is shown in Figure~\ref{fig:sensratfarnewsnr}.
We see in this plot a larger sensitivity increase when including higher-order modes compared to that seen in
Figure~\ref{fig:sensratfar} due to the effect of the $\chi^2$
test. We also now see a larger sensitivity increase for the design Advanced LIGO sensitivity curve, whereas previously
the larger sensitivity increase was seen with the early Advanced LIGO sensitivity curve. This indicates that the $\chi^2$ test
is misclassifying more systems for the design sensitivity curve than for the early curve. These results qualitatively match
the results in \cite{Capano:2013raa} where the authors made predictions of how the sensitivity should increase if one
were able to include higher-order mode waveforms as filters in a search. As with Figure~\ref{fig:sensratfar}, we also show
results considering only injections aligned close to edge-on to the observer in the lower panels. As before, we observe much
larger sensitivity increases, and again this increase is larger when including the effects of the $\chi^2$ test. In
Figure~\ref{fig:sensratfarnewsnr} we chose to set the maximum value on the colorbar to 2 to allow the reader to observe
improvements in sensitivity that are less than 1.5. However there are some values
that are much larger than this, up to a value of 4---indicating a 300\% increase in the number of observed
signals---for the points with the largest mass and largest mass ratio in the
lower right panel. In such regions our new higher-order mode search is especially needed.

While it would be beneficial
to work on exploring and tuning various signal-based consistency tests and classifiers to improve the performance of
searching for higher-order-mode waveforms with template waveforms that do not include higher-order modes, such tests
and classifiers will be more powerful if the template waveforms being used match well to the signals in the data.
Some more work is needed to improve the separation of noise transients from real signals, in the intermediate-mass black
hole binary parameter space, and in the case where the template waveforms match well. However, we recommend that such work
is performed while using waveforms that include higher-order mode waveforms, as described in this work. 

\section{Conclusions}
\label{sec:conclusions}

In this work we have presented a new method for searching for compact binary coalescences using filter waveforms which
include higher-order mode gravitational-wave emission. This method will allow for the first time searches using higher-order
mode filter waveforms to be performed in ongoing analysis of data from second-generation gravitational-wave observatories.
We have demonstrated the sensitivity improvement this method would allow. This improvement is as much as a 100\% improvement for systems
with mass ratios of 10 and total mass of 400 $M_{\odot}$, but is much more modest, $< 10\%$, for systems with equal mass or
with total mass of 50 $M_{\odot}$.
In the cases where the improvement is modest, it implies that the efficiency of the
current, non-higher order mode search, is already good in these areas.
The improvement in sensitivity is largest---as much as 300\%---for systems oriented edge-on
to the observer, which are intrinsically fainter in gravitational-waves than face-on systems, but for which higher-order modes
are especially important. The detection of such signals is key for testing fundamental aspects of General Relativity.
For instance, a clear observation of at least two ringdown modes is needed for testing
the no-hair theorem~\cite{Dreyer:2003bv,Maselli:2017kvl}.
The method we present is also fully generic and could be applied also in searches
for eccentric, or precessing, compact binary mergers.

Using this method to search for higher-order mode signals in the latest Advanced LIGO and Advanced Virgo data would require
waveform models that include both higher-order mode emission and model the effect of the components' spins. At the time of
writing such waveform models are not available, but are currently in rapid development~\cite{London:2017bcn,COTESTA}.
When these waveform models are available it is trivial to extend the results shown here to include the component spins,
although the size of both higher-order mode and non-higher-order mode template banks will increase when including this
additional freedom. Nevertheless, we see no reason why one wouldn't expect the same relative sensitivity improvement
as seen here with non-spinning waveform models, when using spinning higher-order mode waveforms.

A current problem with searches for intermediate-mass black hole binary mergers is that the gravitational emission
from such systems is only observed for a very short time and can be confused with non-Gaussian noise transients.
Developing better techniques to distinguish between noise transients and genuine gravitational-wave signals would
be very beneficial in this search space, although this task is orthogonal to the problem addressed in this paper. We
have also demonstrated that the performance of current signal-based consistency tests is improved significantly by
including higher-order mode effects in the search parameter space.

\section{Acknowledgements}

The authors would like to thank
Stas Babak, Alejandro Boh\'e, Alessandra Buonanno, Collin Capano, Sylvain Marsat, Stephen Privitera and Vivien Raymond
for helpful discussions. The authors also thank Collin Capano for reading through the manuscript and providing useful
feedback and comments. The authors would also like to thank the anonymous referees for carefully reading this
manuscript and providing useful comments, suggestions and feedback.
IH and AN would like to thank the Max Planck Gesellschaft for support.
JCB gratefully  acknowledges support from the NSF grants 1505824, 1505524 and 1333360.
Computations used in
this work were performed on the ``Vulcan'' and ``Atlas'' high-throughput computing clusters operated by
the Max Planck Institute for Gravitational Physics.

\bibliography{manuscript}

\begin{thebibliography}{112}%
\makeatletter
\providecommand \@ifxundefined [1]{%
 \@ifx{#1\undefined}
}%
\providecommand \@ifnum [1]{%
 \ifnum #1\expandafter \@firstoftwo
 \else \expandafter \@secondoftwo
 \fi
}%
\providecommand \@ifx [1]{%
 \ifx #1\expandafter \@firstoftwo
 \else \expandafter \@secondoftwo
 \fi
}%
\providecommand \natexlab [1]{#1}%
\providecommand \enquote  [1]{``#1''}%
\providecommand \bibnamefont  [1]{#1}%
\providecommand \bibfnamefont [1]{#1}%
\providecommand \citenamefont [1]{#1}%
\providecommand \href@noop [0]{\@secondoftwo}%
\providecommand \href [0]{\begingroup \@sanitize@url \@href}%
\providecommand \@href[1]{\@@startlink{#1}\@@href}%
\providecommand \@@href[1]{\endgroup#1\@@endlink}%
\providecommand \@sanitize@url [0]{\catcode `\\12\catcode `\$12\catcode
  `\&12\catcode `\#12\catcode `\^12\catcode `\_12\catcode `\%12\relax}%
\providecommand \@@startlink[1]{}%
\providecommand \@@endlink[0]{}%
\providecommand \url  [0]{\begingroup\@sanitize@url \@url }%
\providecommand \@url [1]{\endgroup\@href {#1}{\urlprefix }}%
\providecommand \urlprefix  [0]{URL }%
\providecommand \Eprint [0]{\href }%
\providecommand \doibase [0]{http://dx.doi.org/}%
\providecommand \selectlanguage [0]{\@gobble}%
\providecommand \bibinfo  [0]{\@secondoftwo}%
\providecommand \bibfield  [0]{\@secondoftwo}%
\providecommand \translation [1]{[#1]}%
\providecommand \BibitemOpen [0]{}%
\providecommand \bibitemStop [0]{}%
\providecommand \bibitemNoStop [0]{.\EOS\space}%
\providecommand \EOS [0]{\spacefactor3000\relax}%
\providecommand \BibitemShut  [1]{\csname bibitem#1\endcsname}%
\let\auto@bib@innerbib\@empty
\bibitem [{\citenamefont {Aasi}\ \emph
  {et~al.}(2015{\natexlab{a}})\citenamefont {Aasi} \emph
  {et~al.}}]{TheLIGOScientific:2014jea}%
  \BibitemOpen
  \bibfield  {author} {\bibinfo {author} {\bibfnamefont {J.}~\bibnamefont
  {Aasi}} \emph {et~al.} (\bibinfo {collaboration} {LIGO Scientific}),\ }\href
  {\doibase 10.1088/0264-9381/32/7/074001} {\bibfield  {journal} {\bibinfo
  {journal} {Class. Quant. Grav.}\ }\textbf {\bibinfo {volume} {32}},\ \bibinfo
  {pages} {074001} (\bibinfo {year} {2015}{\natexlab{a}})},\ \Eprint
  {http://arxiv.org/abs/1411.4547} {arXiv:1411.4547 [gr-qc]} \BibitemShut
  {NoStop}%
\bibitem [{\citenamefont {Abbott}\ \emph
  {et~al.}(2016{\natexlab{a}})\citenamefont {Abbott} \emph
  {et~al.}}]{Abbott:2016blz}%
  \BibitemOpen
  \bibfield  {author} {\bibinfo {author} {\bibfnamefont {B.~P.}\ \bibnamefont
  {Abbott}} \emph {et~al.} (\bibinfo {collaboration} {Virgo, LIGO
  Scientific}),\ }\href {\doibase 10.1103/PhysRevLett.116.061102} {\bibfield
  {journal} {\bibinfo  {journal} {Phys. Rev. Lett.}\ }\textbf {\bibinfo
  {volume} {116}},\ \bibinfo {pages} {061102} (\bibinfo {year}
  {2016}{\natexlab{a}})},\ \Eprint {http://arxiv.org/abs/1602.03837}
  {arXiv:1602.03837 [gr-qc]} \BibitemShut {NoStop}%
\bibitem [{\citenamefont {Abbott}\ \emph
  {et~al.}(2016{\natexlab{b}})\citenamefont {Abbott} \emph
  {et~al.}}]{Abbott:2016nmj}%
  \BibitemOpen
  \bibfield  {author} {\bibinfo {author} {\bibfnamefont {B.}~\bibnamefont
  {Abbott}} \emph {et~al.} (\bibinfo {collaboration} {Virgo, LIGO
  Scientific}),\ }\href {\doibase 10.1103/PhysRevLett.116.241103} {\bibfield
  {journal} {\bibinfo  {journal} {Phys. Rev. Lett.}\ }\textbf {\bibinfo
  {volume} {116}},\ \bibinfo {pages} {241103} (\bibinfo {year}
  {2016}{\natexlab{b}})},\ \Eprint {http://arxiv.org/abs/1606.04855}
  {arXiv:1606.04855 [gr-qc]} \BibitemShut {NoStop}%
\bibitem [{\citenamefont {Abbott}\ \emph
  {et~al.}(2017{\natexlab{a}})\citenamefont {Abbott} \emph
  {et~al.}}]{Abbott:2017vtc}%
  \BibitemOpen
  \bibfield  {author} {\bibinfo {author} {\bibfnamefont {B.}~\bibnamefont
  {Abbott}} \emph {et~al.} (\bibinfo {collaboration} {VIRGO, LIGO
  Scientific}),\ }\href {\doibase 10.1103/PhysRevLett.118.221101} {\bibfield
  {journal} {\bibinfo  {journal} {Phys. Rev. Lett.}\ }\textbf {\bibinfo
  {volume} {118}},\ \bibinfo {pages} {221101} (\bibinfo {year}
  {2017}{\natexlab{a}})}\BibitemShut {NoStop}%
\bibitem [{\citenamefont {Acernese}\ \emph {et~al.}(2015)\citenamefont
  {Acernese} \emph {et~al.}}]{TheVirgo:2014hva}%
  \BibitemOpen
  \bibfield  {author} {\bibinfo {author} {\bibfnamefont {F.}~\bibnamefont
  {Acernese}} \emph {et~al.} (\bibinfo {collaboration} {VIRGO}),\ }\href
  {\doibase 10.1088/0264-9381/32/2/024001} {\bibfield  {journal} {\bibinfo
  {journal} {Class. Quant. Grav.}\ }\textbf {\bibinfo {volume} {32}},\ \bibinfo
  {pages} {024001} (\bibinfo {year} {2015})},\ \Eprint
  {http://arxiv.org/abs/1408.3978} {arXiv:1408.3978 [gr-qc]} \BibitemShut
  {NoStop}%
\bibitem [{\citenamefont {Aso}\ \emph {et~al.}(2013)\citenamefont {Aso},
  \citenamefont {Michimura}, \citenamefont {Somiya}, \citenamefont {Ando},
  \citenamefont {Miyakawa}, \citenamefont {Sekiguchi}, \citenamefont
  {Tatsumi},\ and\ \citenamefont {Yamamoto}}]{Aso:2013eba}%
  \BibitemOpen
  \bibfield  {author} {\bibinfo {author} {\bibfnamefont {Y.}~\bibnamefont
  {Aso}}, \bibinfo {author} {\bibfnamefont {Y.}~\bibnamefont {Michimura}},
  \bibinfo {author} {\bibfnamefont {K.}~\bibnamefont {Somiya}}, \bibinfo
  {author} {\bibfnamefont {M.}~\bibnamefont {Ando}}, \bibinfo {author}
  {\bibfnamefont {O.}~\bibnamefont {Miyakawa}}, \bibinfo {author}
  {\bibfnamefont {T.}~\bibnamefont {Sekiguchi}}, \bibinfo {author}
  {\bibfnamefont {D.}~\bibnamefont {Tatsumi}}, \ and\ \bibinfo {author}
  {\bibfnamefont {H.}~\bibnamefont {Yamamoto}} (\bibinfo {collaboration}
  {KAGRA}),\ }\href {\doibase 10.1103/PhysRevD.88.043007} {\bibfield  {journal}
  {\bibinfo  {journal} {Phys. Rev.}\ }\textbf {\bibinfo {volume} {D88}},\
  \bibinfo {pages} {043007} (\bibinfo {year} {2013})},\ \Eprint
  {http://arxiv.org/abs/1306.6747} {arXiv:1306.6747 [gr-qc]} \BibitemShut
  {NoStop}%
\bibitem [{\citenamefont {Iyer}\ \emph {et~al.}(2011)\citenamefont {Iyer} \emph
  {et~al.}}]{LigoIndia}%
  \BibitemOpen
  \bibfield  {author} {\bibinfo {author} {\bibfnamefont {B.}~\bibnamefont
  {Iyer}} \emph {et~al.},\ }\href@noop {} {\enquote {\bibinfo {title}
  {{LIGO-India, Proposal of the Consortium for Indian Initiative in
  Gravitational-wave Observations (IndIGO)}},}\ } (\bibinfo {year} {2011}),\
  \bibinfo {note}
  {{\url{https://dcc.ligo.org/LIGO-M1100296/public}}}\BibitemShut {NoStop}%
\bibitem [{\citenamefont {Abbott}\ \emph
  {et~al.}(2016{\natexlab{c}})\citenamefont {Abbott} \emph
  {et~al.}}]{Aasi:2013wya}%
  \BibitemOpen
  \bibfield  {author} {\bibinfo {author} {\bibfnamefont {B.~P.}\ \bibnamefont
  {Abbott}} \emph {et~al.} (\bibinfo {collaboration} {VIRGO, LIGO
  Scientific}),\ }\href {\doibase 10.1007/lrr-2016-1} {\bibfield  {journal}
  {\bibinfo  {journal} {Living Rev. Rel.}\ }\textbf {\bibinfo {volume} {19}},\
  \bibinfo {pages} {1} (\bibinfo {year} {2016}{\natexlab{c}})},\ \Eprint
  {http://arxiv.org/abs/1304.0670} {arXiv:1304.0670 [gr-qc]} \BibitemShut
  {NoStop}%
\bibitem [{\citenamefont {Abbott}\ \emph
  {et~al.}(2016{\natexlab{d}})\citenamefont {Abbott} \emph
  {et~al.}}]{Abbott:2016nhf}%
  \BibitemOpen
  \bibfield  {author} {\bibinfo {author} {\bibfnamefont {B.~P.}\ \bibnamefont
  {Abbott}} \emph {et~al.} (\bibinfo {collaboration} {Virgo, LIGO
  Scientific}),\ }\href {\doibase 10.3847/2041-8205/833/1/L1} {\bibfield
  {journal} {\bibinfo  {journal} {Astrophys. J.}\ }\textbf {\bibinfo {volume}
  {833}},\ \bibinfo {pages} {L1} (\bibinfo {year} {2016}{\natexlab{d}})},\
  \Eprint {http://arxiv.org/abs/1602.03842} {arXiv:1602.03842 [astro-ph.HE]}
  \BibitemShut {NoStop}%
\bibitem [{\citenamefont {Abbott}\ \emph
  {et~al.}(2016{\natexlab{e}})\citenamefont {Abbott} \emph
  {et~al.}}]{TheLIGOScientific:2016wfe}%
  \BibitemOpen
  \bibfield  {author} {\bibinfo {author} {\bibfnamefont {B.~P.}\ \bibnamefont
  {Abbott}} \emph {et~al.} (\bibinfo {collaboration} {Virgo, LIGO
  Scientific}),\ }\href {\doibase 10.1103/PhysRevLett.116.241102} {\bibfield
  {journal} {\bibinfo  {journal} {Phys. Rev. Lett.}\ }\textbf {\bibinfo
  {volume} {116}},\ \bibinfo {pages} {241102} (\bibinfo {year}
  {2016}{\natexlab{e}})},\ \Eprint {http://arxiv.org/abs/1602.03840}
  {arXiv:1602.03840 [gr-qc]} \BibitemShut {NoStop}%
\bibitem [{\citenamefont {Abbott}\ \emph
  {et~al.}(2016{\natexlab{f}})\citenamefont {Abbott} \emph
  {et~al.}}]{TheLIGOScientific:2016pea}%
  \BibitemOpen
  \bibfield  {author} {\bibinfo {author} {\bibfnamefont {B.~P.}\ \bibnamefont
  {Abbott}} \emph {et~al.} (\bibinfo {collaboration} {Virgo, LIGO
  Scientific}),\ }\href@noop {} {\  (\bibinfo {year} {2016}{\natexlab{f}})},\
  \Eprint {http://arxiv.org/abs/1606.04856} {arXiv:1606.04856 [gr-qc]}
  \BibitemShut {NoStop}%
\bibitem [{\citenamefont {Amaro-Seoane}\ and\ \citenamefont
  {Santamaria}(2010)}]{AmaroSeoane:2009ui}%
  \BibitemOpen
  \bibfield  {author} {\bibinfo {author} {\bibfnamefont {P.}~\bibnamefont
  {Amaro-Seoane}}\ and\ \bibinfo {author} {\bibfnamefont {L.}~\bibnamefont
  {Santamaria}},\ }\href {\doibase 10.1088/0004-637X/722/2/1197} {\bibfield
  {journal} {\bibinfo  {journal} {Astrophys. J.}\ }\textbf {\bibinfo {volume}
  {722}},\ \bibinfo {pages} {1197} (\bibinfo {year} {2010})},\ \Eprint
  {http://arxiv.org/abs/0910.0254} {arXiv:0910.0254 [astro-ph.CO]} \BibitemShut
  {NoStop}%
\bibitem [{\citenamefont {Amaro-Seoane}(2012)}]{AmaroSeoane:2012tx}%
  \BibitemOpen
  \bibfield  {author} {\bibinfo {author} {\bibfnamefont {P.}~\bibnamefont
  {Amaro-Seoane}},\ }\href@noop {} {\  (\bibinfo {year} {2012})},\ \Eprint
  {http://arxiv.org/abs/1205.5240} {arXiv:1205.5240 [astro-ph.CO]} \BibitemShut
  {NoStop}%
\bibitem [{\citenamefont {Gair}\ \emph {et~al.}(2011)\citenamefont {Gair},
  \citenamefont {Mandel}, \citenamefont {Miller},\ and\ \citenamefont
  {Volonteri}}]{Gair:2010dx}%
  \BibitemOpen
  \bibfield  {author} {\bibinfo {author} {\bibfnamefont {J.~R.}\ \bibnamefont
  {Gair}}, \bibinfo {author} {\bibfnamefont {I.}~\bibnamefont {Mandel}},
  \bibinfo {author} {\bibfnamefont {M.~C.}\ \bibnamefont {Miller}}, \ and\
  \bibinfo {author} {\bibfnamefont {M.}~\bibnamefont {Volonteri}},\ }\href
  {\doibase 10.1007/s10714-010-1104-3} {\bibfield  {journal} {\bibinfo
  {journal} {Gen. Rel. Grav.}\ }\textbf {\bibinfo {volume} {43}},\ \bibinfo
  {pages} {485} (\bibinfo {year} {2011})},\ \Eprint
  {http://arxiv.org/abs/0907.5450} {arXiv:0907.5450 [astro-ph.CO]} \BibitemShut
  {NoStop}%
\bibitem [{\citenamefont {Sesana}\ \emph {et~al.}(2009)\citenamefont {Sesana},
  \citenamefont {Gair}, \citenamefont {Mandel},\ and\ \citenamefont
  {Vecchio}}]{Sesana:2009wg}%
  \BibitemOpen
  \bibfield  {author} {\bibinfo {author} {\bibfnamefont {A.}~\bibnamefont
  {Sesana}}, \bibinfo {author} {\bibfnamefont {J.}~\bibnamefont {Gair}},
  \bibinfo {author} {\bibfnamefont {I.}~\bibnamefont {Mandel}}, \ and\ \bibinfo
  {author} {\bibfnamefont {A.}~\bibnamefont {Vecchio}},\ }\href {\doibase
  10.1088/0004-637X/698/2/L129} {\bibfield  {journal} {\bibinfo  {journal}
  {Astrophys. J.}\ }\textbf {\bibinfo {volume} {698}},\ \bibinfo {pages} {L129}
  (\bibinfo {year} {2009})},\ \Eprint {http://arxiv.org/abs/0903.4177}
  {arXiv:0903.4177 [astro-ph.CO]} \BibitemShut {NoStop}%
\bibitem [{\citenamefont {Volonteri}(2012{\natexlab{a}})}]{Volonteri:2012ig}%
  \BibitemOpen
  \bibfield  {author} {\bibinfo {author} {\bibfnamefont {M.}~\bibnamefont
  {Volonteri}},\ }\bibfield  {booktitle} {\emph {\bibinfo {booktitle}
  {{Proceedings, 4th Meeting on First stars IV: From Hayashi to the future:
  Kyoto, Japan, May 21-25, 2012}}},\ }\href {\doibase 10.1063/1.4754370}
  {\bibfield  {journal} {\bibinfo  {journal} {AIP Conf. Proc.}\ }\textbf
  {\bibinfo {volume} {1480}},\ \bibinfo {pages} {289} (\bibinfo {year}
  {2012}{\natexlab{a}})},\ \Eprint {http://arxiv.org/abs/1209.1195}
  {arXiv:1209.1195 [astro-ph.CO]} \BibitemShut {NoStop}%
\bibitem [{\citenamefont {Volonteri}(2012{\natexlab{b}})}]{Volonteri:2012tp}%
  \BibitemOpen
  \bibfield  {author} {\bibinfo {author} {\bibfnamefont {M.}~\bibnamefont
  {Volonteri}},\ }\href {\doibase 10.1126/science.1220843} {\bibfield
  {journal} {\bibinfo  {journal} {Science}\ }\textbf {\bibinfo {volume}
  {337}},\ \bibinfo {pages} {544} (\bibinfo {year} {2012}{\natexlab{b}})},\
  \Eprint {http://arxiv.org/abs/1208.1106} {arXiv:1208.1106 [astro-ph.CO]}
  \BibitemShut {NoStop}%
\bibitem [{\citenamefont {Nitz}\ \emph
  {et~al.}(2017{\natexlab{a}})\citenamefont {Nitz}, \citenamefont {Dent},
  \citenamefont {Dal~Canton}, \citenamefont {Fairhurst},\ and\ \citenamefont
  {Brown}}]{Nitz:2017svb}%
  \BibitemOpen
  \bibfield  {author} {\bibinfo {author} {\bibfnamefont {A.~H.}\ \bibnamefont
  {Nitz}}, \bibinfo {author} {\bibfnamefont {T.}~\bibnamefont {Dent}}, \bibinfo
  {author} {\bibfnamefont {T.}~\bibnamefont {Dal~Canton}}, \bibinfo {author}
  {\bibfnamefont {S.}~\bibnamefont {Fairhurst}}, \ and\ \bibinfo {author}
  {\bibfnamefont {D.~A.}\ \bibnamefont {Brown}},\ }\href@noop {} {\  (\bibinfo
  {year} {2017}{\natexlab{a}})},\ \Eprint {http://arxiv.org/abs/1705.01513}
  {arXiv:1705.01513 [gr-qc]} \BibitemShut {NoStop}%
\bibitem [{\citenamefont {Usman}\ \emph {et~al.}(2015)\citenamefont {Usman}
  \emph {et~al.}}]{Usman:2015kfa}%
  \BibitemOpen
  \bibfield  {author} {\bibinfo {author} {\bibfnamefont {S.~A.}\ \bibnamefont
  {Usman}} \emph {et~al.},\ }\href@noop {} {\  (\bibinfo {year} {2015})},\
  \Eprint {http://arxiv.org/abs/1508.02357} {arXiv:1508.02357 [gr-qc]}
  \BibitemShut {NoStop}%
\bibitem [{\citenamefont {Dal~Canton}\ \emph {et~al.}(2014)\citenamefont
  {Dal~Canton}, \citenamefont {Nitz}, \citenamefont {Lundgren}, \citenamefont
  {Nielsen}, \citenamefont {Brown} \emph {et~al.}}]{Canton:2014ena}%
  \BibitemOpen
  \bibfield  {author} {\bibinfo {author} {\bibfnamefont {T.}~\bibnamefont
  {Dal~Canton}}, \bibinfo {author} {\bibfnamefont {A.~H.}\ \bibnamefont
  {Nitz}}, \bibinfo {author} {\bibfnamefont {A.~P.}\ \bibnamefont {Lundgren}},
  \bibinfo {author} {\bibfnamefont {A.~B.}\ \bibnamefont {Nielsen}}, \bibinfo
  {author} {\bibfnamefont {D.~A.}\ \bibnamefont {Brown}},  \emph {et~al.},\
  }\href {\doibase 10.1103/PhysRevD.90.082004} {\bibfield  {journal} {\bibinfo
  {journal} {Phys.Rev.}\ }\textbf {\bibinfo {volume} {D90}},\ \bibinfo {pages}
  {082004} (\bibinfo {year} {2014})},\ \Eprint {http://arxiv.org/abs/1405.6731}
  {arXiv:1405.6731 [gr-qc]} \BibitemShut {NoStop}%
\bibitem [{\citenamefont {Cannon}\ \emph {et~al.}(2012)\citenamefont {Cannon},
  \citenamefont {Cariou}, \citenamefont {Chapman}, \citenamefont
  {Crispin-Ortuzar}, \citenamefont {Fotopoulos} \emph
  {et~al.}}]{Cannon:2011vi}%
  \BibitemOpen
  \bibfield  {author} {\bibinfo {author} {\bibfnamefont {K.}~\bibnamefont
  {Cannon}}, \bibinfo {author} {\bibfnamefont {R.}~\bibnamefont {Cariou}},
  \bibinfo {author} {\bibfnamefont {A.}~\bibnamefont {Chapman}}, \bibinfo
  {author} {\bibfnamefont {M.}~\bibnamefont {Crispin-Ortuzar}}, \bibinfo
  {author} {\bibfnamefont {N.}~\bibnamefont {Fotopoulos}},  \emph {et~al.},\
  }\href {\doibase 10.1088/0004-637X/748/2/136} {\bibfield  {journal} {\bibinfo
   {journal} {Astrophys.J.}\ }\textbf {\bibinfo {volume} {748}},\ \bibinfo
  {pages} {136} (\bibinfo {year} {2012})},\ \Eprint
  {http://arxiv.org/abs/1107.2665} {arXiv:1107.2665 [astro-ph.IM]} \BibitemShut
  {NoStop}%
\bibitem [{\citenamefont {Cannon}\ \emph {et~al.}(2013)\citenamefont {Cannon},
  \citenamefont {Hanna},\ and\ \citenamefont {Keppel}}]{Cannon:2012zt}%
  \BibitemOpen
  \bibfield  {author} {\bibinfo {author} {\bibfnamefont {K.}~\bibnamefont
  {Cannon}}, \bibinfo {author} {\bibfnamefont {C.}~\bibnamefont {Hanna}}, \
  and\ \bibinfo {author} {\bibfnamefont {D.}~\bibnamefont {Keppel}},\ }\href
  {\doibase 10.1103/PhysRevD.88.024025} {\bibfield  {journal} {\bibinfo
  {journal} {Phys.Rev.}\ }\textbf {\bibinfo {volume} {D88}},\ \bibinfo {pages}
  {024025} (\bibinfo {year} {2013})},\ \Eprint {http://arxiv.org/abs/1209.0718}
  {arXiv:1209.0718 [gr-qc]} \BibitemShut {NoStop}%
\bibitem [{\citenamefont {Sathyaprakash}\ and\ \citenamefont
  {Dhurandhar}(1991)}]{Sathyaprakash:1991mt}%
  \BibitemOpen
  \bibfield  {author} {\bibinfo {author} {\bibfnamefont {B.~S.}\ \bibnamefont
  {Sathyaprakash}}\ and\ \bibinfo {author} {\bibfnamefont {S.~V.}\ \bibnamefont
  {Dhurandhar}},\ }\href {\doibase 10.1103/PhysRevD.44.3819} {\bibfield
  {journal} {\bibinfo  {journal} {Phys. Rev.}\ }\textbf {\bibinfo {volume}
  {D44}},\ \bibinfo {pages} {3819} (\bibinfo {year} {1991})}\BibitemShut
  {NoStop}%
\bibitem [{\citenamefont {Babak}\ \emph {et~al.}(2006)\citenamefont {Babak},
  \citenamefont {Balasubramanian}, \citenamefont {Churches}, \citenamefont
  {Cokelaer},\ and\ \citenamefont {Sathyaprakash}}]{Babak:2006ty}%
  \BibitemOpen
  \bibfield  {author} {\bibinfo {author} {\bibfnamefont {S.}~\bibnamefont
  {Babak}}, \bibinfo {author} {\bibfnamefont {R.}~\bibnamefont
  {Balasubramanian}}, \bibinfo {author} {\bibfnamefont {D.}~\bibnamefont
  {Churches}}, \bibinfo {author} {\bibfnamefont {T.}~\bibnamefont {Cokelaer}},
  \ and\ \bibinfo {author} {\bibfnamefont {B.~S.}\ \bibnamefont
  {Sathyaprakash}},\ }\href {\doibase 10.1088/0264-9381/23/18/002} {\bibfield
  {journal} {\bibinfo  {journal} {Class. Quant. Grav.}\ }\textbf {\bibinfo
  {volume} {23}},\ \bibinfo {pages} {5477} (\bibinfo {year} {2006})},\ \Eprint
  {http://arxiv.org/abs/gr-qc/0604037} {arXiv:gr-qc/0604037 [gr-qc]}
  \BibitemShut {NoStop}%
\bibitem [{\citenamefont {Ajith}\ \emph {et~al.}(2014)\citenamefont {Ajith},
  \citenamefont {Fotopoulos}, \citenamefont {Privitera}, \citenamefont
  {Neunzert},\ and\ \citenamefont {Weinstein}}]{Ajith:2012mn}%
  \BibitemOpen
  \bibfield  {author} {\bibinfo {author} {\bibfnamefont {P.}~\bibnamefont
  {Ajith}}, \bibinfo {author} {\bibfnamefont {N.}~\bibnamefont {Fotopoulos}},
  \bibinfo {author} {\bibfnamefont {S.}~\bibnamefont {Privitera}}, \bibinfo
  {author} {\bibfnamefont {A.}~\bibnamefont {Neunzert}}, \ and\ \bibinfo
  {author} {\bibfnamefont {A.~J.}\ \bibnamefont {Weinstein}},\ }\href {\doibase
  10.1103/PhysRevD.89.084041} {\bibfield  {journal} {\bibinfo  {journal} {Phys.
  Rev.}\ }\textbf {\bibinfo {volume} {D89}},\ \bibinfo {pages} {084041}
  (\bibinfo {year} {2014})},\ \Eprint {http://arxiv.org/abs/1210.6666}
  {arXiv:1210.6666 [gr-qc]} \BibitemShut {NoStop}%
\bibitem [{\citenamefont {Brown}\ \emph {et~al.}(2012)\citenamefont {Brown},
  \citenamefont {Harry}, \citenamefont {Lundgren},\ and\ \citenamefont
  {Nitz}}]{Brown:2012qf}%
  \BibitemOpen
  \bibfield  {author} {\bibinfo {author} {\bibfnamefont {D.~A.}\ \bibnamefont
  {Brown}}, \bibinfo {author} {\bibfnamefont {I.}~\bibnamefont {Harry}},
  \bibinfo {author} {\bibfnamefont {A.}~\bibnamefont {Lundgren}}, \ and\
  \bibinfo {author} {\bibfnamefont {A.~H.}\ \bibnamefont {Nitz}},\ }\href
  {\doibase 10.1103/PhysRevD.86.084017} {\bibfield  {journal} {\bibinfo
  {journal} {Phys. Rev.}\ }\textbf {\bibinfo {volume} {D86}},\ \bibinfo {pages}
  {084017} (\bibinfo {year} {2012})},\ \Eprint {http://arxiv.org/abs/1207.6406}
  {arXiv:1207.6406 [gr-qc]} \BibitemShut {NoStop}%
\bibitem [{\citenamefont {Capano}\ \emph {et~al.}(2016)\citenamefont {Capano},
  \citenamefont {Harry}, \citenamefont {Privitera},\ and\ \citenamefont
  {Buonanno}}]{Capano:2016dsf}%
  \BibitemOpen
  \bibfield  {author} {\bibinfo {author} {\bibfnamefont {C.}~\bibnamefont
  {Capano}}, \bibinfo {author} {\bibfnamefont {I.}~\bibnamefont {Harry}},
  \bibinfo {author} {\bibfnamefont {S.}~\bibnamefont {Privitera}}, \ and\
  \bibinfo {author} {\bibfnamefont {A.}~\bibnamefont {Buonanno}},\ }\href
  {\doibase 10.1103/PhysRevD.93.124007} {\bibfield  {journal} {\bibinfo
  {journal} {Phys. Rev.}\ }\textbf {\bibinfo {volume} {D93}},\ \bibinfo {pages}
  {124007} (\bibinfo {year} {2016})},\ \Eprint
  {http://arxiv.org/abs/1602.03509} {arXiv:1602.03509 [gr-qc]} \BibitemShut
  {NoStop}%
\bibitem [{\citenamefont {Pretorius}(2005)}]{Pretorius:2005gq}%
  \BibitemOpen
  \bibfield  {author} {\bibinfo {author} {\bibfnamefont {F.}~\bibnamefont
  {Pretorius}},\ }\href {\doibase 10.1103/PhysRevLett.95.121101} {\bibfield
  {journal} {\bibinfo  {journal} {Phys.Rev.Lett.}\ }\textbf {\bibinfo {volume}
  {95}},\ \bibinfo {pages} {121101} (\bibinfo {year} {2005})},\ \Eprint
  {http://arxiv.org/abs/gr-qc/0507014} {arXiv:gr-qc/0507014 [gr-qc]}
  \BibitemShut {NoStop}%
\bibitem [{\citenamefont {Szilagyi}\ \emph {et~al.}(2009)\citenamefont
  {Szilagyi}, \citenamefont {Lindblom},\ and\ \citenamefont
  {Scheel}}]{Szilagyi:2009qz}%
  \BibitemOpen
  \bibfield  {author} {\bibinfo {author} {\bibfnamefont {B.}~\bibnamefont
  {Szilagyi}}, \bibinfo {author} {\bibfnamefont {L.}~\bibnamefont {Lindblom}},
  \ and\ \bibinfo {author} {\bibfnamefont {M.~A.}\ \bibnamefont {Scheel}},\
  }\href {\doibase 10.1103/PhysRevD.80.124010} {\bibfield  {journal} {\bibinfo
  {journal} {Phys.Rev.}\ }\textbf {\bibinfo {volume} {D80}},\ \bibinfo {pages}
  {124010} (\bibinfo {year} {2009})},\ \Eprint {http://arxiv.org/abs/0909.3557}
  {arXiv:0909.3557 [gr-qc]} \BibitemShut {NoStop}%
\bibitem [{\citenamefont {Scheel}\ \emph {et~al.}(2009)\citenamefont {Scheel},
  \citenamefont {Boyle}, \citenamefont {Chu}, \citenamefont {Kidder},
  \citenamefont {Matthews} \emph {et~al.}}]{Scheel:2008rj}%
  \BibitemOpen
  \bibfield  {author} {\bibinfo {author} {\bibfnamefont {M.~A.}\ \bibnamefont
  {Scheel}}, \bibinfo {author} {\bibfnamefont {M.}~\bibnamefont {Boyle}},
  \bibinfo {author} {\bibfnamefont {T.}~\bibnamefont {Chu}}, \bibinfo {author}
  {\bibfnamefont {L.~E.}\ \bibnamefont {Kidder}}, \bibinfo {author}
  {\bibfnamefont {K.~D.}\ \bibnamefont {Matthews}},  \emph {et~al.},\ }\href
  {\doibase 10.1103/PhysRevD.79.024003} {\bibfield  {journal} {\bibinfo
  {journal} {Phys.Rev.}\ }\textbf {\bibinfo {volume} {D79}},\ \bibinfo {pages}
  {024003} (\bibinfo {year} {2009})},\ \Eprint {http://arxiv.org/abs/0810.1767}
  {arXiv:0810.1767 [gr-qc]} \BibitemShut {NoStop}%
\bibitem [{\citenamefont {Frauendiener}(2011)}]{Frauendiener:2011zz}%
  \BibitemOpen
  \bibfield  {author} {\bibinfo {author} {\bibfnamefont {J.}~\bibnamefont
  {Frauendiener}},\ }\href {\doibase 10.1007/s10714-011-1195-5} {\bibfield
  {journal} {\bibinfo  {journal} {Gen.Rel.Grav.}\ }\textbf {\bibinfo {volume}
  {43}},\ \bibinfo {pages} {2931} (\bibinfo {year} {2011})}\BibitemShut
  {NoStop}%
\bibitem [{\citenamefont {Hannam}\ \emph {et~al.}(2010)\citenamefont {Hannam},
  \citenamefont {Husa}, \citenamefont {Ohme}, \citenamefont {Muller},\ and\
  \citenamefont {Bruegmann}}]{Hannam:2010ec}%
  \BibitemOpen
  \bibfield  {author} {\bibinfo {author} {\bibfnamefont {M.}~\bibnamefont
  {Hannam}}, \bibinfo {author} {\bibfnamefont {S.}~\bibnamefont {Husa}},
  \bibinfo {author} {\bibfnamefont {F.}~\bibnamefont {Ohme}}, \bibinfo {author}
  {\bibfnamefont {D.}~\bibnamefont {Muller}}, \ and\ \bibinfo {author}
  {\bibfnamefont {B.}~\bibnamefont {Bruegmann}},\ }\href {\doibase
  10.1103/PhysRevD.82.124008} {\bibfield  {journal} {\bibinfo  {journal}
  {Phys.Rev.}\ }\textbf {\bibinfo {volume} {D82}},\ \bibinfo {pages} {124008}
  (\bibinfo {year} {2010})},\ \Eprint {http://arxiv.org/abs/1007.4789}
  {arXiv:1007.4789 [gr-qc]} \BibitemShut {NoStop}%
\bibitem [{\citenamefont {Chu}(2011)}]{ChuThesis}%
  \BibitemOpen
  \bibfield  {author} {\bibinfo {author} {\bibfnamefont {T.}~\bibnamefont
  {Chu}},\ }\emph {\bibinfo {title} {{Numerical simulations of black-hole
  spacetimes}}},\ \href {http://thesis.library.caltech.edu/6556/} {Ph.D.
  thesis},\ \bibinfo  {school} {Caltech} (\bibinfo {year} {2011})\BibitemShut
  {NoStop}%
\bibitem [{\citenamefont {Vaishnav}\ \emph {et~al.}(2007)\citenamefont
  {Vaishnav}, \citenamefont {Hinder}, \citenamefont {Herrmann},\ and\
  \citenamefont {Shoemaker}}]{Vaishnav:2007nm}%
  \BibitemOpen
  \bibfield  {author} {\bibinfo {author} {\bibfnamefont {B.}~\bibnamefont
  {Vaishnav}}, \bibinfo {author} {\bibfnamefont {I.}~\bibnamefont {Hinder}},
  \bibinfo {author} {\bibfnamefont {F.}~\bibnamefont {Herrmann}}, \ and\
  \bibinfo {author} {\bibfnamefont {D.}~\bibnamefont {Shoemaker}},\ }\href
  {\doibase 10.1103/PhysRevD.76.084020} {\bibfield  {journal} {\bibinfo
  {journal} {Phys. Rev.}\ }\textbf {\bibinfo {volume} {D76}},\ \bibinfo {pages}
  {084020} (\bibinfo {year} {2007})},\ \Eprint {http://arxiv.org/abs/0705.3829}
  {arXiv:0705.3829 [gr-qc]} \BibitemShut {NoStop}%
\bibitem [{\citenamefont {Herrmann}\ \emph {et~al.}(2007)\citenamefont
  {Herrmann}, \citenamefont {Hinder}, \citenamefont {Shoemaker},\ and\
  \citenamefont {Laguna}}]{Herrmann:2006ks}%
  \BibitemOpen
  \bibfield  {author} {\bibinfo {author} {\bibfnamefont {F.}~\bibnamefont
  {Herrmann}}, \bibinfo {author} {\bibfnamefont {I.}~\bibnamefont {Hinder}},
  \bibinfo {author} {\bibfnamefont {D.}~\bibnamefont {Shoemaker}}, \ and\
  \bibinfo {author} {\bibfnamefont {P.}~\bibnamefont {Laguna}},\ }\bibfield
  {booktitle} {\emph {\bibinfo {booktitle} {{New frontiers in numerical
  relativity. Proceedings, International Meeting, NFNR 2006, Potsdam, Germany,
  July 17-21, 2006}}},\ }\href {\doibase 10.1088/0264-9381/24/12/S04}
  {\bibfield  {journal} {\bibinfo  {journal} {Class. Quant. Grav.}\ }\textbf
  {\bibinfo {volume} {24}},\ \bibinfo {pages} {S33} (\bibinfo {year} {2007})},\
  \Eprint {http://arxiv.org/abs/gr-qc/0601026} {arXiv:gr-qc/0601026 [gr-qc]}
  \BibitemShut {NoStop}%
\bibitem [{\citenamefont {Healy}\ \emph {et~al.}(2009)\citenamefont {Healy},
  \citenamefont {Levin},\ and\ \citenamefont {Shoemaker}}]{Healy:2009zm}%
  \BibitemOpen
  \bibfield  {author} {\bibinfo {author} {\bibfnamefont {J.}~\bibnamefont
  {Healy}}, \bibinfo {author} {\bibfnamefont {J.}~\bibnamefont {Levin}}, \ and\
  \bibinfo {author} {\bibfnamefont {D.}~\bibnamefont {Shoemaker}},\ }\href
  {\doibase 10.1103/PhysRevLett.103.131101} {\bibfield  {journal} {\bibinfo
  {journal} {Phys. Rev. Lett.}\ }\textbf {\bibinfo {volume} {103}},\ \bibinfo
  {pages} {131101} (\bibinfo {year} {2009})},\ \Eprint
  {http://arxiv.org/abs/0907.0671} {arXiv:0907.0671 [gr-qc]} \BibitemShut
  {NoStop}%
\bibitem [{\citenamefont {Blanchet}\ \emph {et~al.}(1995)\citenamefont
  {Blanchet}, \citenamefont {Damour}, \citenamefont {Iyer}, \citenamefont
  {Will},\ and\ \citenamefont {Wiseman}}]{Blanchet:1995ez}%
  \BibitemOpen
  \bibfield  {author} {\bibinfo {author} {\bibfnamefont {L.}~\bibnamefont
  {Blanchet}}, \bibinfo {author} {\bibfnamefont {T.}~\bibnamefont {Damour}},
  \bibinfo {author} {\bibfnamefont {B.~R.}\ \bibnamefont {Iyer}}, \bibinfo
  {author} {\bibfnamefont {C.~M.}\ \bibnamefont {Will}}, \ and\ \bibinfo
  {author} {\bibfnamefont {A.}~\bibnamefont {Wiseman}},\ }\href {\doibase
  10.1103/PhysRevLett.74.3515} {\bibfield  {journal} {\bibinfo  {journal}
  {Phys.Rev.Lett.}\ }\textbf {\bibinfo {volume} {74}},\ \bibinfo {pages} {3515}
  (\bibinfo {year} {1995})},\ \Eprint {http://arxiv.org/abs/gr-qc/9501027}
  {arXiv:gr-qc/9501027 [gr-qc]} \BibitemShut {NoStop}%
\bibitem [{\citenamefont {Buonanno}\ and\ \citenamefont
  {Damour}(1999)}]{Buonanno:1998gg}%
  \BibitemOpen
  \bibfield  {author} {\bibinfo {author} {\bibfnamefont {A.}~\bibnamefont
  {Buonanno}}\ and\ \bibinfo {author} {\bibfnamefont {T.}~\bibnamefont
  {Damour}},\ }\href {\doibase 10.1103/PhysRevD.59.084006} {\bibfield
  {journal} {\bibinfo  {journal} {Phys.Rev.}\ }\textbf {\bibinfo {volume}
  {D59}},\ \bibinfo {pages} {084006} (\bibinfo {year} {1999})},\ \Eprint
  {http://arxiv.org/abs/gr-qc/9811091} {arXiv:gr-qc/9811091 [gr-qc]}
  \BibitemShut {NoStop}%
\bibitem [{\citenamefont {Taracchini}\ \emph {et~al.}(2012)\citenamefont
  {Taracchini}, \citenamefont {Pan}, \citenamefont {Buonanno}, \citenamefont
  {Barausse}, \citenamefont {Boyle} \emph {et~al.}}]{Taracchini:2012ig}%
  \BibitemOpen
  \bibfield  {author} {\bibinfo {author} {\bibfnamefont {A.}~\bibnamefont
  {Taracchini}}, \bibinfo {author} {\bibfnamefont {Y.}~\bibnamefont {Pan}},
  \bibinfo {author} {\bibfnamefont {A.}~\bibnamefont {Buonanno}}, \bibinfo
  {author} {\bibfnamefont {E.}~\bibnamefont {Barausse}}, \bibinfo {author}
  {\bibfnamefont {M.}~\bibnamefont {Boyle}},  \emph {et~al.},\ }\href {\doibase
  10.1103/PhysRevD.86.024011} {\bibfield  {journal} {\bibinfo  {journal}
  {Phys.Rev.}\ }\textbf {\bibinfo {volume} {D86}},\ \bibinfo {pages} {024011}
  (\bibinfo {year} {2012})},\ \Eprint {http://arxiv.org/abs/1202.0790}
  {arXiv:1202.0790 [gr-qc]} \BibitemShut {NoStop}%
\bibitem [{\citenamefont {Boh{\'e}}\ \emph {et~al.}(2017)\citenamefont
  {Boh{\'e}} \emph {et~al.}}]{Bohe:2016gbl}%
  \BibitemOpen
  \bibfield  {author} {\bibinfo {author} {\bibfnamefont {A.}~\bibnamefont
  {Boh{\'e}}} \emph {et~al.},\ }\href {\doibase 10.1103/PhysRevD.95.044028}
  {\bibfield  {journal} {\bibinfo  {journal} {Phys. Rev.}\ }\textbf {\bibinfo
  {volume} {D95}},\ \bibinfo {pages} {044028} (\bibinfo {year} {2017})},\
  \Eprint {http://arxiv.org/abs/1611.03703} {arXiv:1611.03703 [gr-qc]}
  \BibitemShut {NoStop}%
\bibitem [{\citenamefont {Ajith}\ \emph {et~al.}(2008)\citenamefont {Ajith},
  \citenamefont {Babak}, \citenamefont {Chen}, \citenamefont {Hewitson},
  \citenamefont {Krishnan} \emph {et~al.}}]{Ajith:2007kx}%
  \BibitemOpen
  \bibfield  {author} {\bibinfo {author} {\bibfnamefont {P.}~\bibnamefont
  {Ajith}}, \bibinfo {author} {\bibfnamefont {S.}~\bibnamefont {Babak}},
  \bibinfo {author} {\bibfnamefont {Y.}~\bibnamefont {Chen}}, \bibinfo {author}
  {\bibfnamefont {M.}~\bibnamefont {Hewitson}}, \bibinfo {author}
  {\bibfnamefont {B.}~\bibnamefont {Krishnan}},  \emph {et~al.},\ }\href
  {\doibase 10.1103/PhysRevD.79.129901, 10.1103/PhysRevD.77.104017} {\bibfield
  {journal} {\bibinfo  {journal} {Phys.Rev.}\ }\textbf {\bibinfo {volume}
  {D77}},\ \bibinfo {pages} {104017} (\bibinfo {year} {2008})},\ \Eprint
  {http://arxiv.org/abs/0710.2335} {arXiv:0710.2335 [gr-qc]} \BibitemShut
  {NoStop}%
\bibitem [{\citenamefont {Khan}\ \emph {et~al.}(2016)\citenamefont {Khan},
  \citenamefont {Husa}, \citenamefont {Hannam}, \citenamefont {Ohme},
  \citenamefont {P{\"u}rrer}, \citenamefont {Jim{\'e}nez~Forteza},\ and\
  \citenamefont {Boh{\'e}}}]{Khan:2015jqa}%
  \BibitemOpen
  \bibfield  {author} {\bibinfo {author} {\bibfnamefont {S.}~\bibnamefont
  {Khan}}, \bibinfo {author} {\bibfnamefont {S.}~\bibnamefont {Husa}}, \bibinfo
  {author} {\bibfnamefont {M.}~\bibnamefont {Hannam}}, \bibinfo {author}
  {\bibfnamefont {F.}~\bibnamefont {Ohme}}, \bibinfo {author} {\bibfnamefont
  {M.}~\bibnamefont {P{\"u}rrer}}, \bibinfo {author} {\bibfnamefont
  {X.}~\bibnamefont {Jim{\'e}nez~Forteza}}, \ and\ \bibinfo {author}
  {\bibfnamefont {A.}~\bibnamefont {Boh{\'e}}},\ }\href {\doibase
  10.1103/PhysRevD.93.044007} {\bibfield  {journal} {\bibinfo  {journal} {Phys.
  Rev.}\ }\textbf {\bibinfo {volume} {D93}},\ \bibinfo {pages} {044007}
  (\bibinfo {year} {2016})},\ \Eprint {http://arxiv.org/abs/1508.07253}
  {arXiv:1508.07253 [gr-qc]} \BibitemShut {NoStop}%
\bibitem [{\citenamefont {Apostolatos}\ \emph {et~al.}(1994)\citenamefont
  {Apostolatos}, \citenamefont {Cutler}, \citenamefont {Sussman},\ and\
  \citenamefont {Thorne}}]{Apostolatos:1994mx}%
  \BibitemOpen
  \bibfield  {author} {\bibinfo {author} {\bibfnamefont {T.~A.}\ \bibnamefont
  {Apostolatos}}, \bibinfo {author} {\bibfnamefont {C.}~\bibnamefont {Cutler}},
  \bibinfo {author} {\bibfnamefont {G.~J.}\ \bibnamefont {Sussman}}, \ and\
  \bibinfo {author} {\bibfnamefont {K.~S.}\ \bibnamefont {Thorne}},\ }\href
  {\doibase 10.1103/PhysRevD.49.6274} {\bibfield  {journal} {\bibinfo
  {journal} {Phys. Rev.}\ }\textbf {\bibinfo {volume} {D49}},\ \bibinfo {pages}
  {6274} (\bibinfo {year} {1994})}\BibitemShut {NoStop}%
\bibitem [{\citenamefont {Damour}\ \emph {et~al.}(2004)\citenamefont {Damour},
  \citenamefont {Gopakumar},\ and\ \citenamefont {Iyer}}]{Damour:2004bz}%
  \BibitemOpen
  \bibfield  {author} {\bibinfo {author} {\bibfnamefont {T.}~\bibnamefont
  {Damour}}, \bibinfo {author} {\bibfnamefont {A.}~\bibnamefont {Gopakumar}}, \
  and\ \bibinfo {author} {\bibfnamefont {B.~R.}\ \bibnamefont {Iyer}},\ }\href
  {\doibase 10.1103/PhysRevD.70.064028} {\bibfield  {journal} {\bibinfo
  {journal} {Phys. Rev.}\ }\textbf {\bibinfo {volume} {D70}},\ \bibinfo {pages}
  {064028} (\bibinfo {year} {2004})},\ \Eprint
  {http://arxiv.org/abs/gr-qc/0404128} {arXiv:gr-qc/0404128 [gr-qc]}
  \BibitemShut {NoStop}%
\bibitem [{\citenamefont {Flanagan}\ and\ \citenamefont
  {Hinderer}(2008)}]{Flanagan:2007ix}%
  \BibitemOpen
  \bibfield  {author} {\bibinfo {author} {\bibfnamefont {E.~E.}\ \bibnamefont
  {Flanagan}}\ and\ \bibinfo {author} {\bibfnamefont {T.}~\bibnamefont
  {Hinderer}},\ }\href {\doibase 10.1103/PhysRevD.77.021502} {\bibfield
  {journal} {\bibinfo  {journal} {Phys. Rev.}\ }\textbf {\bibinfo {volume}
  {D77}},\ \bibinfo {pages} {021502} (\bibinfo {year} {2008})},\ \Eprint
  {http://arxiv.org/abs/0709.1915} {arXiv:0709.1915 [astro-ph]} \BibitemShut
  {NoStop}%
\bibitem [{\citenamefont {Read}\ \emph {et~al.}(2013)\citenamefont {Read},
  \citenamefont {Baiotti}, \citenamefont {Creighton}, \citenamefont {Friedman},
  \citenamefont {Giacomazzo}, \citenamefont {Kyutoku}, \citenamefont
  {Markakis}, \citenamefont {Rezzolla}, \citenamefont {Shibata},\ and\
  \citenamefont {Taniguchi}}]{Read:2013zra}%
  \BibitemOpen
  \bibfield  {author} {\bibinfo {author} {\bibfnamefont {J.~S.}\ \bibnamefont
  {Read}}, \bibinfo {author} {\bibfnamefont {L.}~\bibnamefont {Baiotti}},
  \bibinfo {author} {\bibfnamefont {J.~D.~E.}\ \bibnamefont {Creighton}},
  \bibinfo {author} {\bibfnamefont {J.~L.}\ \bibnamefont {Friedman}}, \bibinfo
  {author} {\bibfnamefont {B.}~\bibnamefont {Giacomazzo}}, \bibinfo {author}
  {\bibfnamefont {K.}~\bibnamefont {Kyutoku}}, \bibinfo {author} {\bibfnamefont
  {C.}~\bibnamefont {Markakis}}, \bibinfo {author} {\bibfnamefont
  {L.}~\bibnamefont {Rezzolla}}, \bibinfo {author} {\bibfnamefont
  {M.}~\bibnamefont {Shibata}}, \ and\ \bibinfo {author} {\bibfnamefont
  {K.}~\bibnamefont {Taniguchi}},\ }\href {\doibase 10.1103/PhysRevD.88.044042}
  {\bibfield  {journal} {\bibinfo  {journal} {Phys. Rev.}\ }\textbf {\bibinfo
  {volume} {D88}},\ \bibinfo {pages} {044042} (\bibinfo {year} {2013})},\
  \Eprint {http://arxiv.org/abs/1306.4065} {arXiv:1306.4065 [gr-qc]}
  \BibitemShut {NoStop}%
\bibitem [{\citenamefont {Blanchet}\ \emph {et~al.}(2008)\citenamefont
  {Blanchet}, \citenamefont {Faye}, \citenamefont {Iyer},\ and\ \citenamefont
  {Sinha}}]{Blanchet2008}%
  \BibitemOpen
  \bibfield  {author} {\bibinfo {author} {\bibfnamefont {L.}~\bibnamefont
  {Blanchet}}, \bibinfo {author} {\bibfnamefont {G.}~\bibnamefont {Faye}},
  \bibinfo {author} {\bibfnamefont {B.~R.}\ \bibnamefont {Iyer}}, \ and\
  \bibinfo {author} {\bibfnamefont {S.}~\bibnamefont {Sinha}},\ }\href
  {\doibase 10.1088/0264-9381/25/16/165003} {\bibfield  {journal} {\bibinfo
  {journal} {Class.Quant.Grav.}\ }\textbf {\bibinfo {volume} {25}},\ \bibinfo
  {pages} {165003} (\bibinfo {year} {2008})},\ \Eprint
  {http://arxiv.org/abs/0802.1249} {arXiv:0802.1249 [gr-qc]} \BibitemShut
  {NoStop}%
\bibitem [{\citenamefont {O'Shaughnessy}\ \emph {et~al.}(2014)\citenamefont
  {O'Shaughnessy}, \citenamefont {Farr}, \citenamefont {Ochsner}, \citenamefont
  {Cho}, \citenamefont {Raymond}, \citenamefont {Kim},\ and\ \citenamefont
  {Lee}}]{OShaughnessy:2014shr}%
  \BibitemOpen
  \bibfield  {author} {\bibinfo {author} {\bibfnamefont {R.}~\bibnamefont
  {O'Shaughnessy}}, \bibinfo {author} {\bibfnamefont {B.}~\bibnamefont {Farr}},
  \bibinfo {author} {\bibfnamefont {E.}~\bibnamefont {Ochsner}}, \bibinfo
  {author} {\bibfnamefont {H.-S.}\ \bibnamefont {Cho}}, \bibinfo {author}
  {\bibfnamefont {V.}~\bibnamefont {Raymond}}, \bibinfo {author} {\bibfnamefont
  {C.}~\bibnamefont {Kim}}, \ and\ \bibinfo {author} {\bibfnamefont {C.-H.}\
  \bibnamefont {Lee}},\ }\href {\doibase 10.1103/PhysRevD.89.102005} {\bibfield
   {journal} {\bibinfo  {journal} {Phys. Rev.}\ }\textbf {\bibinfo {volume}
  {D89}},\ \bibinfo {pages} {102005} (\bibinfo {year} {2014})},\ \Eprint
  {http://arxiv.org/abs/1403.0544} {arXiv:1403.0544 [gr-qc]} \BibitemShut
  {NoStop}%
\bibitem [{\citenamefont {McKechan}(2011)}]{McKechan:2011ps}%
  \BibitemOpen
  \bibfield  {author} {\bibinfo {author} {\bibfnamefont {D.~J.~A.}\
  \bibnamefont {McKechan}},\ }\emph {\bibinfo {title} {{On the use of higher
  order waveforms in the search for gravitational waves emitted by compact
  binary coalescences}}},\ \href
  {http://inspirehep.net/record/889087/files/arXiv:1102.1749.pdf} {Ph.D.
  thesis},\ \bibinfo  {school} {Cardiff U.} (\bibinfo {year} {2011}),\ \Eprint
  {http://arxiv.org/abs/1102.1749} {arXiv:1102.1749 [gr-qc]} \BibitemShut
  {NoStop}%
\bibitem [{\citenamefont {Pekowsky}\ \emph {et~al.}(2013)\citenamefont
  {Pekowsky}, \citenamefont {Healy}, \citenamefont {Shoemaker},\ and\
  \citenamefont {Laguna}}]{Pekowsky:2012sr}%
  \BibitemOpen
  \bibfield  {author} {\bibinfo {author} {\bibfnamefont {L.}~\bibnamefont
  {Pekowsky}}, \bibinfo {author} {\bibfnamefont {J.}~\bibnamefont {Healy}},
  \bibinfo {author} {\bibfnamefont {D.}~\bibnamefont {Shoemaker}}, \ and\
  \bibinfo {author} {\bibfnamefont {P.}~\bibnamefont {Laguna}},\ }\href
  {\doibase 10.1103/PhysRevD.87.084008} {\bibfield  {journal} {\bibinfo
  {journal} {Phys.Rev.}\ }\textbf {\bibinfo {volume} {D87}},\ \bibinfo {pages}
  {084008} (\bibinfo {year} {2013})},\ \Eprint {http://arxiv.org/abs/1210.1891}
  {arXiv:1210.1891 [gr-qc]} \BibitemShut {NoStop}%
\bibitem [{\citenamefont {Capano}\ \emph {et~al.}(2014)\citenamefont {Capano},
  \citenamefont {Pan},\ and\ \citenamefont {Buonanno}}]{Capano:2013raa}%
  \BibitemOpen
  \bibfield  {author} {\bibinfo {author} {\bibfnamefont {C.}~\bibnamefont
  {Capano}}, \bibinfo {author} {\bibfnamefont {Y.}~\bibnamefont {Pan}}, \ and\
  \bibinfo {author} {\bibfnamefont {A.}~\bibnamefont {Buonanno}},\ }\href
  {\doibase 10.1103/PhysRevD.89.102003} {\bibfield  {journal} {\bibinfo
  {journal} {Phys.Rev.}\ }\textbf {\bibinfo {volume} {D89}},\ \bibinfo {pages}
  {102003} (\bibinfo {year} {2014})},\ \Eprint {http://arxiv.org/abs/1311.1286}
  {arXiv:1311.1286 [gr-qc]} \BibitemShut {NoStop}%
\bibitem [{\citenamefont {Varma}\ \emph {et~al.}(2014)\citenamefont {Varma},
  \citenamefont {Ajith}, \citenamefont {Husa}, \citenamefont {Bustillo},
  \citenamefont {Hannam} \emph {et~al.}}]{Varma:2014jxa}%
  \BibitemOpen
  \bibfield  {author} {\bibinfo {author} {\bibfnamefont {V.}~\bibnamefont
  {Varma}}, \bibinfo {author} {\bibfnamefont {P.}~\bibnamefont {Ajith}},
  \bibinfo {author} {\bibfnamefont {S.}~\bibnamefont {Husa}}, \bibinfo {author}
  {\bibfnamefont {J.~C.}\ \bibnamefont {Bustillo}}, \bibinfo {author}
  {\bibfnamefont {M.}~\bibnamefont {Hannam}},  \emph {et~al.},\ }\href@noop {}
  {\  (\bibinfo {year} {2014})},\ \Eprint {http://arxiv.org/abs/1409.2349}
  {arXiv:1409.2349 [gr-qc]} \BibitemShut {NoStop}%
\bibitem [{\citenamefont {Calder{\'o}n~Bustillo}\ \emph
  {et~al.}(2016)\citenamefont {Calder{\'o}n~Bustillo}, \citenamefont {Husa},
  \citenamefont {Sintes},\ and\ \citenamefont {P{\"u}rrer}}]{Bustillo:2015qty}%
  \BibitemOpen
  \bibfield  {author} {\bibinfo {author} {\bibfnamefont {J.}~\bibnamefont
  {Calder{\'o}n~Bustillo}}, \bibinfo {author} {\bibfnamefont {S.}~\bibnamefont
  {Husa}}, \bibinfo {author} {\bibfnamefont {A.~M.}\ \bibnamefont {Sintes}}, \
  and\ \bibinfo {author} {\bibfnamefont {M.}~\bibnamefont {P{\"u}rrer}},\
  }\href {\doibase 10.1103/PhysRevD.93.084019} {\bibfield  {journal} {\bibinfo
  {journal} {Phys. Rev.}\ }\textbf {\bibinfo {volume} {D93}},\ \bibinfo {pages}
  {084019} (\bibinfo {year} {2016})},\ \Eprint
  {http://arxiv.org/abs/1511.02060} {arXiv:1511.02060 [gr-qc]} \BibitemShut
  {NoStop}%
\bibitem [{\citenamefont {Calder{\'o}n~Bustillo}\ \emph
  {et~al.}(2017)\citenamefont {Calder{\'o}n~Bustillo}, \citenamefont {Laguna},\
  and\ \citenamefont {Shoemaker}}]{Bustillo:2016gid}%
  \BibitemOpen
  \bibfield  {author} {\bibinfo {author} {\bibfnamefont {J.}~\bibnamefont
  {Calder{\'o}n~Bustillo}}, \bibinfo {author} {\bibfnamefont {P.}~\bibnamefont
  {Laguna}}, \ and\ \bibinfo {author} {\bibfnamefont {D.}~\bibnamefont
  {Shoemaker}},\ }\href {\doibase 10.1103/PhysRevD.95.104038} {\bibfield
  {journal} {\bibinfo  {journal} {Phys. Rev.}\ }\textbf {\bibinfo {volume}
  {D95}},\ \bibinfo {pages} {104038} (\bibinfo {year} {2017})},\ \Eprint
  {http://arxiv.org/abs/1612.02340} {arXiv:1612.02340 [gr-qc]} \BibitemShut
  {NoStop}%
\bibitem [{\citenamefont {Klimenko}\ \emph {et~al.}(2008)\citenamefont
  {Klimenko}, \citenamefont {Yakushin}, \citenamefont {Mercer},\ and\
  \citenamefont {Mitselmakher}}]{Klimenko:2008fu}%
  \BibitemOpen
  \bibfield  {author} {\bibinfo {author} {\bibfnamefont {S.}~\bibnamefont
  {Klimenko}}, \bibinfo {author} {\bibfnamefont {I.}~\bibnamefont {Yakushin}},
  \bibinfo {author} {\bibfnamefont {A.}~\bibnamefont {Mercer}}, \ and\ \bibinfo
  {author} {\bibfnamefont {G.}~\bibnamefont {Mitselmakher}},\ }\bibfield
  {booktitle} {\emph {\bibinfo {booktitle} {{Proceedings, 18th International
  Conference on General Relativity and Gravitation (GRG18) and 7th Edoardo
  Amaldi Conference on Gravitational Waves (Amaldi7), Sydney, Australia, July
  2007}}},\ }\href {\doibase 10.1088/0264-9381/25/11/114029} {\bibfield
  {journal} {\bibinfo  {journal} {Class. Quant. Grav.}\ }\textbf {\bibinfo
  {volume} {25}},\ \bibinfo {pages} {114029} (\bibinfo {year} {2008})},\
  \Eprint {http://arxiv.org/abs/0802.3232} {arXiv:0802.3232 [gr-qc]}
  \BibitemShut {NoStop}%
\bibitem [{\citenamefont {Sutton}\ \emph {et~al.}(2010)\citenamefont {Sutton}
  \emph {et~al.}}]{Sutton:2009gi}%
  \BibitemOpen
  \bibfield  {author} {\bibinfo {author} {\bibfnamefont {P.~J.}\ \bibnamefont
  {Sutton}} \emph {et~al.},\ }\href {\doibase 10.1088/1367-2630/12/5/053034}
  {\bibfield  {journal} {\bibinfo  {journal} {New J. Phys.}\ }\textbf {\bibinfo
  {volume} {12}},\ \bibinfo {pages} {053034} (\bibinfo {year} {2010})},\
  \Eprint {http://arxiv.org/abs/0908.3665} {arXiv:0908.3665 [gr-qc]}
  \BibitemShut {NoStop}%
\bibitem [{\citenamefont {Cornish}\ and\ \citenamefont
  {Littenberg}(2015)}]{Cornish:2014kda}%
  \BibitemOpen
  \bibfield  {author} {\bibinfo {author} {\bibfnamefont {N.~J.}\ \bibnamefont
  {Cornish}}\ and\ \bibinfo {author} {\bibfnamefont {T.~B.}\ \bibnamefont
  {Littenberg}},\ }\href {\doibase 10.1088/0264-9381/32/13/135012} {\bibfield
  {journal} {\bibinfo  {journal} {Class. Quant. Grav.}\ }\textbf {\bibinfo
  {volume} {32}},\ \bibinfo {pages} {135012} (\bibinfo {year} {2015})},\
  \Eprint {http://arxiv.org/abs/1410.3835} {arXiv:1410.3835 [gr-qc]}
  \BibitemShut {NoStop}%
\bibitem [{\citenamefont {Lynch}\ \emph {et~al.}(2017)\citenamefont {Lynch},
  \citenamefont {Vitale}, \citenamefont {Essick}, \citenamefont
  {Katsavounidis},\ and\ \citenamefont {Robinet}}]{Lynch:2015yin}%
  \BibitemOpen
  \bibfield  {author} {\bibinfo {author} {\bibfnamefont {R.}~\bibnamefont
  {Lynch}}, \bibinfo {author} {\bibfnamefont {S.}~\bibnamefont {Vitale}},
  \bibinfo {author} {\bibfnamefont {R.}~\bibnamefont {Essick}}, \bibinfo
  {author} {\bibfnamefont {E.}~\bibnamefont {Katsavounidis}}, \ and\ \bibinfo
  {author} {\bibfnamefont {F.}~\bibnamefont {Robinet}},\ }\href {\doibase
  10.1103/PhysRevD.95.104046} {\bibfield  {journal} {\bibinfo  {journal} {Phys.
  Rev.}\ }\textbf {\bibinfo {volume} {D95}},\ \bibinfo {pages} {104046}
  (\bibinfo {year} {2017})},\ \Eprint {http://arxiv.org/abs/1511.05955}
  {arXiv:1511.05955 [gr-qc]} \BibitemShut {NoStop}%
\bibitem [{\citenamefont {Klimenko}\ \emph {et~al.}(2016)\citenamefont
  {Klimenko} \emph {et~al.}}]{Klimenko:2015ypf}%
  \BibitemOpen
  \bibfield  {author} {\bibinfo {author} {\bibfnamefont {S.}~\bibnamefont
  {Klimenko}} \emph {et~al.},\ }\href {\doibase 10.1103/PhysRevD.93.042004}
  {\bibfield  {journal} {\bibinfo  {journal} {Phys. Rev.}\ }\textbf {\bibinfo
  {volume} {D93}},\ \bibinfo {pages} {042004} (\bibinfo {year} {2016})},\
  \Eprint {http://arxiv.org/abs/1511.05999} {arXiv:1511.05999 [gr-qc]}
  \BibitemShut {NoStop}%
\bibitem [{\citenamefont {Mazzolo}\ \emph {et~al.}(2014)\citenamefont {Mazzolo}
  \emph {et~al.}}]{Mazzolo:2014kta}%
  \BibitemOpen
  \bibfield  {author} {\bibinfo {author} {\bibfnamefont {G.}~\bibnamefont
  {Mazzolo}} \emph {et~al.},\ }\href {\doibase 10.1103/PhysRevD.90.063002}
  {\bibfield  {journal} {\bibinfo  {journal} {Phys. Rev.}\ }\textbf {\bibinfo
  {volume} {D90}},\ \bibinfo {pages} {063002} (\bibinfo {year} {2014})},\
  \Eprint {http://arxiv.org/abs/1404.7757} {arXiv:1404.7757 [gr-qc]}
  \BibitemShut {NoStop}%
\bibitem [{\citenamefont {Pan}\ \emph {et~al.}(2011)\citenamefont {Pan},
  \citenamefont {Buonanno}, \citenamefont {Boyle}, \citenamefont {Buchman},
  \citenamefont {Kidder}, \citenamefont {Pfeiffer},\ and\ \citenamefont
  {Scheel}}]{Pan:2011gk}%
  \BibitemOpen
  \bibfield  {author} {\bibinfo {author} {\bibfnamefont {Y.}~\bibnamefont
  {Pan}}, \bibinfo {author} {\bibfnamefont {A.}~\bibnamefont {Buonanno}},
  \bibinfo {author} {\bibfnamefont {M.}~\bibnamefont {Boyle}}, \bibinfo
  {author} {\bibfnamefont {L.~T.}\ \bibnamefont {Buchman}}, \bibinfo {author}
  {\bibfnamefont {L.~E.}\ \bibnamefont {Kidder}}, \bibinfo {author}
  {\bibfnamefont {H.~P.}\ \bibnamefont {Pfeiffer}}, \ and\ \bibinfo {author}
  {\bibfnamefont {M.~A.}\ \bibnamefont {Scheel}},\ }\href {\doibase
  10.1103/PhysRevD.84.124052} {\bibfield  {journal} {\bibinfo  {journal} {Phys.
  Rev.}\ }\textbf {\bibinfo {volume} {D84}},\ \bibinfo {pages} {124052}
  (\bibinfo {year} {2011})},\ \Eprint {http://arxiv.org/abs/1106.1021}
  {arXiv:1106.1021 [gr-qc]} \BibitemShut {NoStop}%
\bibitem [{\citenamefont {London}\ \emph {et~al.}(2017)\citenamefont {London},
  \citenamefont {Khan}, \citenamefont {Fauchon-Jones}, \citenamefont {Forteza},
  \citenamefont {Hannam}, \citenamefont {Husa}, \citenamefont {Kalaghatgi},
  \citenamefont {Ohme},\ and\ \citenamefont {Pannarale}}]{London:2017bcn}%
  \BibitemOpen
  \bibfield  {author} {\bibinfo {author} {\bibfnamefont {L.}~\bibnamefont
  {London}}, \bibinfo {author} {\bibfnamefont {S.}~\bibnamefont {Khan}},
  \bibinfo {author} {\bibfnamefont {E.}~\bibnamefont {Fauchon-Jones}}, \bibinfo
  {author} {\bibfnamefont {X.~J.}\ \bibnamefont {Forteza}}, \bibinfo {author}
  {\bibfnamefont {M.}~\bibnamefont {Hannam}}, \bibinfo {author} {\bibfnamefont
  {S.}~\bibnamefont {Husa}}, \bibinfo {author} {\bibfnamefont {C.}~\bibnamefont
  {Kalaghatgi}}, \bibinfo {author} {\bibfnamefont {F.}~\bibnamefont {Ohme}}, \
  and\ \bibinfo {author} {\bibfnamefont {F.}~\bibnamefont {Pannarale}},\
  }\href@noop {} {\  (\bibinfo {year} {2017})},\ \Eprint
  {http://arxiv.org/abs/1708.00404} {arXiv:1708.00404 [gr-qc]} \BibitemShut
  {NoStop}%
\bibitem [{\citenamefont {Cotesta}\ \emph {et~al.}(2017)\citenamefont {Cotesta}
  \emph {et~al.}}]{COTESTA}%
  \BibitemOpen
  \bibfield  {author} {\bibinfo {author} {\bibfnamefont {R.}~\bibnamefont
  {Cotesta}} \emph {et~al.},\ }\href@noop {} {\bibfield  {journal} {\bibinfo
  {journal} {In preparation}\ } (\bibinfo {year} {2017})}\BibitemShut {NoStop}%
\bibitem [{\citenamefont {Gossan}\ \emph {et~al.}(2012)\citenamefont {Gossan},
  \citenamefont {Veitch},\ and\ \citenamefont {Sathyaprakash}}]{Gossan:2011ha}%
  \BibitemOpen
  \bibfield  {author} {\bibinfo {author} {\bibfnamefont {S.}~\bibnamefont
  {Gossan}}, \bibinfo {author} {\bibfnamefont {J.}~\bibnamefont {Veitch}}, \
  and\ \bibinfo {author} {\bibfnamefont {B.~S.}\ \bibnamefont
  {Sathyaprakash}},\ }\href {\doibase 10.1103/PhysRevD.85.124056} {\bibfield
  {journal} {\bibinfo  {journal} {Phys. Rev.}\ }\textbf {\bibinfo {volume}
  {D85}},\ \bibinfo {pages} {124056} (\bibinfo {year} {2012})},\ \Eprint
  {http://arxiv.org/abs/1111.5819} {arXiv:1111.5819 [gr-qc]} \BibitemShut
  {NoStop}%
\bibitem [{\citenamefont {Yang}\ \emph {et~al.}(2017)\citenamefont {Yang},
  \citenamefont {Yagi}, \citenamefont {Blackman}, \citenamefont {Lehner},
  \citenamefont {Paschalidis}, \citenamefont {Pretorius},\ and\ \citenamefont
  {Yunes}}]{Yang:2017zxs}%
  \BibitemOpen
  \bibfield  {author} {\bibinfo {author} {\bibfnamefont {H.}~\bibnamefont
  {Yang}}, \bibinfo {author} {\bibfnamefont {K.}~\bibnamefont {Yagi}}, \bibinfo
  {author} {\bibfnamefont {J.}~\bibnamefont {Blackman}}, \bibinfo {author}
  {\bibfnamefont {L.}~\bibnamefont {Lehner}}, \bibinfo {author} {\bibfnamefont
  {V.}~\bibnamefont {Paschalidis}}, \bibinfo {author} {\bibfnamefont
  {F.}~\bibnamefont {Pretorius}}, \ and\ \bibinfo {author} {\bibfnamefont
  {N.}~\bibnamefont {Yunes}},\ }\href {\doibase 10.1103/PhysRevLett.118.161101}
  {\bibfield  {journal} {\bibinfo  {journal} {Phys. Rev. Lett.}\ }\textbf
  {\bibinfo {volume} {118}},\ \bibinfo {pages} {161101} (\bibinfo {year}
  {2017})},\ \Eprint {http://arxiv.org/abs/1701.05808} {arXiv:1701.05808
  [gr-qc]} \BibitemShut {NoStop}%
\bibitem [{\citenamefont {Berti}\ \emph {et~al.}(2015)\citenamefont {Berti}
  \emph {et~al.}}]{Berti:2015itd}%
  \BibitemOpen
  \bibfield  {author} {\bibinfo {author} {\bibfnamefont {E.}~\bibnamefont
  {Berti}} \emph {et~al.},\ }\href {\doibase 10.1088/0264-9381/32/24/243001}
  {\bibfield  {journal} {\bibinfo  {journal} {Class. Quant. Grav.}\ }\textbf
  {\bibinfo {volume} {32}},\ \bibinfo {pages} {243001} (\bibinfo {year}
  {2015})},\ \Eprint {http://arxiv.org/abs/1501.07274} {arXiv:1501.07274
  [gr-qc]} \BibitemShut {NoStop}%
\bibitem [{\citenamefont {Sigg}(2016)}]{O1sens}%
  \BibitemOpen
  \bibfield  {author} {\bibinfo {author} {\bibfnamefont {D.}~\bibnamefont
  {Sigg}},\ }\href@noop {} {\enquote {\bibinfo {title} {{Advanced LIGO first
  observing run representative sensitivity}},}\ } (\bibinfo {year} {2016}),\
  \bibinfo {note} {{\url{https://dcc.ligo.org/LIGO-G1600150/public}
  \url{https://dcc.ligo.org/LIGO-G1600151/public}}}\BibitemShut {NoStop}%
\bibitem [{\citenamefont {Shoemaker}(2015)}]{ZDHP}%
  \BibitemOpen
  \bibfield  {author} {\bibinfo {author} {\bibfnamefont {D.}~\bibnamefont
  {Shoemaker}},\ }\href@noop {} {\enquote {\bibinfo {title} {Advanced ligo
  anticipated sensitivity curves},}\ } (\bibinfo {year} {2015}),\ \bibinfo
  {note} {{\url{https://dcc.ligo.org/LIGO-T0900288/public}}}\BibitemShut
  {NoStop}%
\bibitem [{\citenamefont {Finn}\ and\ \citenamefont
  {Chernoff}(1993)}]{Finn:1992xs}%
  \BibitemOpen
  \bibfield  {author} {\bibinfo {author} {\bibfnamefont {L.~S.}\ \bibnamefont
  {Finn}}\ and\ \bibinfo {author} {\bibfnamefont {D.~F.}\ \bibnamefont
  {Chernoff}},\ }\href {\doibase 10.1103/PhysRevD.47.2198} {\bibfield
  {journal} {\bibinfo  {journal} {Phys. Rev.}\ }\textbf {\bibinfo {volume}
  {D47}},\ \bibinfo {pages} {2198} (\bibinfo {year} {1993})},\ \Eprint
  {http://arxiv.org/abs/gr-qc/9301003} {arXiv:gr-qc/9301003 [gr-qc]}
  \BibitemShut {NoStop}%
\bibitem [{\citenamefont {Jaranowski}\ \emph {et~al.}(1998)\citenamefont
  {Jaranowski}, \citenamefont {Krolak},\ and\ \citenamefont
  {Schutz}}]{Jaranowski:1998qm}%
  \BibitemOpen
  \bibfield  {author} {\bibinfo {author} {\bibfnamefont {P.}~\bibnamefont
  {Jaranowski}}, \bibinfo {author} {\bibfnamefont {A.}~\bibnamefont {Krolak}},
  \ and\ \bibinfo {author} {\bibfnamefont {B.~F.}\ \bibnamefont {Schutz}},\
  }\href {\doibase 10.1103/PhysRevD.58.063001} {\bibfield  {journal} {\bibinfo
  {journal} {Phys.Rev.}\ }\textbf {\bibinfo {volume} {D58}},\ \bibinfo {pages}
  {063001} (\bibinfo {year} {1998})},\ \Eprint
  {http://arxiv.org/abs/gr-qc/9804014} {arXiv:gr-qc/9804014 [gr-qc]}
  \BibitemShut {NoStop}%
\bibitem [{\citenamefont {Goldberg}\ \emph {et~al.}(1967)\citenamefont
  {Goldberg} \emph {et~al.}}]{sharm}%
  \BibitemOpen
  \bibfield  {author} {\bibinfo {author} {\bibnamefont {Goldberg}} \emph
  {et~al.},\ }\href@noop {} {\bibfield  {journal} {\bibinfo  {journal} {J.
  Math. Phys. 8:2155--2161}\ } (\bibinfo {year} {1967})}\BibitemShut {NoStop}%
\bibitem [{\citenamefont {Healy}\ \emph {et~al.}(2013)\citenamefont {Healy},
  \citenamefont {Laguna}, \citenamefont {Pekowsky},\ and\ \citenamefont
  {Shoemaker}}]{Healy:2013jza}%
  \BibitemOpen
  \bibfield  {author} {\bibinfo {author} {\bibfnamefont {J.}~\bibnamefont
  {Healy}}, \bibinfo {author} {\bibfnamefont {P.}~\bibnamefont {Laguna}},
  \bibinfo {author} {\bibfnamefont {L.}~\bibnamefont {Pekowsky}}, \ and\
  \bibinfo {author} {\bibfnamefont {D.}~\bibnamefont {Shoemaker}},\ }\href
  {\doibase 10.1103/PhysRevD.88.024034} {\bibfield  {journal} {\bibinfo
  {journal} {Phys. Rev.}\ }\textbf {\bibinfo {volume} {D88}},\ \bibinfo {pages}
  {024034} (\bibinfo {year} {2013})},\ \Eprint {http://arxiv.org/abs/1302.6953}
  {arXiv:1302.6953 [gr-qc]} \BibitemShut {NoStop}%
\bibitem [{\citenamefont {Bustillo}\ \emph {et~al.}(2015)\citenamefont
  {Bustillo}, \citenamefont {Boh{\'e}}, \citenamefont {Husa}, \citenamefont
  {Sintes}, \citenamefont {Hannam} \emph {et~al.}}]{Bustillo:2015ova}%
  \BibitemOpen
  \bibfield  {author} {\bibinfo {author} {\bibfnamefont {J.~C.}\ \bibnamefont
  {Bustillo}}, \bibinfo {author} {\bibfnamefont {A.}~\bibnamefont {Boh{\'e}}},
  \bibinfo {author} {\bibfnamefont {S.}~\bibnamefont {Husa}}, \bibinfo {author}
  {\bibfnamefont {A.~M.}\ \bibnamefont {Sintes}}, \bibinfo {author}
  {\bibfnamefont {M.}~\bibnamefont {Hannam}},  \emph {et~al.},\ }\href@noop {}
  {\  (\bibinfo {year} {2015})},\ \Eprint {http://arxiv.org/abs/1501.00918}
  {arXiv:1501.00918 [gr-qc]} \BibitemShut {NoStop}%
\bibitem [{\citenamefont {Damour}\ \emph {et~al.}(2013)\citenamefont {Damour},
  \citenamefont {Nagar},\ and\ \citenamefont {Bernuzzi}}]{Damour:2012ky}%
  \BibitemOpen
  \bibfield  {author} {\bibinfo {author} {\bibfnamefont {T.}~\bibnamefont
  {Damour}}, \bibinfo {author} {\bibfnamefont {A.}~\bibnamefont {Nagar}}, \
  and\ \bibinfo {author} {\bibfnamefont {S.}~\bibnamefont {Bernuzzi}},\ }\href
  {\doibase 10.1103/PhysRevD.87.084035} {\bibfield  {journal} {\bibinfo
  {journal} {Phys.Rev.}\ }\textbf {\bibinfo {volume} {D87}},\ \bibinfo {pages}
  {084035} (\bibinfo {year} {2013})},\ \Eprint {http://arxiv.org/abs/1212.4357}
  {arXiv:1212.4357 [gr-qc]} \BibitemShut {NoStop}%
\bibitem [{\citenamefont {Santamaria}\ \emph {et~al.}(2010)\citenamefont
  {Santamaria}, \citenamefont {Ohme}, \citenamefont {Ajith}, \citenamefont
  {Bruegmann}, \citenamefont {Dorband} \emph {et~al.}}]{Santamaria:2010yb}%
  \BibitemOpen
  \bibfield  {author} {\bibinfo {author} {\bibfnamefont {L.}~\bibnamefont
  {Santamaria}}, \bibinfo {author} {\bibfnamefont {F.}~\bibnamefont {Ohme}},
  \bibinfo {author} {\bibfnamefont {P.}~\bibnamefont {Ajith}}, \bibinfo
  {author} {\bibfnamefont {B.}~\bibnamefont {Bruegmann}}, \bibinfo {author}
  {\bibfnamefont {N.}~\bibnamefont {Dorband}},  \emph {et~al.},\ }\href
  {\doibase 10.1103/PhysRevD.82.064016} {\bibfield  {journal} {\bibinfo
  {journal} {Phys.Rev.}\ }\textbf {\bibinfo {volume} {D82}},\ \bibinfo {pages}
  {064016} (\bibinfo {year} {2010})},\ \Eprint {http://arxiv.org/abs/1005.3306}
  {arXiv:1005.3306 [gr-qc]} \BibitemShut {NoStop}%
\bibitem [{\citenamefont {Hannam}\ \emph {et~al.}(2014)\citenamefont {Hannam},
  \citenamefont {Schmidt}, \citenamefont {Boh{\'e}}, \citenamefont {Haegel},
  \citenamefont {Husa} \emph {et~al.}}]{Hannam:2013oca}%
  \BibitemOpen
  \bibfield  {author} {\bibinfo {author} {\bibfnamefont {M.}~\bibnamefont
  {Hannam}}, \bibinfo {author} {\bibfnamefont {P.}~\bibnamefont {Schmidt}},
  \bibinfo {author} {\bibfnamefont {A.}~\bibnamefont {Boh{\'e}}}, \bibinfo
  {author} {\bibfnamefont {L.}~\bibnamefont {Haegel}}, \bibinfo {author}
  {\bibfnamefont {S.}~\bibnamefont {Husa}},  \emph {et~al.},\ }\href {\doibase
  10.1103/PhysRevLett.113.151101} {\bibfield  {journal} {\bibinfo  {journal}
  {Phys.Rev.Lett.}\ }\textbf {\bibinfo {volume} {113}},\ \bibinfo {pages}
  {151101} (\bibinfo {year} {2014})},\ \Eprint {http://arxiv.org/abs/1308.3271}
  {arXiv:1308.3271 [gr-qc]} \BibitemShut {NoStop}%
\bibitem [{\citenamefont {P{\"u}rrer}(2014)}]{Purrer:2014fza}%
  \BibitemOpen
  \bibfield  {author} {\bibinfo {author} {\bibfnamefont {M.}~\bibnamefont
  {P{\"u}rrer}},\ }\href {\doibase 10.1088/0264-9381/31/19/195010} {\bibfield
  {journal} {\bibinfo  {journal} {Class.Quant.Grav.}\ }\textbf {\bibinfo
  {volume} {31}},\ \bibinfo {pages} {195010} (\bibinfo {year} {2014})},\
  \Eprint {http://arxiv.org/abs/1402.4146} {arXiv:1402.4146 [gr-qc]}
  \BibitemShut {NoStop}%
\bibitem [{\citenamefont {Varma}\ and\ \citenamefont
  {Ajith}(2016)}]{Varma:2016dnf}%
  \BibitemOpen
  \bibfield  {author} {\bibinfo {author} {\bibfnamefont {V.}~\bibnamefont
  {Varma}}\ and\ \bibinfo {author} {\bibfnamefont {P.}~\bibnamefont {Ajith}},\
  }\href@noop {} {\  (\bibinfo {year} {2016})},\ \Eprint
  {http://arxiv.org/abs/1612.05608} {arXiv:1612.05608 [gr-qc]} \BibitemShut
  {NoStop}%
\bibitem [{\citenamefont {Kumar}\ \emph {et~al.}(2014)\citenamefont {Kumar}
  \emph {et~al.}}]{Kumar:2013gwa}%
  \BibitemOpen
  \bibfield  {author} {\bibinfo {author} {\bibfnamefont {P.}~\bibnamefont
  {Kumar}} \emph {et~al.},\ }\href {\doibase 10.1103/PhysRevD.89.042002}
  {\bibfield  {journal} {\bibinfo  {journal} {Phys. Rev.}\ }\textbf {\bibinfo
  {volume} {D89}},\ \bibinfo {pages} {042002} (\bibinfo {year} {2014})},\
  \Eprint {http://arxiv.org/abs/1310.7949} {arXiv:1310.7949 [gr-qc]}
  \BibitemShut {NoStop}%
\bibitem [{\citenamefont {Szilagyi}\ \emph {et~al.}(2015)\citenamefont
  {Szilagyi}, \citenamefont {Blackman}, \citenamefont {Buonanno}, \citenamefont
  {Taracchini}, \citenamefont {Pfeiffer} \emph {et~al.}}]{Szilagyi:2015rwa}%
  \BibitemOpen
  \bibfield  {author} {\bibinfo {author} {\bibfnamefont {B.}~\bibnamefont
  {Szilagyi}}, \bibinfo {author} {\bibfnamefont {J.}~\bibnamefont {Blackman}},
  \bibinfo {author} {\bibfnamefont {A.}~\bibnamefont {Buonanno}}, \bibinfo
  {author} {\bibfnamefont {A.}~\bibnamefont {Taracchini}}, \bibinfo {author}
  {\bibfnamefont {H.~P.}\ \bibnamefont {Pfeiffer}},  \emph {et~al.},\
  }\href@noop {} {\  (\bibinfo {year} {2015})},\ \Eprint
  {http://arxiv.org/abs/1502.04953} {arXiv:1502.04953 [gr-qc]} \BibitemShut
  {NoStop}%
\bibitem [{\citenamefont {Blackman}\ \emph
  {et~al.}(2017{\natexlab{a}})\citenamefont {Blackman}, \citenamefont {Field},
  \citenamefont {Scheel}, \citenamefont {Galley}, \citenamefont {Hemberger},
  \citenamefont {Schmidt},\ and\ \citenamefont {Smith}}]{Blackman:2017dfb}%
  \BibitemOpen
  \bibfield  {author} {\bibinfo {author} {\bibfnamefont {J.}~\bibnamefont
  {Blackman}}, \bibinfo {author} {\bibfnamefont {S.~E.}\ \bibnamefont {Field}},
  \bibinfo {author} {\bibfnamefont {M.~A.}\ \bibnamefont {Scheel}}, \bibinfo
  {author} {\bibfnamefont {C.~R.}\ \bibnamefont {Galley}}, \bibinfo {author}
  {\bibfnamefont {D.~A.}\ \bibnamefont {Hemberger}}, \bibinfo {author}
  {\bibfnamefont {P.}~\bibnamefont {Schmidt}}, \ and\ \bibinfo {author}
  {\bibfnamefont {R.}~\bibnamefont {Smith}},\ }\href {\doibase
  10.1103/PhysRevD.95.104023} {\bibfield  {journal} {\bibinfo  {journal} {Phys.
  Rev.}\ }\textbf {\bibinfo {volume} {D95}},\ \bibinfo {pages} {104023}
  (\bibinfo {year} {2017}{\natexlab{a}})},\ \Eprint
  {http://arxiv.org/abs/1701.00550} {arXiv:1701.00550 [gr-qc]} \BibitemShut
  {NoStop}%
\bibitem [{\citenamefont {Blackman}\ \emph
  {et~al.}(2017{\natexlab{b}})\citenamefont {Blackman}, \citenamefont {Field},
  \citenamefont {Scheel}, \citenamefont {Galley}, \citenamefont {Ott},
  \citenamefont {Boyle}, \citenamefont {Kidder}, \citenamefont {Pfeiffer},\
  and\ \citenamefont {Szilágyi}}]{Blackman:2017pcm}%
  \BibitemOpen
  \bibfield  {author} {\bibinfo {author} {\bibfnamefont {J.}~\bibnamefont
  {Blackman}}, \bibinfo {author} {\bibfnamefont {S.~E.}\ \bibnamefont {Field}},
  \bibinfo {author} {\bibfnamefont {M.~A.}\ \bibnamefont {Scheel}}, \bibinfo
  {author} {\bibfnamefont {C.~R.}\ \bibnamefont {Galley}}, \bibinfo {author}
  {\bibfnamefont {C.~D.}\ \bibnamefont {Ott}}, \bibinfo {author} {\bibfnamefont
  {M.}~\bibnamefont {Boyle}}, \bibinfo {author} {\bibfnamefont {L.~E.}\
  \bibnamefont {Kidder}}, \bibinfo {author} {\bibfnamefont {H.~P.}\
  \bibnamefont {Pfeiffer}}, \ and\ \bibinfo {author} {\bibfnamefont
  {B.}~\bibnamefont {Szilágyi}},\ }\href {\doibase 10.1103/PhysRevD.96.024058}
  {\bibfield  {journal} {\bibinfo  {journal} {Phys. Rev.}\ }\textbf {\bibinfo
  {volume} {D96}},\ \bibinfo {pages} {024058} (\bibinfo {year}
  {2017}{\natexlab{b}})},\ \Eprint {http://arxiv.org/abs/1705.07089}
  {arXiv:1705.07089 [gr-qc]} \BibitemShut {NoStop}%
\bibitem [{\citenamefont {Babak}\ \emph {et~al.}(2013)\citenamefont {Babak},
  \citenamefont {Biswas}, \citenamefont {Brady}, \citenamefont {Brown},
  \citenamefont {Cannon} \emph {et~al.}}]{Babak:2012zx}%
  \BibitemOpen
  \bibfield  {author} {\bibinfo {author} {\bibfnamefont {S.}~\bibnamefont
  {Babak}}, \bibinfo {author} {\bibfnamefont {R.}~\bibnamefont {Biswas}},
  \bibinfo {author} {\bibfnamefont {P.}~\bibnamefont {Brady}}, \bibinfo
  {author} {\bibfnamefont {D.}~\bibnamefont {Brown}}, \bibinfo {author}
  {\bibfnamefont {K.}~\bibnamefont {Cannon}},  \emph {et~al.},\ }\href
  {\doibase 10.1103/PhysRevD.87.024033} {\bibfield  {journal} {\bibinfo
  {journal} {Phys.Rev.}\ }\textbf {\bibinfo {volume} {D87}},\ \bibinfo {pages}
  {024033} (\bibinfo {year} {2013})},\ \Eprint {http://arxiv.org/abs/1208.3491}
  {arXiv:1208.3491 [gr-qc]} \BibitemShut {NoStop}%
\bibitem [{\citenamefont {Adams}\ \emph {et~al.}(2016)\citenamefont {Adams},
  \citenamefont {Buskulic}, \citenamefont {Germain}, \citenamefont {Guidi},
  \citenamefont {Marion}, \citenamefont {Montani}, \citenamefont {Mours},
  \citenamefont {Piergiovanni},\ and\ \citenamefont {Wang}}]{Adams:2015ulm}%
  \BibitemOpen
  \bibfield  {author} {\bibinfo {author} {\bibfnamefont {T.}~\bibnamefont
  {Adams}}, \bibinfo {author} {\bibfnamefont {D.}~\bibnamefont {Buskulic}},
  \bibinfo {author} {\bibfnamefont {V.}~\bibnamefont {Germain}}, \bibinfo
  {author} {\bibfnamefont {G.~M.}\ \bibnamefont {Guidi}}, \bibinfo {author}
  {\bibfnamefont {F.}~\bibnamefont {Marion}}, \bibinfo {author} {\bibfnamefont
  {M.}~\bibnamefont {Montani}}, \bibinfo {author} {\bibfnamefont
  {B.}~\bibnamefont {Mours}}, \bibinfo {author} {\bibfnamefont
  {F.}~\bibnamefont {Piergiovanni}}, \ and\ \bibinfo {author} {\bibfnamefont
  {G.}~\bibnamefont {Wang}},\ }\href {\doibase 10.1088/0264-9381/33/17/175012}
  {\bibfield  {journal} {\bibinfo  {journal} {Class. Quant. Grav.}\ }\textbf
  {\bibinfo {volume} {33}},\ \bibinfo {pages} {175012} (\bibinfo {year}
  {2016})},\ \Eprint {http://arxiv.org/abs/1512.02864} {arXiv:1512.02864
  [gr-qc]} \BibitemShut {NoStop}%
\bibitem [{\citenamefont {Allen}\ \emph {et~al.}(2012)\citenamefont {Allen},
  \citenamefont {Anderson}, \citenamefont {Brady}, \citenamefont {Brown},\ and\
  \citenamefont {Creighton}}]{Allen:2005fk}%
  \BibitemOpen
  \bibfield  {author} {\bibinfo {author} {\bibfnamefont {B.}~\bibnamefont
  {Allen}}, \bibinfo {author} {\bibfnamefont {W.~G.}\ \bibnamefont {Anderson}},
  \bibinfo {author} {\bibfnamefont {P.~R.}\ \bibnamefont {Brady}}, \bibinfo
  {author} {\bibfnamefont {D.~A.}\ \bibnamefont {Brown}}, \ and\ \bibinfo
  {author} {\bibfnamefont {J.~D.~E.}\ \bibnamefont {Creighton}},\ }\href
  {\doibase 10.1103/PhysRevD.85.122006} {\bibfield  {journal} {\bibinfo
  {journal} {Phys. Rev.}\ }\textbf {\bibinfo {volume} {D85}},\ \bibinfo {pages}
  {122006} (\bibinfo {year} {2012})},\ \Eprint
  {http://arxiv.org/abs/gr-qc/0509116} {arXiv:gr-qc/0509116 [gr-qc]}
  \BibitemShut {NoStop}%
\bibitem [{\citenamefont {Pan}\ \emph {et~al.}(2004)\citenamefont {Pan},
  \citenamefont {Buonanno}, \citenamefont {Chen},\ and\ \citenamefont
  {Vallisneri}}]{Pan:2003qt}%
  \BibitemOpen
  \bibfield  {author} {\bibinfo {author} {\bibfnamefont {Y.}~\bibnamefont
  {Pan}}, \bibinfo {author} {\bibfnamefont {A.}~\bibnamefont {Buonanno}},
  \bibinfo {author} {\bibfnamefont {Y.-b.}\ \bibnamefont {Chen}}, \ and\
  \bibinfo {author} {\bibfnamefont {M.}~\bibnamefont {Vallisneri}},\ }\href
  {\doibase 10.1103/PhysRevD.69.104017, 10.1103/PhysRevD.74.029905} {\bibfield
  {journal} {\bibinfo  {journal} {Phys. Rev.}\ }\textbf {\bibinfo {volume}
  {D69}},\ \bibinfo {pages} {104017} (\bibinfo {year} {2004})},\ \bibinfo
  {note} {[Erratum: Phys. Rev.D74,029905(2006)]},\ \Eprint
  {http://arxiv.org/abs/gr-qc/0310034} {arXiv:gr-qc/0310034 [gr-qc]}
  \BibitemShut {NoStop}%
\bibitem [{\citenamefont {Harry}\ \emph {et~al.}(2016)\citenamefont {Harry},
  \citenamefont {Privitera}, \citenamefont {Boh{\'e}},\ and\ \citenamefont
  {Buonanno}}]{Harry:2016ijz}%
  \BibitemOpen
  \bibfield  {author} {\bibinfo {author} {\bibfnamefont {I.}~\bibnamefont
  {Harry}}, \bibinfo {author} {\bibfnamefont {S.}~\bibnamefont {Privitera}},
  \bibinfo {author} {\bibfnamefont {A.}~\bibnamefont {Boh{\'e}}}, \ and\
  \bibinfo {author} {\bibfnamefont {A.}~\bibnamefont {Buonanno}},\ }\href
  {\doibase 10.1103/PhysRevD.94.024012} {\bibfield  {journal} {\bibinfo
  {journal} {Phys. Rev.}\ }\textbf {\bibinfo {volume} {D94}},\ \bibinfo {pages}
  {024012} (\bibinfo {year} {2016})},\ \Eprint
  {http://arxiv.org/abs/1603.02444} {arXiv:1603.02444 [gr-qc]} \BibitemShut
  {NoStop}%
\bibitem [{\citenamefont {Schmidt}\ \emph {et~al.}(2015)\citenamefont
  {Schmidt}, \citenamefont {Ohme},\ and\ \citenamefont
  {Hannam}}]{Schmidt:2014iyl}%
  \BibitemOpen
  \bibfield  {author} {\bibinfo {author} {\bibfnamefont {P.}~\bibnamefont
  {Schmidt}}, \bibinfo {author} {\bibfnamefont {F.}~\bibnamefont {Ohme}}, \
  and\ \bibinfo {author} {\bibfnamefont {M.}~\bibnamefont {Hannam}},\ }\href
  {\doibase 10.1103/PhysRevD.91.024043} {\bibfield  {journal} {\bibinfo
  {journal} {Phys. Rev.}\ }\textbf {\bibinfo {volume} {D91}},\ \bibinfo {pages}
  {024043} (\bibinfo {year} {2015})},\ \Eprint {http://arxiv.org/abs/1408.1810}
  {arXiv:1408.1810 [gr-qc]} \BibitemShut {NoStop}%
\bibitem [{\citenamefont {Maggiore}(2007)}]{Maggiore:1900zz}%
  \BibitemOpen
  \bibfield  {author} {\bibinfo {author} {\bibfnamefont {M.}~\bibnamefont
  {Maggiore}},\ }\href {http://www.oup.com/uk/catalogue/?ci=9780198570745}
  {\emph {\bibinfo {title} {{Gravitational Waves. Vol. 1: Theory and
  Experiments}}}},\ Oxford Master Series in Physics\ (\bibinfo  {publisher}
  {Oxford University Press},\ \bibinfo {year} {2007})\BibitemShut {NoStop}%
\bibitem [{\citenamefont {Creighton}\ and\ \citenamefont
  {Anderson}(2011)}]{Creighton:2011zz}%
  \BibitemOpen
  \bibfield  {author} {\bibinfo {author} {\bibfnamefont {J.~D.~E.}\
  \bibnamefont {Creighton}}\ and\ \bibinfo {author} {\bibfnamefont {W.~G.}\
  \bibnamefont {Anderson}},\ }\href
  {http://www.wiley-vch.de/publish/dt/books/ISBN3-527-40886-X} {\emph {\bibinfo
  {title} {{Gravitational-wave physics and astronomy: An introduction to
  theory, experiment and data analysis}}}}\ (\bibinfo {year}
  {2011})\BibitemShut {NoStop}%
\bibitem [{\citenamefont {Dhurandhar}\ \emph {et~al.}(2017)\citenamefont
  {Dhurandhar}, \citenamefont {Krishnan},\ and\ \citenamefont
  {Willis}}]{Dhurandhar:2017rlr}%
  \BibitemOpen
  \bibfield  {author} {\bibinfo {author} {\bibfnamefont {S.}~\bibnamefont
  {Dhurandhar}}, \bibinfo {author} {\bibfnamefont {B.}~\bibnamefont
  {Krishnan}}, \ and\ \bibinfo {author} {\bibfnamefont {J.~L.}\ \bibnamefont
  {Willis}},\ }\href@noop {} {\  (\bibinfo {year} {2017})},\ \Eprint
  {http://arxiv.org/abs/1707.08163} {arXiv:1707.08163 [gr-qc]} \BibitemShut
  {NoStop}%
\bibitem [{\citenamefont {Nitz}\ \emph
  {et~al.}(2017{\natexlab{b}})\citenamefont {Nitz}, \citenamefont {Harry},
  \citenamefont {Biwer}, \citenamefont {Brown}, \citenamefont {Willis} \emph
  {et~al.}}]{alex_nitz_2017_826910}%
  \BibitemOpen
  \bibfield  {author} {\bibinfo {author} {\bibfnamefont {A.}~\bibnamefont
  {Nitz}}, \bibinfo {author} {\bibfnamefont {I.}~\bibnamefont {Harry}},
  \bibinfo {author} {\bibfnamefont {C.~M.}\ \bibnamefont {Biwer}}, \bibinfo
  {author} {\bibfnamefont {D.}~\bibnamefont {Brown}}, \bibinfo {author}
  {\bibfnamefont {J.}~\bibnamefont {Willis}},  \emph {et~al.},\ }\href
  {\doibase 10.5281/zenodo.826910} {\enquote {\bibinfo {title} {ligo-cbc/pycbc:
  O2 production release 14},}\ } (\bibinfo {year}
  {2017}{\natexlab{b}})\BibitemShut {NoStop}%
\bibitem [{\citenamefont {Dal~Canton}\ and\ \citenamefont
  {Harry}(2017)}]{DalCanton:2017ala}%
  \BibitemOpen
  \bibfield  {author} {\bibinfo {author} {\bibfnamefont {T.}~\bibnamefont
  {Dal~Canton}}\ and\ \bibinfo {author} {\bibfnamefont {I.~W.}\ \bibnamefont
  {Harry}},\ }\href@noop {} {\  (\bibinfo {year} {2017})},\ \Eprint
  {http://arxiv.org/abs/1705.01845} {arXiv:1705.01845 [gr-qc]} \BibitemShut
  {NoStop}%
\bibitem [{\citenamefont {Harry}\ \emph {et~al.}(2009)\citenamefont {Harry},
  \citenamefont {Allen},\ and\ \citenamefont {Sathyaprakash}}]{Harry:2009ea}%
  \BibitemOpen
  \bibfield  {author} {\bibinfo {author} {\bibfnamefont {I.~W.}\ \bibnamefont
  {Harry}}, \bibinfo {author} {\bibfnamefont {B.}~\bibnamefont {Allen}}, \ and\
  \bibinfo {author} {\bibfnamefont {B.~S.}\ \bibnamefont {Sathyaprakash}},\
  }\href {\doibase 10.1103/PhysRevD.80.104014} {\bibfield  {journal} {\bibinfo
  {journal} {Phys. Rev.}\ }\textbf {\bibinfo {volume} {D80}},\ \bibinfo {pages}
  {104014} (\bibinfo {year} {2009})},\ \Eprint {http://arxiv.org/abs/0908.2090}
  {arXiv:0908.2090 [gr-qc]} \BibitemShut {NoStop}%
\bibitem [{\citenamefont {Privitera}\ \emph {et~al.}(2014)\citenamefont
  {Privitera}, \citenamefont {Mohapatra}, \citenamefont {Ajith}, \citenamefont
  {Cannon}, \citenamefont {Fotopoulos} \emph {et~al.}}]{Privitera:2013xza}%
  \BibitemOpen
  \bibfield  {author} {\bibinfo {author} {\bibfnamefont {S.}~\bibnamefont
  {Privitera}}, \bibinfo {author} {\bibfnamefont {S.~R.~P.}\ \bibnamefont
  {Mohapatra}}, \bibinfo {author} {\bibfnamefont {P.}~\bibnamefont {Ajith}},
  \bibinfo {author} {\bibfnamefont {K.}~\bibnamefont {Cannon}}, \bibinfo
  {author} {\bibfnamefont {N.}~\bibnamefont {Fotopoulos}},  \emph {et~al.},\
  }\href {\doibase 10.1103/PhysRevD.89.024003} {\bibfield  {journal} {\bibinfo
  {journal} {Phys.Rev.}\ }\textbf {\bibinfo {volume} {D89}},\ \bibinfo {pages}
  {024003} (\bibinfo {year} {2014})},\ \Eprint {http://arxiv.org/abs/1310.5633}
  {arXiv:1310.5633 [gr-qc]} \BibitemShut {NoStop}%
\bibitem [{\citenamefont {Harry}\ \emph {et~al.}(2014)\citenamefont {Harry},
  \citenamefont {Nitz}, \citenamefont {Brown}, \citenamefont {Lundgren},
  \citenamefont {Ochsner},\ and\ \citenamefont {Keppel}}]{Harry:2013tca}%
  \BibitemOpen
  \bibfield  {author} {\bibinfo {author} {\bibfnamefont {I.~W.}\ \bibnamefont
  {Harry}}, \bibinfo {author} {\bibfnamefont {A.~H.}\ \bibnamefont {Nitz}},
  \bibinfo {author} {\bibfnamefont {D.~A.}\ \bibnamefont {Brown}}, \bibinfo
  {author} {\bibfnamefont {A.~P.}\ \bibnamefont {Lundgren}}, \bibinfo {author}
  {\bibfnamefont {E.}~\bibnamefont {Ochsner}}, \ and\ \bibinfo {author}
  {\bibfnamefont {D.}~\bibnamefont {Keppel}},\ }\href {\doibase
  10.1103/PhysRevD.89.024010} {\bibfield  {journal} {\bibinfo  {journal} {Phys.
  Rev.}\ }\textbf {\bibinfo {volume} {D89}},\ \bibinfo {pages} {024010}
  (\bibinfo {year} {2014})},\ \Eprint {http://arxiv.org/abs/1307.3562}
  {arXiv:1307.3562 [gr-qc]} \BibitemShut {NoStop}%
\bibitem [{\citenamefont {Buonanno}\ \emph {et~al.}(2003)\citenamefont
  {Buonanno}, \citenamefont {Chen},\ and\ \citenamefont
  {Vallisneri}}]{Buonanno:2002fy}%
  \BibitemOpen
  \bibfield  {author} {\bibinfo {author} {\bibfnamefont {A.}~\bibnamefont
  {Buonanno}}, \bibinfo {author} {\bibfnamefont {Y.-b.}\ \bibnamefont {Chen}},
  \ and\ \bibinfo {author} {\bibfnamefont {M.}~\bibnamefont {Vallisneri}},\
  }\href {\doibase 10.1103/PhysRevD.67.104025, 10.1103/PhysRevD.67.104025
  10.1103/PhysRevD.74.029904, 10.1103/PhysRevD.74.029904} {\bibfield  {journal}
  {\bibinfo  {journal} {Phys.Rev.}\ }\textbf {\bibinfo {volume} {D67}},\
  \bibinfo {pages} {104025} (\bibinfo {year} {2003})},\ \Eprint
  {http://arxiv.org/abs/gr-qc/0211087} {arXiv:gr-qc/0211087 [gr-qc]}
  \BibitemShut {NoStop}%
\bibitem [{\citenamefont {Graff}\ \emph {et~al.}(2015)\citenamefont {Graff},
  \citenamefont {Buonanno},\ and\ \citenamefont
  {Sathyaprakash}}]{Graff:2015bba}%
  \BibitemOpen
  \bibfield  {author} {\bibinfo {author} {\bibfnamefont {P.~B.}\ \bibnamefont
  {Graff}}, \bibinfo {author} {\bibfnamefont {A.}~\bibnamefont {Buonanno}}, \
  and\ \bibinfo {author} {\bibfnamefont {B.}~\bibnamefont {Sathyaprakash}},\
  }\href {\doibase 10.1103/PhysRevD.92.022002} {\bibfield  {journal} {\bibinfo
  {journal} {Phys. Rev.}\ }\textbf {\bibinfo {volume} {D92}},\ \bibinfo {pages}
  {022002} (\bibinfo {year} {2015})},\ \Eprint
  {http://arxiv.org/abs/1504.04766} {arXiv:1504.04766 [gr-qc]} \BibitemShut
  {NoStop}%
\bibitem [{\citenamefont {Dent}\ and\ \citenamefont
  {Veitch}(2014)}]{Dent:2013cva}%
  \BibitemOpen
  \bibfield  {author} {\bibinfo {author} {\bibfnamefont {T.}~\bibnamefont
  {Dent}}\ and\ \bibinfo {author} {\bibfnamefont {J.}~\bibnamefont {Veitch}},\
  }\href {\doibase 10.1103/PhysRevD.89.062002} {\bibfield  {journal} {\bibinfo
  {journal} {Phys. Rev.}\ }\textbf {\bibinfo {volume} {D89}},\ \bibinfo {pages}
  {062002} (\bibinfo {year} {2014})},\ \Eprint {http://arxiv.org/abs/1311.7174}
  {arXiv:1311.7174 [gr-qc]} \BibitemShut {NoStop}%
\bibitem [{\citenamefont {Aasi}\ \emph {et~al.}(2012)\citenamefont {Aasi} \emph
  {et~al.}}]{Aasi:2012wd}%
  \BibitemOpen
  \bibfield  {author} {\bibinfo {author} {\bibfnamefont {J.}~\bibnamefont
  {Aasi}} \emph {et~al.} (\bibinfo {collaboration} {VIRGO}),\ }\href {\doibase
  10.1088/0264-9381/29/15/155002} {\bibfield  {journal} {\bibinfo  {journal}
  {Class. Quant. Grav.}\ }\textbf {\bibinfo {volume} {29}},\ \bibinfo {pages}
  {155002} (\bibinfo {year} {2012})},\ \Eprint {http://arxiv.org/abs/1203.5613}
  {arXiv:1203.5613 [gr-qc]} \BibitemShut {NoStop}%
\bibitem [{\citenamefont {Aasi}\ \emph
  {et~al.}(2015{\natexlab{b}})\citenamefont {Aasi} \emph
  {et~al.}}]{Aasi:2014mqd}%
  \BibitemOpen
  \bibfield  {author} {\bibinfo {author} {\bibfnamefont {J.}~\bibnamefont
  {Aasi}} \emph {et~al.} (\bibinfo {collaboration} {VIRGO, LIGO Scientific}),\
  }\href {\doibase 10.1088/0264-9381/32/11/115012} {\bibfield  {journal}
  {\bibinfo  {journal} {Class. Quant. Grav.}\ }\textbf {\bibinfo {volume}
  {32}},\ \bibinfo {pages} {115012} (\bibinfo {year} {2015}{\natexlab{b}})},\
  \Eprint {http://arxiv.org/abs/1410.7764} {arXiv:1410.7764 [gr-qc]}
  \BibitemShut {NoStop}%
\bibitem [{\citenamefont {Abbott}\ \emph
  {et~al.}(2016{\natexlab{g}})\citenamefont {Abbott} \emph
  {et~al.}}]{TheLIGOScientific:2016zmo}%
  \BibitemOpen
  \bibfield  {author} {\bibinfo {author} {\bibfnamefont {B.~P.}\ \bibnamefont
  {Abbott}} \emph {et~al.} (\bibinfo {collaboration} {Virgo, LIGO
  Scientific}),\ }\href {\doibase 10.1088/0264-9381/33/13/134001} {\bibfield
  {journal} {\bibinfo  {journal} {Class. Quant. Grav.}\ }\textbf {\bibinfo
  {volume} {33}},\ \bibinfo {pages} {134001} (\bibinfo {year}
  {2016}{\natexlab{g}})},\ \Eprint {http://arxiv.org/abs/1602.03844}
  {arXiv:1602.03844 [gr-qc]} \BibitemShut {NoStop}%
\bibitem [{\citenamefont {Harry}\ and\ \citenamefont
  {Fairhurst}(2011)}]{Harry:2010fr}%
  \BibitemOpen
  \bibfield  {author} {\bibinfo {author} {\bibfnamefont {I.~W.}\ \bibnamefont
  {Harry}}\ and\ \bibinfo {author} {\bibfnamefont {S.}~\bibnamefont
  {Fairhurst}},\ }\href {\doibase 10.1103/PhysRevD.83.084002} {\bibfield
  {journal} {\bibinfo  {journal} {Phys. Rev.}\ }\textbf {\bibinfo {volume}
  {D83}},\ \bibinfo {pages} {084002} (\bibinfo {year} {2011})},\ \Eprint
  {http://arxiv.org/abs/1012.4939} {arXiv:1012.4939 [gr-qc]} \BibitemShut
  {NoStop}%
\bibitem [{\citenamefont {Biswas}\ \emph {et~al.}(2013)\citenamefont {Biswas}
  \emph {et~al.}}]{Biswas:2013wfa}%
  \BibitemOpen
  \bibfield  {author} {\bibinfo {author} {\bibfnamefont {R.}~\bibnamefont
  {Biswas}} \emph {et~al.},\ }\href {\doibase 10.1103/PhysRevD.88.062003}
  {\bibfield  {journal} {\bibinfo  {journal} {Phys. Rev.}\ }\textbf {\bibinfo
  {volume} {D88}},\ \bibinfo {pages} {062003} (\bibinfo {year} {2013})},\
  \Eprint {http://arxiv.org/abs/1303.6984} {arXiv:1303.6984 [astro-ph.IM]}
  \BibitemShut {NoStop}%
\bibitem [{\citenamefont {Messick}\ \emph {et~al.}(2017)\citenamefont {Messick}
  \emph {et~al.}}]{Messick:2016aqy}%
  \BibitemOpen
  \bibfield  {author} {\bibinfo {author} {\bibfnamefont {C.}~\bibnamefont
  {Messick}} \emph {et~al.},\ }\href {\doibase 10.1103/PhysRevD.95.042001}
  {\bibfield  {journal} {\bibinfo  {journal} {Phys. Rev.}\ }\textbf {\bibinfo
  {volume} {D95}},\ \bibinfo {pages} {042001} (\bibinfo {year} {2017})},\
  \Eprint {http://arxiv.org/abs/1604.04324} {arXiv:1604.04324 [astro-ph.IM]}
  \BibitemShut {NoStop}%
\bibitem [{\citenamefont {Abbott}\ \emph
  {et~al.}(2017{\natexlab{b}})\citenamefont {Abbott} \emph
  {et~al.}}]{Abbott:2017iws}%
  \BibitemOpen
  \bibfield  {author} {\bibinfo {author} {\bibfnamefont {B.~P.}\ \bibnamefont
  {Abbott}} \emph {et~al.} (\bibinfo {collaboration} {Virgo, LIGO
  Scientific}),\ }\href {\doibase 10.1103/PhysRevD.96.022001} {\bibfield
  {journal} {\bibinfo  {journal} {Phys. Rev.}\ }\textbf {\bibinfo {volume}
  {D96}},\ \bibinfo {pages} {022001} (\bibinfo {year} {2017}{\natexlab{b}})},\
  \Eprint {http://arxiv.org/abs/1704.04628} {arXiv:1704.04628 [gr-qc]}
  \BibitemShut {NoStop}%
\bibitem [{\citenamefont {Nitz}(2017)}]{Nitz:2017lco}%
  \BibitemOpen
  \bibfield  {author} {\bibinfo {author} {\bibfnamefont {A.~H.}\ \bibnamefont
  {Nitz}},\ }\href@noop {} {\  (\bibinfo {year} {2017})},\ \Eprint
  {http://arxiv.org/abs/1709.08974} {arXiv:1709.08974 [gr-qc]} \BibitemShut
  {NoStop}%
\bibitem [{\citenamefont {Mohapatra}\ \emph {et~al.}(2014)\citenamefont
  {Mohapatra}, \citenamefont {Cadonati}, \citenamefont {Caudill}, \citenamefont
  {Clark}, \citenamefont {Hanna}, \citenamefont {Klimenko}, \citenamefont
  {Pankow}, \citenamefont {Vaulin}, \citenamefont {Vedovato},\ and\
  \citenamefont {Vitale}}]{Mohapatra:2014rda}%
  \BibitemOpen
  \bibfield  {author} {\bibinfo {author} {\bibfnamefont {S.}~\bibnamefont
  {Mohapatra}}, \bibinfo {author} {\bibfnamefont {L.}~\bibnamefont {Cadonati}},
  \bibinfo {author} {\bibfnamefont {S.}~\bibnamefont {Caudill}}, \bibinfo
  {author} {\bibfnamefont {J.}~\bibnamefont {Clark}}, \bibinfo {author}
  {\bibfnamefont {C.}~\bibnamefont {Hanna}}, \bibinfo {author} {\bibfnamefont
  {S.}~\bibnamefont {Klimenko}}, \bibinfo {author} {\bibfnamefont
  {C.}~\bibnamefont {Pankow}}, \bibinfo {author} {\bibfnamefont
  {R.}~\bibnamefont {Vaulin}}, \bibinfo {author} {\bibfnamefont
  {G.}~\bibnamefont {Vedovato}}, \ and\ \bibinfo {author} {\bibfnamefont
  {S.}~\bibnamefont {Vitale}},\ }\href {\doibase 10.1103/PhysRevD.90.022001}
  {\bibfield  {journal} {\bibinfo  {journal} {Phys. Rev.}\ }\textbf {\bibinfo
  {volume} {D90}},\ \bibinfo {pages} {022001} (\bibinfo {year} {2014})},\
  \Eprint {http://arxiv.org/abs/1405.6589} {arXiv:1405.6589 [gr-qc]}
  \BibitemShut {NoStop}%
\bibitem [{\citenamefont {Calderon~Bustillo}\ \emph {et~al.}(2017)\citenamefont
  {Calderon~Bustillo}, \citenamefont {Salemi}, \citenamefont {Dal~Canton},\
  and\ \citenamefont {Jani}}]{Bustillo:pycbccwb}%
  \BibitemOpen
  \bibfield  {author} {\bibinfo {author} {\bibfnamefont {J.}~\bibnamefont
  {Calderon~Bustillo}}, \bibinfo {author} {\bibfnamefont {F.}~\bibnamefont
  {Salemi}}, \bibinfo {author} {\bibfnamefont {T.}~\bibnamefont {Dal~Canton}},
  \ and\ \bibinfo {author} {\bibfnamefont {K.~P.}\ \bibnamefont {Jani}},\
  }\href {https://dcc.ligo.org/LIGO-P1700204} {\emph {\bibinfo {title}
  {{Sensitivity of gravitational wave searches to the full signal of
  intermediate mass black hole binaries during the LIGO O1 Science Run}}}},\
  \bibinfo {type} {Tech. Rep.}\ \bibinfo {number} {{LIGO}-P1700204}\ (\bibinfo
  {institution} {{LIGO} Project},\ \bibinfo {year} {2017})\BibitemShut
  {NoStop}%
\bibitem [{\citenamefont {Abadie}\ \emph {et~al.}(2012)\citenamefont {Abadie}
  \emph {et~al.}}]{Colaboration:2011np}%
  \BibitemOpen
  \bibfield  {author} {\bibinfo {author} {\bibfnamefont {J.}~\bibnamefont
  {Abadie}} \emph {et~al.} (\bibinfo {collaboration} {LIGO Collaboration, Virgo
  Collaboration}),\ }\href {\doibase 10.1103/PhysRevD.85.082002} {\bibfield
  {journal} {\bibinfo  {journal} {Phys.Rev.}\ }\textbf {\bibinfo {volume}
  {D85}},\ \bibinfo {pages} {082002} (\bibinfo {year} {2012})},\ \Eprint
  {http://arxiv.org/abs/1111.7314} {arXiv:1111.7314 [gr-qc]} \BibitemShut
  {NoStop}%
\bibitem [{\citenamefont {Dreyer}\ \emph {et~al.}(2004)\citenamefont {Dreyer},
  \citenamefont {Kelly}, \citenamefont {Krishnan}, \citenamefont {Finn},
  \citenamefont {Garrison},\ and\ \citenamefont
  {Lopez-Aleman}}]{Dreyer:2003bv}%
  \BibitemOpen
  \bibfield  {author} {\bibinfo {author} {\bibfnamefont {O.}~\bibnamefont
  {Dreyer}}, \bibinfo {author} {\bibfnamefont {B.~J.}\ \bibnamefont {Kelly}},
  \bibinfo {author} {\bibfnamefont {B.}~\bibnamefont {Krishnan}}, \bibinfo
  {author} {\bibfnamefont {L.~S.}\ \bibnamefont {Finn}}, \bibinfo {author}
  {\bibfnamefont {D.}~\bibnamefont {Garrison}}, \ and\ \bibinfo {author}
  {\bibfnamefont {R.}~\bibnamefont {Lopez-Aleman}},\ }\href {\doibase
  10.1088/0264-9381/21/4/003} {\bibfield  {journal} {\bibinfo  {journal}
  {Class. Quant. Grav.}\ }\textbf {\bibinfo {volume} {21}},\ \bibinfo {pages}
  {787} (\bibinfo {year} {2004})},\ \Eprint
  {http://arxiv.org/abs/gr-qc/0309007} {arXiv:gr-qc/0309007 [gr-qc]}
  \BibitemShut {NoStop}%
\bibitem [{\citenamefont {Maselli}\ \emph {et~al.}(2017)\citenamefont
  {Maselli}, \citenamefont {Kokkotas},\ and\ \citenamefont
  {Laguna}}]{Maselli:2017kvl}%
  \BibitemOpen
  \bibfield  {author} {\bibinfo {author} {\bibfnamefont {A.}~\bibnamefont
  {Maselli}}, \bibinfo {author} {\bibfnamefont {K.}~\bibnamefont {Kokkotas}}, \
  and\ \bibinfo {author} {\bibfnamefont {P.}~\bibnamefont {Laguna}},\ }\href
  {\doibase 10.1103/PhysRevD.95.104026} {\bibfield  {journal} {\bibinfo
  {journal} {Phys. Rev.}\ }\textbf {\bibinfo {volume} {D95}},\ \bibinfo {pages}
  {104026} (\bibinfo {year} {2017})},\ \Eprint
  {http://arxiv.org/abs/1702.01110} {arXiv:1702.01110 [gr-qc]} \BibitemShut
  {NoStop}%
\end{thebibliography}%

\end{document}